\documentclass[12pt]{article}
\usepackage{comment}
\usepackage{amsmath}
\usepackage{amsthm}
\usepackage{amssymb}
\usepackage{appendix}
\usepackage{mathrsfs}
\usepackage{enumerate}
\usepackage{mathtools}
\usepackage{multirow}
\usepackage{dsfont}
\usepackage{mdframed}
\usepackage{algorithm}
\usepackage[noend]{algpseudocode}
\usepackage{bm}
\usepackage{bbm}
\usepackage{natbib}
\setcitestyle{aysep={}} 
\usepackage{url}
\usepackage[dvipsnames]{xcolor}
\usepackage[colorlinks,linkcolor=Magenta,citecolor=Magenta,urlcolor = Magenta]{hyperref}
\hypersetup{colorlinks,allcolors=blue}
\usepackage[labelsep=period, font=sc, skip=0pt,justification=centering]{caption}
\usepackage{graphicx}
\usepackage{pdfpages}
\usepackage{setspace}
\usepackage{float}
\usepackage[top=0.9in, bottom=0.9in, left=1.0in, right=1.0in]{geometry}
\usepackage{makecell}
\usepackage{subcaption}
\usepackage[hang,flushmargin]{footmisc}
\usepackage{titling}
\usepackage{tikz}
\usetikzlibrary{shapes,positioning}

\tikzset{ell/.style={rectangle,draw,minimum height=0.65cm,minimum width=1cm,inner sep=0.25cm}}

\usepackage{titlesec}
\titlelabel{\thetitle.\ }

\newtheorem{theorem}{Theorem}

\newtheorem{corollary}{Corollary}
\newtheorem{condition}{Condition}

\newtheorem{lemma}{Lemma}

\newtheorem{remark}{Remark}

\newcommand{\be}{\begin{eqnarray*}}
\newcommand{\ee}{\end{eqnarray*}}

\graphicspath{{../code/R/graphs/paper/}}%

\usepackage{stackengine}
\usepackage{blindtext}
\usepackage{authblk}

\begin{document}
\title{Inference on High Dimensional \\ Selective Labeling Models}

\author[1]{Shakeeb Khan}
\author[2]{Elie Tamer}
\author[3]{Qingsong Yao}

\affil[1]{Department of Economics, Boston College}
\affil[2]{Deptartment of Economics, Harvard University}
  \affil[3]{Department of Economics, Louisiana State University}

\date{  \today}

\maketitle
\vskip -.5in
{\footnotesize \singlespace

\begin{center}
    {\bf Abstract}
\end{center} 

The paper reconsiders the problem of inference on parameters in binary outcome models when these outcomes are subject to possibly endogenous censoring.  Recently, these models have gained increasing interest in the computer science and machine learning literatures where the issue of endogenous sample selection is referred to as the {selective labels } problem. Such models are relevant in diverse empirical settings, including criminal justice, healthcare, and insurance. Notable recent studies in this area include \citet{lakkarujuetal2017}, \citet{kleinbergetal}, and \citet*{asheshsl}, which examine judicial bail decisions—where the outcome of whether a defendant fails to appear in court is observed only if the judge grants bail. Inference on such model parameters can be computationally challenging for two reasons. One is the nonconcavity of the bivariate likelihood function,
and the other is the large number of covariates in each equation. Despite these hurdles,  we propose a novel  distribution free estimation procedure
that is computationally friendly especially in the many covariates settings. The new method combines the semiparametric batched gradient descent algorithm introduced in \citet*{khanetal2021} with a novel sorting algorithm incorporated  to control for selection bias. Asymptotic properties of the new procedure are established under increasing dimension conditions in both equations, and its finite sample properties are explored through  a simulation study and an application using  Stanford Open Policing Project data set \citep{pierson2020large}. Extensions to models with endogenous treatment are also proposed.

\noindent



\newpage

\setcounter{page}{1}

\setlength{\abovedisplayshortskip}{5pt}
\setlength{\belowdisplayshortskip}{5pt} 
\setlength{\abovedisplayskip}{5pt}
\setlength{\belowdisplayskip}{5pt}

\normalsize

\setcounter{equation}{0}

\begin{spacing}{1.25}
\section{Introduction}

This paper addresses the challenge of inference on large-dimensional selective labeling models with binary outcomes. These models emerge in numerous domains where the observed binary outcomes result from the choices made by one of the agents within the system. Recently, they have garnered significant attention in the fields of computer science and machine learning, where this issue is known as the ``selective labels problem'', or as an  endogenous sample selection problem. Applications of these models span a wide range of areas, including criminal justice, healthcare, and insurance. For important recent work in this area, see for example \cite{lakkarujuetal2017}, \cite{kleinbergetal} and \cite{asheshsl}. These authors  focus on judicial bail decisions, where one observes the outcome of whether a defendant filed to return for their court appearance only if the judge in the case decides to release the defendant on bail. Letting $D_i$ denote the binary
decision to grant bail, and $Y_i$ denote the binary outcome of the defendant returning for court appearance, they consider a model of the form
\begin{equation}
Y_i=
\begin{cases}
0 \mbox{ or } 1, \ \ \ \ \ \ \ \ \ \ \ \ \ \ \ \ \ \ \ \ \ \  \mbox{ if } D_i=1 \\
\mbox{not observed (NA)}, \ \ \  \ \  \mbox{ otherwise }
\end{cases}
\end{equation}
This process and the ensuing model can be best explained with the diagram below.
The top node indicates the decision made by the agent which corresponds to a {\em yes} ($D_i=1$) or {\em no} ($D_i=0$) on individual $i$.
The other observed dependent variable, corresponding to the two nodes beneath the top one, is denoted by $Y_i$, where $Y_i\in \{ 0,1,NA\}$ and denotes the resulting outcome (return to court in our example). The selective labels problem occurs because the observation of outcome $Y_i$ is constrained by the decision $D_i$ made by the agent:

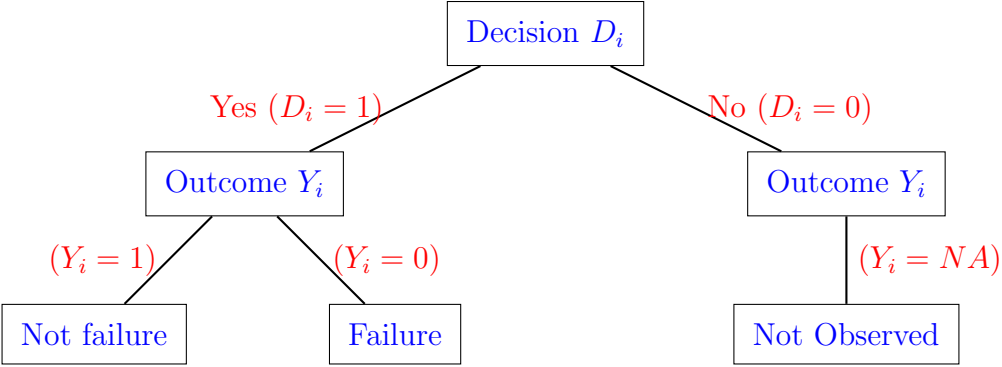
\begin{center}
\begin{tikzpicture}[>=stealth]
\node[ell] (e4)at (4,4) {\color{blue}Outcome $Y_i$};
\node[ell] (e3)at (-4,4) {\color{blue}Outcome $Y_i$};
\node[ell] (e5)at (-0,6) {\color{blue}Decision $D_i$};
\node[ell] (e6) at (-6,2) {\color{blue} Not failure};
\node[ell] (e7) at (-2,2) {\color{blue}  Failure};
\node[ell] (e8) at (4,2) {\color{blue} Not Observed};

\draw [thick] (e5) to node[left]{\color{red}Yes $(D_i=1)$} (e3);
\draw [thick] (e5) to node[right]{\color{red}No $(D_i=0)$} (e4);
\draw[thick] (e3) to node[left]{\color{red} $(Y_i=1)$} (e6);
\draw[thick] (e3) to node[right]{\color{red} $(Y_i=0)$} (e7);
\draw[thick] (e4) to node[right]{\color{red} $(Y_i=NA)$} (e8);

\end{tikzpicture}
\end{center}

Of course controlling for selection bias has a rich history in the econometrics literature, but usually for models where the outcome variable after selection is continuous.
Seminal work include \cite{gronau}, \cite{heckman1974} and in the semiparametric literature see, e.g.  \cite{ahn1993} and 
\cite{neweyej}\footnote{For work on a nonparametric sample selection model, see \cite{neweyetal}}, who identify and estimate the model parameters as a semilinear or partially linear model for selected observations.\footnote{The partially linear model itself includes an expansive literature  in the econometrics and statistics literature. Seminal work includes \cite{robinson1988}, \cite{speckman1988}. Recent developments for estimating this class of models includes, for large dimensional models, \cite{bellonietal2014}, \cite{chernozhukhovetal2018}, \cite{hsiaozhou2024}, and for data combination, \cite{xaviermaurel2025}.For Bayesian analysis of a wide class of simultaneous equation models, some of which can be nonlinear like in the Heckman model, see \cite{chib1}, \cite{chib2}. In these models unobserved components are assumed to be normally distributed, though finite mixtures of thereof can be considered. Nonparametric functions are considered there as is the case in  \cite{robinson1988}
but the dimension of the regressors  both parametric and nonparmatric components of their semilinear model is restricted to be finite.}

More formally, the econometric model we consider:
\begin{eqnarray}
D_i&=&I\left(z_{0,i}+\boldsymbol{Z}_i^{\mathrm{T}}\boldsymbol{\delta}_0-U_i>0\right) \label{sel_eq}\\
Y_i&=&D_iY^*_i = D_i\cdot I\left( x_{0,i}+\boldsymbol{X}_i^{\mathrm{T}}\boldsymbol{\beta}_0-V_i>0\right) \label{out_eq}
\end{eqnarray}
The above system of equations is of a similar structure to 
that used in the classical sample selection model introduced
in \cite{heckman1974}. $I(\cdot)$ is the indicator function, $(z_{0,i}, \boldsymbol{Z}_i^{\mathrm{T}})^{\mathrm{T}}$ and $(x_{0,i}, \boldsymbol{X}_i^{\mathrm{T}})^{\mathrm{T}}$ denote vectors of observed regressors in selection and outcome equations, respectively. 
$D_i$ and $Y_i$ denote observed {\em binary} outcomes while $Y^*_i$ is only observed if $D_i=1.$  Also,  $U_i$ and $V_i$ are unobserved random shocks while  $\boldsymbol{\delta}_0$ and $\boldsymbol{\beta}_0$ are the parameters of interest. 
As we have mentioned above, this model  differs from the classical selection model mainly in that we focus on the case where the outcome equation (\ref{out_eq}) is also binary (and here we allow for large dimensional regressors).

The structure of the rest of the paper is organized as follows.
In the next section we define our new (algorithmic) estimation procedures for the unknown regression coefficients in (\ref{sel_eq}) and (\ref{out_eq}), which are designed to be computationally efficient, and hence, suitable to implement for models where the dimensions of $\boldsymbol{Z}_i$ and/or $\boldsymbol{X}_i$ are large. 
Section \ref{asymptotics} explores the asymptotic properties of the new methods, establishing their  limiting distribution theories 
for  models of  increasing dimensions, which are gaining widespread and growing interest in the big data and machine learning literature, but have yet to be studied for this selective labeling model. Section \ref{mcsimulations} explores the finite sample properties of our procedures by means of a simulation study and Section \ref{section5} provides some empirical applications of our method. In Section \ref{section6}, we investigate extensions of the selective labeling models by allowing for endogenous treatment.  
Section \ref{section7} concludes by summarizing our results and suggesting areas for future work. An Appendix collects tabular results from the simulation study and  all the proofs of the main theorems.

\section{Algorithmic Estimation Procedures}\label{algorithms}
\setcounter{equation}{0}

This section introduces algorithmic estimation procedures for models (\ref{sel_eq}) and (\ref{out_eq}), 
where  $\boldsymbol{Z}_{e,i} \equiv (z_{0,i}, \boldsymbol{Z}_i^{\mathrm{T}})^{\mathrm{T}}\in \mathscr{Z}_e \subseteq R^{p_{Z}+1}$, $\boldsymbol{\delta}_0\in\mathscr{D}\subseteq R^{p_{Z}}$, $\boldsymbol{X}_{e,i} \equiv(x_{0,i}, \boldsymbol{X}_i^{\mathrm{T}})^{\mathrm{T}}\in \mathscr{X}_e \subseteq R^{p_{X}+1}$, and $\boldsymbol{\beta}_0\in\mathscr{B}\subseteq R^{p_{X} }$. $\boldsymbol{Z}_{e,i}$ and $\boldsymbol{X}_{e,i}$ are observed  vectors of regressors in the selection and outcome equations, whose dimensions $p_Z$ and $p_X$ may increase with sample size $n$ but satisfy $\max\{p_Z, p_X\} \leq n$. $\boldsymbol{\delta}_0$ and $\boldsymbol{\beta}_0$ are unknown parameter vectors.  $U_i, V_i$ are unobserved random shocks with joint distribution function $F(u,v)$, whose marginal distributions are given by $F_U(u)$ and $F_V(v)$. 

We impose the following condition over the data set we observe. 
\begin{condition}
\label{condition1}$\{D_i,Y_i, \boldsymbol{Z}_{e,i},\boldsymbol{X}_{e,i},U_{i},V_{i}\}$ are iid over $i$ and satisfy
	(\ref{sel_eq}) and (\ref{out_eq}). $U_{i}$ and $V_{i}$
	are jointly independent of $\boldsymbol{Z}_{e,i}$ and $\boldsymbol{X}_{e,i}$. We observe the data set $\mathcal{S}_{n}=\left\{ D_{i},Y_{i},\boldsymbol{Z}_{e,i},\boldsymbol{X}_{e,i}\right\} _{i=1}^{n}$. 
\end{condition}

The remainder  of this section proposes two novel computationally efficient algorithms for estimating $\boldsymbol{\beta}_0$. The proposed methods are  to first estimate the parameter vector $\boldsymbol{\delta}_0$ in the selection equation,  and with that, use matching as in \cite{ahn1993} or series expansion as in \cite{neweyetal} and \cite{neweyej} to estimate the selection correction function, and finally estimate 
$\boldsymbol{\beta}_0$. We will not use rank estimation in either step because  the dimension of $\boldsymbol{Z}_i$ and $\boldsymbol{X}_i$ potentially can be  large and the resulting computational burden could be extremely heavy. Instead, we adopt iteration-based methods which feature simple implementation and fast computation speed.

\subsection{The First-Step Estimator}

We first introduce the algorithm for estimating $\boldsymbol{\delta}_0$. Define $\boldsymbol{\phi}_{\boldsymbol{\delta},q_{\boldsymbol{\delta}}}(\cdot) = \left(\phi_0(\cdot), \cdots, \phi_{q_{\boldsymbol{\delta}}}(\cdot)\right)^{\mathrm{T}}$, where $\phi_0(\cdot), \phi_1(\cdot), \cdots$ are a sequence of basis functions, and $q_{\boldsymbol{\delta}}$ is the order of sieve approximation used in the first-step estimation. For the choice of basis functions, see \citet{chen2007large} and \citet{khanetal2021}. The algorithm is described as follows. 

\fbox{\begin{minipage}{\textwidth}
\textbf{Algorithm 0 for estimating $ \boldsymbol{\delta}_0$} 
\begin{enumerate}
\item Start with $k=0$ and  $\widehat {\boldsymbol{\delta}}^0,  \widehat{ \boldsymbol{\pi}}^0, \widehat F_{U}^0(\cdot)$, where $\widehat {\boldsymbol{\delta}}^0\in R^{p_{Z}}$ is the initial guess  of $\boldsymbol{\delta}_0$,  $\widehat{ \boldsymbol{\pi}}^0 \in R^{q_{\delta}+1}$ is the initial guess of the pseudo true sieve coefficient, and $\widehat F_{U}^0(\cdot)$ is the initial guess of $F_U(\cdot)$, the CDF of $U_i$.

\item With $\widehat {\boldsymbol{\delta}}^k$, define $\widehat{Z}_{i,k} = z_{0,i}+\boldsymbol{Z}_i^{\mathrm{T}}\widehat{\boldsymbol{\delta}}^k$, and update the sieve coefficient to $\widehat{\boldsymbol{\pi}}^{k+1}$ using the following 
\[
\widehat {\boldsymbol{\pi}}^{k+1} = \left(\sum_{i=1}^n \boldsymbol{\phi}_{\delta,q_{\delta}}(\widehat{Z}_{i,k})\boldsymbol{\phi}_{\delta,q_{\delta}}(\widehat{Z}_{i,k})^{\mathrm{T}}\right)^{-1}\left(\sum_{i=1}^n \boldsymbol{\phi}_{\delta,q_{\delta}}(\widehat{Z}_{i,k}) D_i\right)
\]
\item With $\widehat {\boldsymbol{\pi}}^{k+1}$, update $\widehat F_{U}^k(\cdot)$ to $\widehat F_{U}^{k+1}(\cdot)$ by $\widehat{F}_{U}^{k+1}(\cdot) = \boldsymbol{\phi}_{\delta,q_{\delta}}(\cdot)^{\mathrm{T}}\widehat {\boldsymbol{\pi}}^{k+1}$.
\item With $\widehat F_{U}^{k+1}(\cdot)$, update $\widehat {\boldsymbol{\delta}}^k$ to $\widehat {\boldsymbol{\delta}}^{k+1}$ using 
\[\widehat {\boldsymbol{\delta}}^{k+1}=\widehat {\boldsymbol{\delta}}^k-\frac{\gamma_k}{n}  \sum_{i=1}^n\left(\widehat F_{U}^{k+1}\left(\widehat{Z}_{i,k}\right) - D_i\right) \boldsymbol{Z}_i\]
where $\gamma_k >0$ is learning rate.
\item Set $k = k+1$ and go back to Step 2 unless some terminating conditions are satisfied.
\end{enumerate}
\end{minipage}}

Above algorithm is the sieve-based gradient descent estimator (SBGD) proposed by  \citet{khanetal2021}. Under some regularity conditions\footnote{The regularity conditions ensured point identification of $\boldsymbol{\delta}_0$. Recent work in \cite{khantamerwei2025} consider models where $\boldsymbol{\delta}_0$ is only partially identified and propose a two step procedure which converges to the identified set.}, \citet{khanetal2021} show that for $k$ sufficiently large, $\widehat{ \boldsymbol{\delta}}^k$ is consistent and asymptotically normally distributed under increasing dimensions. 

\subsection{The Second-Step Estimator}
Denote the first-step estimator as $\widehat{\boldsymbol{\delta}}$. With  $\widehat{\boldsymbol{\delta}}$ in hand, we now consider estimating $\boldsymbol{\beta}_0$. As in this SBGD, we will need to  control for selection bias by explicitly estimating the selection correction function. To provide some intuition, suppose that we know the joint CDF of $U_i$ and $ V_i$, then the probability of $V_i<v$ conditioned on $U_i<u$ is given by 
\[
P(V_i<v|U_i<u)  = \frac{F(u, v)}{F_U(u)} \equiv G(u, v).
\]
Define $Z_{0,i} = z_{0,i}+\boldsymbol{Z}_i^{\mathrm{T}}\boldsymbol{\delta}_0$ and $X_{0,i} = x_{0,i}+\boldsymbol{X}_i^{\mathrm{T}}\boldsymbol{\beta}_0$, we have that  \[E(Y_i|\boldsymbol{Z}_{e,i}, \boldsymbol{X}_{e,i}, D_i = 1) = G(Z_{0,i},X_{0,i}).\]  
The key observation here is that $G(\cdot, \cdot)$ is increasing with respect to its second argument. Suppose further that we also know  $\boldsymbol{\delta}_0$. Define $\widehat{X}_{i,k} = x_{0,i}+\boldsymbol{X}_i^{\mathrm{T}}\widehat {\boldsymbol{\beta}}^k$, then batch gradient descent algorithm based on the loss function in \citet{khanetal2021} immediately leads to the following iterative algorithm for estimating $\boldsymbol{\beta}_0$ using only observations with $D_i = 1$,
\begin{equation}\label{BGDupdate}
\widehat{\boldsymbol{\beta}}^{k+1} = \widehat {\boldsymbol{\beta}}^{k} - \frac{\gamma_k }{S_n} \sum_{i=1}^nD_i\left( G\left(Z_{0,i},\widehat{X}_{i,k}\right) - Y_i\right)\boldsymbol{X}_i,
\end{equation}
where $S_n=\sum_{i=1}^n D_i$ is the number of observations whose first-step outcome is 1.  Obviously, update (\ref{BGDupdate}) takes the index in the selection equation into consideration, so   effectively controls for the selection bias. 

However, since both $F(u,v)$ and $\boldsymbol{\delta}_0$ are unknown, the above algorithm is indeed infeasible. Note that the second issue can be easily resolved by plugging in our first-step estimator $\widehat{\boldsymbol{\delta}}$, while the first remains unsolved. An intuitive solution to such issue is to obtain an estimator for the conditional expectation $G(u, v)$ and  then plug such estimator into update (\ref{BGDupdate}).  In this section we propose two methods to estimate such conditional expectation, one being local in nature and the other  global.

The first local estimator uses matching to control for selection bias similar to   \cite{ahn1993}. To provide some intuition, suppose that the first-step estimator $\widehat {\boldsymbol{\delta}}$ is consistent  and $\widehat{\boldsymbol{\beta}}^k$, the starting point in the $k$-th iteration, is close to $\boldsymbol{\beta}_0$. Define 
 $\widehat{Z}_{i} = z_{0,i}+\boldsymbol{Z}_i^{\mathrm{T}}\widehat{\boldsymbol{\delta}}$, so long as $G$ is smooth enough, we have that 
\[
E\left(\left.Y_i\right|\boldsymbol{Z}_{e,i}, \boldsymbol{X}_{e,i}, D_i=1\right) = G(Z_{0,i}, X_{0,i}) \approx G(\widehat Z_{i}, \widehat X_{i,k}).
\]
The above result implies that for arbitrary $j\neq i$ such that  $(\widehat Z_{j}, \widehat X_{j,k})$ is close enough to $(\widehat Z_{i}, \widehat X_{i,k})$, $Y_j$ can be used as a (noisy) replacement for $E\left(\left.Y_i\right|\boldsymbol{Z}_{e,i}, \boldsymbol{X}_{e,i}, D_i=1\right)$. This implies that $E\left(\left.Y_i\right|\boldsymbol{Z}_{e,i}, \boldsymbol{X}_{e,i}, D_i=1\right)$ can be estimated by  a weighted combination of $Y_j$'s, where decreasing weight is assigned to each  $Y_j$ as the distance between $(\widehat Z_{j}, \widehat X_{j,k})$ and $(\widehat Z_{i}, \widehat X_{i,k})$ increases. Such idea is similar to that in \cite{ahn1993}. 

To improve computational efficiency of the  algorithm,  in this paper we consider a nearest neighbor-type\footnote{Kernel-based weights are also easy to construct, which can be similarly done as in \citet{khanetal2021}. However, constructing the weights involves $O(n^2)$ computational burdens in each round, which may cause heavy computation burdens, see Yao (2024).} weighting scheme. Define the Euclidean distance between $(\widehat Z_{j}, \widehat X_{j,k})$ and $(\widehat Z_{i}, \widehat X_{i,k})$ as 
\[d_{ij}^k = \left\Vert (\widehat Z_{j}, \widehat X_{j,k}) - (\widehat Z_{i}, \widehat X_{i,k})\right\Vert=\sqrt{\left(\widehat Z_{j} - \widehat Z_{i}\right)^2+ \left(\widehat{X}_{j,k} - \widehat{X}_{i,k}\right)^2}.\] 
For any $i$ with $D_i = 1$, rearrange the indices of $Y_j$ with $j\neq i$ and $D_j=1$
as $\varrho^k(i,1),  \cdots, \varrho^k(i,S_n -1)$ 
such that 
$
d_{i,\varrho^k(i,1)}^k \leq  \cdots \leq d_{i, \varrho^k(i,S_n -1)}^k$\footnote{If there is a tie then follow the original order of the indices.}. 
Then the weights based on \textit{$m$-nearest neighbor} is given by 

\begin{equation}\label{weight_update}
    w_{ij}^{k+1} = 
    \begin{cases}
        1/m \ \ \ \ \ \  \mbox {if}  \ j \in\{ \varrho^k(i,1),  \cdots, \varrho^k(i,m)\}\\
        0  \ \ \ \ \ \ \ \ \ \   \mbox {otherwise}
    \end{cases}
\end{equation}


\begin{remark}
    Constructing weights based on $m$ nearest neighbor has computational complexity of order $O(mn\log(n))$, which is much faster than constructing kernel-based weights as long as $m$ is small. In Section \ref{asymptotics} we show that the our resulting estimator will be consistent as long as $m/\log(n)\rightarrow \infty$. This implies that the minimal computational complexity required is roughly of order $O(n\log^2(n))$. 
\end{remark}

Given the above nearest-neighbor weighting scheme, the algorithm for estimating $\boldsymbol{\beta}_0$ based on the idea of matching is provided as follows.

\fbox{\begin{minipage}\textwidth
    
\textbf{Algorithm 1 for estimating $\boldsymbol{\beta}_0$:}
\begin{enumerate}
\item Start with $k=0$, first-step estimator $\widehat {\boldsymbol{\delta}} $, initial guess of weights $\{w_{ij}^0\}_{i,j=1}^n$ and initial guess $\widehat{\boldsymbol{ \beta}}_0$.

\item With $\widehat {\boldsymbol{\beta}}_{k}$, update the weights $\{w_{ij}^k\}_{i,j=1}^n$ to $\{w_{ij}^{k+1}\}_{i,j=1}^n$ using (\ref{weight_update}).

\item With $\{w_{ij}^{k+1}\}_{i,j=1}^n$, update $\widehat {\boldsymbol{\beta}}_{k}$ to $\widehat {\boldsymbol{\beta}}_{k+1}$ using 
\[
\widehat {\boldsymbol{\beta}}_{k+1} = \widehat {\boldsymbol{\beta}}_{k} - \frac{\gamma_k }{S_n} \sum_{i=1}^n\sum_{j=1}^n w_{ij}^{k+1}D_iD_j \left( Y_j - Y_i\right)\boldsymbol{X}_i,
\]
where $\gamma_k >0$ is the learning rate.

\item Set $k = k+1$ and go back to Step 2 unless some terminating conditions are satisfied.
\end{enumerate}

\end{minipage}}

We next propose an algorithmic approach which controls for selection bias by nonparametrically estimating the selection correction function globally 
using method of sieves. This was done for the standard selection model with linear outcome equation in \citet*{neweyetal} and \cite{neweyej}.  Let $\boldsymbol{\Phi}_q(\cdot, \cdot)$
be a $(q+1)^2$-dimensional vector of basis functions.  Note that we can decompose $D_{i}Y_{i}$ as follows 
\begin{equation}
	D_{i}Y_{i}= D_{i}\boldsymbol{\Phi}_q\left(\widehat{Z}_i,\widehat{X}_{i,k}\right)^{\mathrm{T}}\boldsymbol{\Pi}_{q}+\mathcal{E}_{i,q,k},\label{truncation}
\end{equation}
where $\boldsymbol{\Pi}_q$ is the unknown pseudo true sieve parameter vector\footnote{Note that for any sequence of sieve functions $\{\Phi_{st}(u,v)\}_{s,t=0}^{\infty}$ that is complete in $C(R^2)$ space and any function $G(u,v)\in C(R^2)$, there exists a sequence of sieve coefficients $\{\pi_{st}\}_{s,t=0}^{\infty}$ such that $G(u,v)=\sum_{s,t=0}^{\infty}\pi_{s,t}\Phi_{st}(u,v)$. Then $\boldsymbol{\Pi}_q$ is the vector of first $(q+1)^2$   sieve coefficients. See \citet{chen2007large} for more detailed discussion.}, and $\mathcal{E}_{i,q,k}$ is the error that can be decomposed as follows
\begin{align*}
	\mathcal{E}_{q,k,i} & =\underset{\text{Error due to truncation of the sieve space}}{\underbrace{D_{i}\left(G\left(\widehat{Z}_i,\widehat{X}_{i,k}\right)-\boldsymbol{\Phi}_{q}\left(\widehat{Z}_i,\widehat{X}_{i,k}\right)^{\mathrm{T}}\boldsymbol{\Pi}_{q}\right)}}+\underset{\text{Error due to first-step estimation}}{\underbrace{D_{i}\left(G\left(Z_{0,i},\widehat{X}_{i,k}\right) - G\left(\widehat{Z}_i,\widehat{X}_{i,k}\right)\right)}}\\
	& +\underset{\text{Error due to second-step estimation of the \ensuremath{k}-th round}}{\underbrace{D_{i}\left(G\left(Z_{0,i},X_{0,i}\right)-G\left(Z_{0,i},\widehat{X}_{i,k}\right)\right)}}  +\underset{\text{Sampling randomness}}{\underbrace{D_{i}\left(Y_i - G\left(Z_{0,i},X_{0,i}\right) \right)}}.
\end{align*}

When $G(u, v)$ is smooth enough, the first term on the right side of the above equation will be small as long as $q$ is large. Moreover, suppose again that the first-step estimator $\widehat {\boldsymbol{\delta}}$ is consistent   and $\widehat{\boldsymbol{\beta}}^k$  is close to $\boldsymbol{\beta}_0$, then both second and third terms are small.  Finally, the expectation of the last term conditioned on $\boldsymbol{Z}_{e,i}, \boldsymbol{X}_{e,i}$ and $D_i =1$ is zero. This naturally leads to an OLS-type estimator for $\boldsymbol{\Pi}_q$ given as follows
\begin{align}\label{sieve_update}
\widehat{\boldsymbol{\Pi}}^{k+1}_q =  \left[\sum_{i=1}^n D_i\boldsymbol{\Phi}_q\left(\widehat{Z}_i,  \widehat{X}_i^k\right)\boldsymbol{\Phi}_q\left(\widehat{Z}_i,  \widehat{X}_i^k\right)^{\mathrm{T}}\right]^{-1}\times
\left[\sum_{i=1}^n D_iY_i\boldsymbol{\Phi}_q\left(\widehat{Z}_i,  \widehat{X}_i^k\right)\right],
\end{align}
and the unknown conditional expectation function $G(u,v)$ can be estimated by $\widehat G^k(u,v)=\boldsymbol{\Phi}_q(u,v)^{\mathrm{T}}\widehat{\boldsymbol{\Pi}}_q^k$. Given the estimator of $G(u,v)$, we can plug it back to (\ref{BGDupdate}), and conduct the update. The algorithm is formally detailed as follows.

\fbox{\begin{minipage}\textwidth
    
\textbf{Algorithm 2 for Estimating $\boldsymbol{\beta}_0$:}
\begin{enumerate}
\item Start with $k=0$, the first-step estimator $\widehat {\boldsymbol{\delta}}$, initial guess of $\boldsymbol{\beta}_0$, $\widehat {\boldsymbol{\beta}}^0$, initial guess of the sieve parameter $\widehat {\boldsymbol{\Pi}}_q^0$, and initial guess of the conditional expectation function $\widehat G^0(u, v)$.  
\item In the $k$-th round,  with $\widehat{\boldsymbol{\beta}}^k$, update $\widehat {\boldsymbol{\Pi}}_q^{k}$ to $\widehat {\boldsymbol{\Pi}}_q^{k+1}$ using 
(\ref{sieve_update}).
\item With $\widehat {\boldsymbol{\Pi}}^{k+1}_q$, update $\widehat G^k(u,v)$ to $\widehat G^{k+1}(u,v)$ using $\widehat G^{k+1} \left(u, 
v\right)= \boldsymbol{\Phi}_q( u, v)^{\mathrm{T}}\widehat{\boldsymbol{\Pi}}_q^{k+1}$.
\item With $\widehat G^{k+1}(u,v)$, update $\widehat {\boldsymbol{\beta}}^k $ to $\widehat {\boldsymbol{\beta}}^{k+1}$ using
\[
\widehat {\boldsymbol{\beta}}^{k+1} = \widehat {\boldsymbol{\beta}}^{k} - \frac{\gamma_k}{S_n} \sum_{i=1}^n D_i\left(\widehat G^{k+1}\left(\widehat{Z}_i, \widehat{X}_{i,k}\right) - Y_i\right)\boldsymbol{X}_i,
\]
where $\gamma_k>0$ is the learning rate.
\item Set $k = k+1$ and go back to Step 2 unless some terminating conditions are satisfied.
\end{enumerate}
\end{minipage}}

\setcounter{equation}{0}

\section{Statistical Properties}\label{asymptotics}

This section formally studies the statistical properties of the proposed iteration-based estimators. We start with introducing the conditions on the  first-step
estimator. In particular, we assume that  $\widehat{\boldsymbol{\delta}}$ has the following asymptotic linear
representation.
\begin{condition}\label{condition2}
	The first-step estimator $\widehat{\boldsymbol{\delta}}$ satisfies
	\[
	\left\Vert\sqrt{n}\left(\widehat{\boldsymbol{\delta}}-\boldsymbol{\delta}_0\right) - \boldsymbol{\Psi}_{\boldsymbol{\delta}}^{-1}\frac{1}{\sqrt{n}}\sum_{i=1}^{n}\boldsymbol{\psi}_{\boldsymbol{\delta}}\left(\boldsymbol{Z}_{e,i}\right)\left(D_{i}-F_{U}\left(Z_{0,i}\right)\right)\right\Vert = o_{P}\left(1\right),
	\]
	where $\boldsymbol{\Psi}_{\boldsymbol{\delta}}$ is a $p_{Z}\times p_{Z}$ invertible
	matrix and $\boldsymbol{\psi}_{\boldsymbol{\delta}}\left(\cdot\right)$ is a $p_{Z}\times1$
	nonrandom function. Furthermore, there hold $0<\inf_{n}\underline{\lambda}\left(\boldsymbol{\Psi}_{\boldsymbol{\delta}}\right)<\sup_{n}\overline{\lambda}\left(\boldsymbol{\Psi}_{\boldsymbol{\delta}}\right)<\infty$ and $ \sup_{n}\Vert\boldsymbol{\psi}_{\boldsymbol{\delta}}\Vert_{\infty}/\sqrt{p_{Z}}<\infty$.
\end{condition}
\begin{remark}
The asymptotic linear representation in \autoref{condition2} simply repeats the results of Theorem 8 in \citet{khanetal2021}; more primitive conditions that guarantee such condition can
be found therein. According to Theorem 8 of \citet{khanetal2021}, we
have that $
\boldsymbol{\psi}_{\boldsymbol{\delta}}\left(\boldsymbol{Z}_e\right)=\boldsymbol{Z} -E_{\widetilde{\boldsymbol{Z}}_e}( \widetilde{\boldsymbol{Z}}| \widetilde{Z}_0 = Z_0)$
and 
$
\boldsymbol{\Psi}_{\delta}=E\left(\nabla F_{U}\left(Z_0\right)\boldsymbol{\psi}_{\boldsymbol{\delta}}\left(\boldsymbol{Z}_{e}\right)\boldsymbol{Z}^{\mathrm{T}}\right)
$, where $Z_0 = z_0 + \boldsymbol{Z}^{\mathrm{T}}\boldsymbol{\delta}_0$, $\widetilde{Z}_0 = \widetilde{z}_0 + \widetilde{\boldsymbol{Z}}^{\mathrm{T}}\boldsymbol{\delta}_0$,  $\widetilde{Z}_e$ is an independent copy of $\boldsymbol{Z}_{e}$, and $E_{\widetilde{\boldsymbol{Z}}_e}$ computes the expectation with respect to $\widetilde{\boldsymbol{Z}}_e$.     Moreover, under \autoref{condition2}, $E\Vert \boldsymbol{\Psi}_{\boldsymbol{\delta}}^{-1}\frac{1}{\sqrt{n}}\sum_{i=1}^{n}\boldsymbol{\psi}_{\boldsymbol{\delta}}\left(\boldsymbol{Z}_{e,i}\right)\left(D_{i}-F_{U}\left(Z_{0,i}\right)\right)\Vert^2 \lesssim p_{Z}$, so $\Vert\widehat{\boldsymbol{\delta}} - \boldsymbol{\delta}_0\Vert=O_P(\sqrt{p_{Z}/n})$.
\end{remark}

The next condition regulates the data generating process. 

\begin{condition}\label{condition3}
	For all $n$, there hold: (i) $\mathscr{Z}_{e}=\left[0,1\right]^{p_{Z}+1}$ and $\mathscr{X}_{e}=\left[0,1\right]^{p_{X}+1}$; 
 (ii) There exists some constant $C_{0}>0$ such that  $\mathscr{D}\subseteq [-C_0, C_0]^{p_{Z}}$ and $\mathscr{B}\subseteq[-C_0, C_0]^{p_{X}}$;
 (iii) There exists a positive constant $C_G>0$ such that $\left\Vert \nabla_{u}G(u,v)\right\Vert_{\infty}, \left\Vert \nabla_{v}G(u,v)\right\Vert _{\infty}$, $ \left\Vert \nabla_{uu}G(u,v)\right\Vert _{\infty}$, $ \left\Vert \nabla_{uv}G(u,v)\right\Vert _{\infty}$, and $ \left\Vert \nabla_{vv}G(u,v)\right\Vert _{\infty}$ are upper bounded by $C_G$; (iv) 
	There exists some constant $C_{D}>0$ such that $P_D \equiv  P(D_i = 1) =E(F_U(Z_{0,i}))\geq C_{D}$.
\end{condition}

\begin{remark}
    \autoref{condition3}(i) simply normalizes the feature space. In principle our results apply to any scenarios with bounded feature space. \autoref{condition3}(iii) requires that the unknown conditional expectation $G(\cdot, \cdot)$ is smooth enough so can be nonparametrically estimated. Finally, \autoref{condition3}(iv) requires that $S_n$ increases roughly at the rate at $O(n)$. In particular, we have that $P(S_n > P_Dn/2) > 1-  1/C_D^2n$. 
\end{remark}


\subsection{Statistical Properties of Matching-Based Estimator}

This subsection studies the statistical properties of the matching-based BGD algorithm. To ease  exposition, define $Z(\boldsymbol{\delta}) = z_0 + \boldsymbol{Z}^{\mathrm{T}}\boldsymbol{\delta}$, $Z_i(\boldsymbol{\delta}) = z_{0,i} + \boldsymbol{Z}_i^{\mathrm{T}}\boldsymbol{\delta}$, $X(\boldsymbol{\beta}) = x_0 + \boldsymbol{X}^{\mathrm{T}}\boldsymbol{\beta}$, and $X_i(\boldsymbol{\beta}) = x_{0,i} + \boldsymbol{X}_i^{\mathrm{T}}\boldsymbol{\beta}$. For any $\boldsymbol{\delta}\in\mathscr{D}$ and $\boldsymbol{\beta}\in\mathscr{B}$, define  $\mathbb{Z}_{\boldsymbol{\delta}} = \{Z(\boldsymbol{\delta}): \boldsymbol{Z}_e\in\mathscr{Z}_e\}$ and $\mathbb{X}_{\boldsymbol{\beta}} = \{X(\boldsymbol{\beta}): \boldsymbol{X}_e\in\mathscr{X}_e\}$. We next introduce some additional technical conditions.

\begin{condition}\label{condition3.1-1}
 For each pair of $\boldsymbol{\delta}$ and $\boldsymbol{\beta}$, denote the joint density of $Z( \boldsymbol{\delta})$ and $X(\boldsymbol{\beta})$ conditioned on event $D=1$ as $\iota \left(\nu_Z, \nu_X, \boldsymbol{\delta}, \boldsymbol{\beta}\right)$. There exists a sequence $\{\alpha_n\}_{n=1}^{\infty}$ such that $\alpha_n>0$ for all $n$, and    $\inf_{\boldsymbol{\delta}\in\mathscr{D}, \boldsymbol{\beta}\in\mathscr{B}}\inf_{\nu_Z\in\mathbb{Z}_{\boldsymbol{\delta}}, \nu_X\in\mathbb{X}_{\boldsymbol{\beta}}}\iota \left(\nu_{Z}, \nu_X, \boldsymbol{\delta}, \boldsymbol{\beta}\right)\geq C_{\iota}\alpha_n$, where $C_{\iota}>0$. 
\end{condition}

\begin{remark}
    \autoref{condition3.1-1} requires that for each sample size $n$, the joint density function of $Z(\boldsymbol{\delta})$ and $X(\boldsymbol{\beta})$ conditioned on $D=1$ is lower bounded by positive constant that may depend on $n$. Such condition guarantees that  with large probability,  for any pair $(Z(\boldsymbol{\delta}), X(\boldsymbol{\beta}))$, we can find  observations whose indices  constructed based on $\boldsymbol{\delta}$ and $\boldsymbol{\beta}$ are sufficiently close  to such pair as long as the sample size is large enough.   For the case of fixed dimensionality, such restrictions are also imposed in \citet{ichimura1993semiparametric}. We note that in some cases, the lower boundedness may fail to hold. But as long as we can find subsets of $\mathscr{Z}_e$ and $\mathscr{X}_e$ such that the condition holds, we can replace the above condition by introducing trimming to our algorithm. See \citet{khanetal2021} for more details. 
\end{remark}

\begin{condition}\label{condition3.1-2}
    For any $\boldsymbol{\delta}\in\mathscr{D}$ and $\boldsymbol{\beta}\in\mathscr{B}$,  define 
    \[ \mathcal{Y}\left(\nu_Z,\nu_X,\boldsymbol{\delta},\boldsymbol{\beta}\right) = 
    E\left(\left. Y\right|Z\left(\boldsymbol{\delta}\right) = \nu_Z, X\left(\boldsymbol{\beta}\right) = \nu_X, D=1\right), \nu_Z\in\mathbb{Z}_{\boldsymbol{\delta}}, \nu_X\in\mathbb{X}_{\boldsymbol{\beta}} 
\]
 For any $\boldsymbol{\delta},\boldsymbol{\delta}^{\prime}\in\mathscr{D}$, $\boldsymbol{\beta},\boldsymbol{\beta}^{\prime}\in\mathscr{B}$,  $\nu_Z\in\mathbb{Z}_{\boldsymbol{\delta}}$, $\nu_Z^{\prime}\in\mathbb{Z}_{\boldsymbol{\delta^{\prime}}}$, $\nu_X\in\mathbb{X}_{\boldsymbol{\beta}}$, and $\nu_X^{\prime}\in\mathbb{X}_{\boldsymbol{\beta^{\prime}}}$, there holds 
 \begin{align*}
& \left|\mathcal{Y}\left(\nu_Z,\nu_X,\boldsymbol{\delta},\boldsymbol{\beta}\right)  
 - \mathcal{Y}\left(\nu_Z^{\prime},\nu_X^{\prime},\boldsymbol{\delta}^{\prime},\boldsymbol{\beta}^{\prime}\right)\right| \\
 & \leq C_{\mathcal{Y}}\cdot\left(\left|\nu_Z-\nu_Z^{\prime}\right| + \left|\nu_X-\nu_X^{\prime}\right| + \sqrt{p_Z}\left\Vert\boldsymbol{\delta} - \boldsymbol{\delta}^{\prime}\right\Vert + \sqrt{p_X}\left\Vert \boldsymbol{\beta} - \boldsymbol{\beta}^{\prime}\right\Vert \right),
 \end{align*}
 where $C_{\mathcal{Y}}$ is a positive constant. 
\end{condition}

\begin{remark}
    We have that 
    \begin{align*}
    & E\left(\left. Y\right|D=1, Z\left(\boldsymbol{\delta}\right) = \nu_Z, X\left(\boldsymbol{\beta}\right) = \nu_X\right)\\
   = & E\left(\left.E\left(\left. Y\right|D=1, Z\left(\boldsymbol{\delta}\right) = \nu_Z, X\left(\boldsymbol{\beta}\right) = \nu_X,\boldsymbol{Z}_e, \boldsymbol{X}_e\right)\right|D=1, Z\left(\boldsymbol{\delta}\right) = \nu_Z, X\left(\boldsymbol{\beta}\right) = \nu_X\right)\\
   =& E\left(\left. G\left(Z_0, X_0\right)\right|D = 1, Z\left(\boldsymbol{\delta}\right)=\nu_Z, X\left(\boldsymbol{\beta}\right)=\nu_X\right).
    \end{align*}
\end{remark}

We finally introduce a condition that can be used as a sufficient condition for contraction mapping.  
\begin{condition}\label{condition3.1-3}
    For any $\boldsymbol{Z}_e, \boldsymbol{X}_e$,  $\boldsymbol{\beta}$, and $\varsigma\in[0,1]$, define  
    \begin{align*}
    &\widetilde{\boldsymbol{\Psi}}_M\left(\boldsymbol{Z}_e,\boldsymbol{X}_e, \boldsymbol{\beta}, \varsigma\right) = \\
&E_{\widetilde{\boldsymbol{Z}}_{e},\widetilde{\boldsymbol{X}}_{e},\widetilde{D}}\left(\left.\nabla_{v}G\left(Z_{0},X_0+\varsigma\left(\boldsymbol{X}-\widetilde{\boldsymbol{X}}\right)^{\mathrm{T}}\Delta\boldsymbol{\beta}\right)\boldsymbol{X}\left(\boldsymbol{X}-\widetilde{\boldsymbol{X}}\right)^{\mathrm{T}}\right|\widetilde{Z}_{0}=Z_{0},\widetilde{X}\left(\boldsymbol{\beta}\right)=X\left(\boldsymbol{\beta}\right),\widetilde{D}=1\right),
    \end{align*}
    and 
    \[
\boldsymbol{\Psi}_M\left(\boldsymbol{\beta}\right) = \int_{0}^1E_{\boldsymbol{Z}_e,\boldsymbol{X}_e, D}\left(\left.\widetilde{\boldsymbol{\Psi}}_M\left(\boldsymbol{Z}_e,\boldsymbol{X}_e, \boldsymbol{\beta},\varsigma\right)\right|D = 1\right)d\varsigma
    \]
There hold $\sup_{\boldsymbol{\beta}\in\mathscr{B}}\overline{\lambda}\left(\boldsymbol{\Psi}_M\left(\boldsymbol{\beta}\right)+\boldsymbol{\Psi}_M^{\mathrm{T}}\left(\boldsymbol{\beta}\right)\right)\leq\overline{\lambda}_{\boldsymbol{\Psi}_M}<\infty$
	and $\inf_{\boldsymbol{\beta}\in\mathscr{B}}\underline{\lambda}\left(\boldsymbol{\Psi}_M\left(\boldsymbol{\beta}\right)+\boldsymbol{\Psi}_M^{\mathrm{T}}\left(\boldsymbol{\beta}\right)\right)\geq\underline{\lambda}_{\boldsymbol{\Psi}_M}>0.$
\end{condition}

\begin{remark}
    \autoref{condition3.1-3} is the key condition that guarantees the validity of iteration-based estimator. Under such assumption, we have that $\Vert\left(\boldsymbol{I}_{p_X}-\gamma \boldsymbol{\Psi}_M(\boldsymbol{\beta})\right)\boldsymbol{a}\Vert\leq C_{\gamma}\Vert \boldsymbol{a} \Vert$  for some $0<C_{\gamma}<1$ and arbitrary $\boldsymbol{\beta}\in\mathscr{B}$ and $\boldsymbol{a}\in R^{p_X}$. This guarantees the contraction map of the proposed algorithm. In general, \autoref{condition3.1-3} is more  likely to hold when at least one regressor is continuous, which excludes the case where all the regressors are discrete.  For examples of data generating processes satisfying \autoref{condition3.1-3},  see Remark 4 in \citet{khanetal2021}. 
\end{remark}

Given the above conditions, we now state our theorem regrading the convergence rate of the matching-based estimator, whose proof is provided in Appendix, Section \ref{proofs}. 

\begin{theorem}\label{theorme3.1-1}
    Suppose that \autoref{condition1} -- \autoref{condition3.1-3} hold,  $n\alpha_n\rightarrow \infty$, and $\mathfrak{P}\leq n$. If we choose $m$ such that $m/\log(n)\rightarrow \infty$ and $m/n\rightarrow \infty$ and a constant learning rate $\gamma_{k}=\gamma < \min\{\overline{\lambda}_{\boldsymbol{\Psi}_M}^{-1}, \underline{\lambda}_{\boldsymbol{\Psi}_M}/2p_{X}^2C_G^2\}$, 
	then there holds 
\[\sup_{k\geq k_M\left(n,m,\gamma\right) + 1}\Vert \Delta\widehat{\boldsymbol{\beta}}_{k}\Vert =O_{P}\left(\sqrt{\frac{\mathfrak{P}(\mathfrak{P}+m)\cdot \log\left(n\right)}{n\alpha_n} + \frac{\mathfrak{P}^2\log (n)}{m}} \right),\]
	where \[k_M(n,m,\gamma) = \frac{\log\left(\sqrt{\frac{\mathfrak{P}(\mathfrak{P}+m)\cdot \log\left(n\right)}{n\alpha_n} + \frac{\mathfrak{P}^2\log (n)}{m}} \right) - \log\left(\left\Vert\Delta\widehat{\boldsymbol{\beta}}^{1}\right\Vert\right)}{\log(1-\underline{\lambda}_{\boldsymbol{\Psi}_M}\gamma/4)}.\] 
\end{theorem}

    \autoref{theorme3.1-1} states that the matching-based estimator will be consistent when the number of nearest-neighbor points $m$ and the learning rate $\gamma_k$ are properly chosen. For the rate of convergence, the first term corresponds to the bias and the second term variance. It's intuitive that the bias increases with $m$, while the variance decreases with $m$. The optimal rate of $m$ is $\sqrt{\mathfrak{P}n\alpha_n}$, and the corresponding convergence rate of the matching-based estimator is $\sqrt{\log(n)\left(\frac{\mathfrak{P}^2}{n\alpha_n} + \frac{\mathfrak{P}\sqrt{\mathfrak{P}}}{\sqrt{n\alpha_n}}\right)}$. Note that due to the sizable bias, the resulting estimator is not guaranteed to be $1/\sqrt{n}$-consistent. This issue will be resolved for the sieve-based estimator. Nevertheless, the matching based estimator is easy to implement so can be used as a computationally efficient first-step estimator. 

\subsection{Statistical Properties of Sieve-Based Estimator}

This section studies the statistical properties of the sieve-based BGD algorithm
proposed in the previous section.  We further introduce some technical conditions. 

\begin{condition}\label{condition4}
	The vector of basis functions $\boldsymbol{\Phi}_q$ satisfies: 
    (i) Let $\Phi_j(u, v)$ denote the $j$-th argument of $\boldsymbol{\Phi}_q$. For each $q$ and all $1\leq j\leq (q+1)^2$, $\Vert\Phi_j\Vert_{\infty} \leq C(0,q)$, $\max\{\Vert\nabla_u\Phi_j\Vert_{\infty}, \Vert\nabla_v\Phi_j\Vert_{\infty}\} \leq C_{\Phi,1, q}$, $\max\{\Vert\nabla_{uu}\Phi_j\Vert_{\infty}, \Vert\nabla_{uv}\Phi_j\Vert_{\infty}, \Vert\nabla_{vv}\Phi_j\Vert_{\infty}\} \leq C_{\Phi,2,q}$, where $C(0,q), C_{\boldsymbol{\Phi}, 1, q}$ and $C_{\boldsymbol{\Phi},2,q}$ are all positive constants that depend on $q$ only, and moreover, $\log(\max\{C(0,q), C_{\boldsymbol{\Phi}, 1, q}, C_{\boldsymbol{\Phi},2,q}\}) = O(\log(n))$; (ii) Define  $\boldsymbol{\Gamma}_q (\boldsymbol{\beta})= E[\boldsymbol{\Phi}_q\left(Z_{0,i}, X_i(\boldsymbol{\beta})\right)\boldsymbol{\Phi}_q\left(Z_{0,i}, X_i(\boldsymbol{\beta})\right)^{\mathrm{T}}|D_i=1].$
	There exist $0<\underline{\lambda}_{\boldsymbol{\Phi}}\leq\overline{\lambda}_{\boldsymbol{\Phi}}<\infty$
	such that $\underline{\lambda}_{\boldsymbol{\Phi}}\leq\inf_{\boldsymbol{\beta}\in\mathscr{B}}\underline{\lambda}\left(\boldsymbol{\Gamma}_{q}\left(\boldsymbol{\beta}\right)\right)\leq\sup_{\boldsymbol{\beta}\in\mathscr{B}}\overline{\lambda}\left(\boldsymbol{\Gamma}_{q}\left(\boldsymbol{\beta}\right)\right)\leq\overline{\lambda}_{\boldsymbol{\Phi}}$
	for all $q$;
 (iii) $\left\Vert G\left(u,v\right)-\boldsymbol{\Phi}_q(u,v)^{\mathrm{T}}\boldsymbol{\Pi}_q\right\Vert _{\infty}\leq\mathscr{R}(q)$.
\end{condition}

\begin{remark}
    \autoref{condition4} provides standard restrictions on the basis functions for sieve estimation. \autoref{condition4}(i) requires that the sieve functions are bounded and are twice continuously differentiable with bounded derivatives. The upper bounds, which are functions of $q$, increasing at best at poly-$n$ rate (polynomials of $n$). \autoref{condition4}(ii) allows us to provide uniform convergence rate for the estimators of sieve coefficients. Finally, \autoref{condition4}(iii) guarantees that the unknown function $G$ can be uniformly approximated by the basis functions. Note that when $q$ is fixed, the sieve approximation error rate $\mathscr{R}(q)$ decreases with the increase of $G$'s degree of smoothness. For more discussion on the properties of sieve approximation, see \citet{chen2007large}. 
\end{remark}

\begin{condition}\label{condition5} For any  $\boldsymbol{\beta}\in \mathscr{B}$, $\nu_Z\in\mathbb{Z}_{\boldsymbol{\delta}_0}$ and  $ \nu_X\in\mathbb{X}_{\boldsymbol{\beta}}$, define 
    $\mathcal{X}(\nu_Z,\nu_X, \boldsymbol{\beta}) = E(\boldsymbol{X}|Z_{0}=\nu_Z,  X(\boldsymbol{\beta}) = \nu_X, D=1 )$. There hold:
     (i) Let $\mathcal{X}_j(\nu_Z,\nu_X,\boldsymbol{\beta})$ denote the $j$-th argument of $\mathcal{X}(\nu_Z,\nu_X,\boldsymbol{\beta})$. There exists a positive constant $C_X$ such that for all $1\leq j\leq p_{X}$,  any $\boldsymbol{\beta},\boldsymbol{\beta}^{\prime}\in\mathscr{B}$,  $\nu_X\in\mathbb{X}_{\boldsymbol{\beta}}$, and $\nu_X^{\prime}\in\mathbb{X}_{\boldsymbol{\beta^{\prime}}}$, there holds 
 \[
 \left|\mathcal{X}_j\left(\nu_Z,\nu_X, \boldsymbol{\beta}\right)  
 - \mathcal{X}_j\left(\nu_Z^{\prime},\nu_X^{\prime}, \boldsymbol{\beta}^{\prime}\right)\right| \leq C_X\cdot\left(\left|\nu_Z-\nu_Z^{\prime}\right| + \left|\nu_X-\nu_X^{\prime}\right|  + \sqrt{p_X}\left\Vert \boldsymbol{\beta} - \boldsymbol{\beta}^{\prime}\right\Vert \right);
\](ii) For each $\boldsymbol{\beta}\in\mathscr{B}$, there exists $\boldsymbol{\Pi}_q^X(\boldsymbol{\beta})\in R^{(q+1)^2}$ such that \[\sup_{\nu_Z, \nu_X, \boldsymbol{\beta}}\Vert \mathcal{X}(\nu_Z,\nu_X,\boldsymbol{\beta}) - \boldsymbol{\Pi}_{q}^X(\boldsymbol{\beta})^{\mathrm{T}} \boldsymbol{\Phi}_q(\nu_Z,\nu_X)\Vert \leq \mathscr{R}_X(q).\]
\end{condition}

\begin{remark}
\autoref{condition5}(i) restricts the smoothness of $\mathcal{X}(\nu_Z,\nu_X,\boldsymbol{\beta})$. \autoref{condition5}(ii) requires that for each $j$,  $\mathcal{X}_j(\nu_Z,\nu_X,\boldsymbol{\beta})$ can be uniformly approximated by linear combinations of sieve functions. Similar to the previous condition, the approximation error $\mathscr{R}_X(q)$ depends on both the order of sieve functions and the smoothness of $\mathcal{X}(\nu_Z,\nu_X,\boldsymbol{\beta})$. 
    
\end{remark}

We finally introduce a condition that is similar to \autoref{condition3.1-3}.

\begin{condition}\label{condition6}
	Define  
\[
\boldsymbol{\Psi}_S\left(\boldsymbol{\beta}\right)=\int_{0}^{1}E\left(\left.\nabla_{v}G\left(Z_{0,i},X_{0,i}+\varsigma\boldsymbol{X}_{i}^{\mathrm{T}}\Delta\boldsymbol{\beta}\right)\left(\boldsymbol{X}_{i}-\mathcal{X}\left(Z_{0,i}, x_{0,i}+\boldsymbol{X}^{\mathrm{T}}_i\boldsymbol{\beta}, \boldsymbol{\beta}\right)\right)\boldsymbol{X}_{i}^{\mathrm{T}}\right|D_i = 1\right)d\varsigma.
	\]
	There hold $\sup_{\boldsymbol{\beta}\in\mathscr{B}}\overline{\lambda}\left(\boldsymbol{\Psi}_S\left(\boldsymbol{\beta}\right)+\boldsymbol{\Psi}_S^{\mathrm{T}}\left(\boldsymbol{\beta}\right)\right)\leq\overline{\lambda}_{\boldsymbol{\Psi}_S}<\infty$
	and $\inf_{\boldsymbol{\beta}\in\mathscr{B}}\underline{\lambda}\left(\boldsymbol{\Psi}_S\left(\boldsymbol{\beta}\right)+\boldsymbol{\Psi}_S^{\mathrm{T}}\left(\boldsymbol{\beta}\right)\right)\geq\underline{\lambda}_{\boldsymbol{\Psi}_S}>0.$
\end{condition}

Define \[\Xi_{1,n} = \sqrt{p_{X}}\mathscr{R}(q) + q^2C(0,q)^2\mathscr{R}_X(q)+\frac{\sqrt{p_{X}}q^4C(0,q)^3(p_{Z}C(1,q) + C(0,q)\sqrt{\log(n)p_{X}}}{\sqrt{n}}.\] 
Under the above conditions, we have the following result, whose proofs are provided in the Appendix, Section \ref{proofs}.

\begin{theorem}\label{theorem3.2-1}
	Suppose that \autoref{condition1}--\autoref{condition3}  and \autoref{condition4}- \autoref{condition6} hold, and that $q$ is chosen such that $\Xi_{1,n}\rightarrow0$. If we choose a constant learning rate $\gamma_{k}=\gamma < \min\{\overline{\lambda}_{\boldsymbol{\Psi}_S}^{-1}, \underline{\lambda}_{\boldsymbol{\Psi}_S}/2p_{X}^2C_G^2\}$, 
	then there holds 
\[\sup_{k\geq k_S\left(n,\gamma\right)}\Vert \Delta\widehat{\boldsymbol{\beta}}_{k}\Vert =O_P\left(\Xi_{1,n}\right),\]
	where \[k_S(n,\gamma) = \frac{\log\left(\Xi_{1,n}\right) - \log\left(\Vert\Delta\widehat{\boldsymbol{\beta}}^{1}\Vert\right)}{\log\left(1-\underline{\lambda}_{\boldsymbol{\Psi}_S}\gamma/4\right)}.\]
\end{theorem}

\autoref{theorem3.2-1} provides asymptotic consistency of the sieve-based estimator for mildly increasing dimensionality as long as $\Xi_{1,n}\rightarrow 0$.
Based on \autoref{theorem3.2-1}, we can further establish the asymptotic linear representation for  the sieve-based estimator. Define $\Xi_{2,n}$ as
\[
\Xi_{2,n} =  p_X\sqrt{p_{X}}q^4C(0,q)^{3}\Xi_{1,n}^{2}\left(q^{2}C(0,q)C(1,q)^2+C(2,q)\right).
\]
We have the following result. 

\begin{theorem}\label{theorem3.2-2}
Let all the requirements in \autoref{theorem3.2-1} hold and $\Xi_{2,n}\rightarrow 0$, then we have that 
\begin{align*}
\Delta\widehat{\boldsymbol{\beta}}_{k+1} & =\left(\boldsymbol{I}_{p_{X}} - \gamma\boldsymbol{\Psi}_S\left(\boldsymbol{\beta}_0\right)\right)\Delta\widehat{\boldsymbol{\beta}}_{k}+\frac{\gamma}{n}\sum_{i=1}^{n}P_D^{-1}D_{i}\left(\boldsymbol{X}_{i}-\mathcal{X}\left(Z_{0,i},X_{0,i},\boldsymbol{\beta}_{0}\right)\right)\varepsilon_{i}\\
 & -\frac{\gamma}{n}\sum_{i=1}^{n}\Sigma_{X,Z}\boldsymbol{\Psi}_{\boldsymbol{\delta}}^{-1}\psi_{\boldsymbol{\delta}}\left(Z_{0,i}\right)\left(D_{i}-F_{U}\left(Z_{0,i}\right)\right)+\boldsymbol{\Omega}_{n,k},
\end{align*}
where $\Sigma_{X,Z}=E\left(\nabla_{u}G\left(Z_{0,i},X_{0,i}\right)\left(\boldsymbol{X}_{i}-\mathcal{X}\left(Z_{0,i},X_{0,i},\boldsymbol{\beta}_{0}\right)\right)\boldsymbol{Z}_{i}^{\mathrm{T}}|D_i=1\right)$
and $\sup_{k\geq k_S\left(n,\gamma\right)}\left\Vert \boldsymbol{\Omega}_{n,k}\right\Vert =O_{P}\left(\Xi_{2,n}\right).$
If further $\sqrt{n}\Xi_{2,n}\rightarrow 0$, we have that 
\begin{align*}
\sup_{k\geq k_S\left(n,\gamma\right)+ \frac{\log\left(\Xi_{2,n}\right)}{\log\left(1-\gamma\underline{\lambda}_{\Psi}/4\right)}} & \left\Vert \sqrt{n}\Delta\widehat{\boldsymbol{\beta}}_{k}-\boldsymbol{\Psi}_S^{-1}\left(\boldsymbol{\beta}_0\right)\frac{1}{\sqrt{n}}\sum_{i=1}^{n}P_D^{-1}D_{i}\left(\boldsymbol{X}_{i}-\mathcal{X}\left(Z_{0,i},X_{0,i},\boldsymbol{\beta}_{0}\right)\right)\varepsilon_{i}\right.\\
 & \left.+ \boldsymbol{\Psi}_S^{-1}\left(\boldsymbol{\beta}_0\right)\Sigma_{X,Z}\boldsymbol{\Psi}_{\boldsymbol{\delta}}^{-1}\frac{1}{\sqrt{n}}\sum_{i=1}^{n}\boldsymbol{\psi}_{\boldsymbol{\delta}}\left(Z_{0,i}\right)\left(D_{i}-F_{U}\left(Z_{0,i}\right)\right)\right\Vert =o_{P}\left(1\right).
\end{align*}
\end{theorem}
\autoref{theorem3.2-2} establishes the asymptotic linear representation for our sieve-based estimator. Such a result is useful in that inference for the unknown parameter can be subsequently conducted. In particular, we have the following two corollaries. 

\begin{corollary}\label{corollary1}
Let all the conditions in \autoref{theorem3.2-2} hold. Define $\widehat{\boldsymbol{\beta}}=\widehat{\boldsymbol{\beta}}_{k}$
for any $k\geq k_S\left(n,\gamma\right)+ \log\left(\Xi_{2,n}\right)/{\log\left(1-\gamma\underline{\lambda}_{\boldsymbol{\Psi}_S}/4\right)}$.
For any $p_{X}\times1$ vector $\mathcal{W}$, if further   \[ \mathcal{W}^{\mathrm{T}}\boldsymbol{\Psi}_S^{-1}\left(\boldsymbol{\beta}_0\right)\left(\boldsymbol{X}-\mathcal{X}\left(Z_{0},X_{0},\boldsymbol{\beta}_{0}\right)\right)\rightarrow_{a.s.}\mathfrak{X}_{\mathcal{W}} \]
and \[\mathcal{W}^{\mathrm{T}}\boldsymbol{\Psi}^{-1}_S\left(\boldsymbol{\beta}_0\right)\Sigma_{X,Z}\boldsymbol{\Psi}_{\boldsymbol{\delta}}^{-1}\boldsymbol{\psi}_{\boldsymbol{\delta}}\left(\boldsymbol{Z}_{e}\right)\rightarrow_{a.s.}\mathfrak{Z}_{\mathcal{W}} \]
hold, where $\mathfrak{X}_{\mathcal{W}}$ and
$\mathfrak{Z}_{\mathcal{W}}$ are fixed random variables with bounded second moments. Then we have that 
\begin{align*}
\sqrt{n}\mathcal{W}^{\mathrm{T}}\Delta\widehat{\boldsymbol{\beta}}=\frac{1}{\sqrt{n}} & \sum_{i=1}^{n}\left(P_D^{-1}D_{i}\mathfrak{X}_{\mathcal{W},i}\varepsilon_{i} -\mathfrak{Z}_{\mathcal{W},i}\left(D_{i}-F_{U}\left(Z_{0,i}\right)\right)\right)+o_{P}\left(1\right) 
\Longrightarrow N \left(0,\Sigma_{\mathcal{W},1} + \Sigma_{\mathcal{W},2}\right).
\end{align*}
where \[\Sigma_{\mathcal{W},1} = P_D^{-1}E\left(\left.G\left(Z_{0,i},X_{0,i}\right)\left(1-G\left(Z_{0,i},X_{0,i}\right)\right) \mathfrak{X}_{\mathcal{W},i}\mathfrak{X}_{\mathcal{W},i}^{\mathrm{T}}\right|D_i = 1\right)\] and \[\Sigma_{\mathcal{W},2} = E\left(F_{U}\left(Z_{0,i}\right)\left(1-F_{U}\left(Z_{0,i}\right)\right)\mathfrak{Z}_{\mathcal{W},i}\mathfrak{Z}_{\mathcal{W},i}^{\mathrm{T}}\right).\]
\end{corollary}

The above corollary states that the  linear  combinations of arguments of $\boldsymbol{\widehat{\beta}}$ is $1/\sqrt{n}$-consistent and $\sqrt{n}\mathcal{W}^{\mathrm{T}}(\boldsymbol{\widehat{\beta}} - \boldsymbol{\beta}_0)$ is asymptotically normally distributed with asymptotic covariance matrix $\Sigma_{\mathcal{W},1} + \Sigma_{\mathcal{W},2}$. To estimate $\Sigma_{\mathcal{W},1} + \Sigma_{\mathcal{W},2}$, define $\widehat{X}_{i} = x_{0,i}+\boldsymbol{X}_i^{\mathrm{T}}\widehat{\boldsymbol{\beta}}$,  \[\widehat {\boldsymbol{\pi}}  = \left(\sum_{i=1}^n \boldsymbol{\phi}_{\delta,q_{\delta}}(\widehat{Z}_{i})\boldsymbol{\phi}_{\delta,q_{\delta}}(\widehat{Z}_{i})^{\mathrm{T}}\right)^{-1}\left(\sum_{i=1}^n \boldsymbol{\phi}_{\delta,q_{\delta}}(\widehat{Z}_{i}) D_i\right),\] \[\widehat{F}_{U}(u) = \boldsymbol{\phi}_{\delta,q_{\delta}}(u)^{\mathrm{T}}\widehat {\boldsymbol{\pi}}, \ \ \widehat{\nabla F}_{U}(u) = \nabla\boldsymbol{\phi}_{\delta,q_{\delta}}(u)^{\mathrm{T}}\widehat {\boldsymbol{\pi}},\]
\[\widehat{\boldsymbol{\psi}}_{\delta}\left(\boldsymbol{Z}_e\right)=\boldsymbol{Z} -\frac{1}{n}\sum_{i=1}^{n}\boldsymbol{Z}_{i}\boldsymbol{\phi}_{\delta,q_{\delta}}\left(\widehat{Z}_{i} \right)^{\mathrm{T}}\left(\frac{1}{n}\sum_{i=1}^n\boldsymbol{\phi}_{\delta,q_{\delta}}\left(\widehat{Z}_{i} \right)\boldsymbol{\phi}_{\delta,q_{\delta}}\left(\widehat{Z}_{i} \right)^{\mathrm{T}}\right)^{-1}\boldsymbol{\phi}_{\delta,q_{\delta}}\left(z_0 + \boldsymbol{Z}^{\mathrm{T}}\widehat{\boldsymbol{\delta}}\right),\] \[\widehat{\boldsymbol{\Psi}}_{\delta}=\frac{1}{n}\sum_{i=1}^n\left(\widehat{\nabla F}_{U}\left(\widehat{Z}_{i}\right)\widehat{\boldsymbol{\psi}}_{\delta}\left(\boldsymbol{Z}_{e,i}\right)\boldsymbol{Z}_i^{\mathrm{T}}\right),\] \[\widehat{\mathcal{X}}_n\left(\nu_{Z},\nu_{X},\widehat{\boldsymbol{\beta}}\right)=\frac{1}{S_n}\sum_{i=1}^{n}D_{i}\boldsymbol{X}_{i}\boldsymbol{\Phi}_{q}\left(\widehat{Z}_{i},\widehat{X}_i\right)^{\mathrm{T}}\widehat{\boldsymbol{\Gamma}}_{n,q}^{-1}\left(\widehat{\boldsymbol{\delta}},\widehat{\boldsymbol{\beta}}\right)\boldsymbol{\Phi}_{q}\left(\nu_{Z},\nu_{X}\right),\] \[\widehat{\boldsymbol{\Pi}} _q =  \left[\sum_{i=1}^n D_i\boldsymbol{\Phi}_q\left(\widehat{Z}_i,  \widehat{X}_i\right)\boldsymbol{\Phi}_q\left(\widehat{Z}_i,  \widehat{X}_i\right)^{\mathrm{T}}\right]^{-1}\times
\left[\sum_{i=1}^n D_iY_i\boldsymbol{\Phi}_q\left(\widehat{Z}_i,  \widehat{X}_i\right)\right],\] \[\widehat{G}(u, v) = \boldsymbol{\Phi}_q(u,v)^{\mathrm{T}}\widehat{\boldsymbol{\Pi}}_q, \widehat{\nabla_uG}(u, v) = \nabla_u\boldsymbol{\Phi}_q(u,v)^{\mathrm{T}}\widehat{\boldsymbol{\Pi}}_q, \widehat{\nabla_vG}(u, v) = \nabla_v\boldsymbol{\Phi}_q(u,v)^{\mathrm{T}}\widehat{\boldsymbol{\Pi}}_{q},\] \[\widehat{\boldsymbol{\Psi}}_S = \frac{1}{S_n}\sum_{i=1}^nD_i \widehat{\nabla_{v}G}\left(\widehat{Z}_{i},\widehat{X}_{i}\right)\left(\boldsymbol{X}_{i}-\widehat{\mathcal{X}}_n\left(\widehat{Z}_i,\widehat{X}_i,\widehat{\boldsymbol{\beta}}\right)\right)\boldsymbol{X}_{i}^{\mathrm{T}},\] \[\widehat{\Sigma}_{X,Z}=\frac{1}{S_n}\sum_{i=1}^nD_i\widehat{\nabla_{u}G}\left(\widehat{Z}_{i},\widehat{X}_{i}\right)\left(\boldsymbol{X}_{i}-\widehat{\mathcal{X}}_n\left(\widehat{Z}_i,\widehat{X}_i,\widehat{\boldsymbol{\beta}}\right)\right)\boldsymbol{Z}_{i}^{\mathrm{T}}.\]
 For any fixed $p_X\times 1$ vector $\mathcal{W}$, further define 
\begin{align*}
&\widehat{\mathfrak{X}}_{\mathcal{W},i} = \mathcal{W}^{\mathrm{T}}\widehat{\boldsymbol{\Psi}}_S^{-1}\left(\boldsymbol{X}_i-\widehat{\mathcal{X}}_n\left(\widehat{Z}_i,\widehat{X}_i,\widehat{\boldsymbol{\beta}}\right) \right), \ \widehat{\mathfrak{Z}}_{\mathcal{W},i} = \mathcal{W}^{\mathrm{T}}\widehat{\boldsymbol{\Psi}}^{-1}_S\widehat{\Sigma}_{X,Z}\widehat{\boldsymbol{\Psi}}_{\boldsymbol{\delta}}^{-1}\widehat{\boldsymbol{\psi}}_{\boldsymbol{\delta}}\left(\boldsymbol{Z}_{e,i}\right),\\
&\widehat{\Sigma}_{\mathcal{W},1} = \frac{n}{S_n^2}\sum_{i=1}^nD_i\widehat{G}\left(\widehat{Z}_{i},\widehat{X}_{i}\right)\left(1-\widehat{G}\left(\widehat{Z}_{i},\widehat{X}_{i}\right)\right) \widehat{\mathfrak{X}}_{\mathcal{W},i}\widehat{\mathfrak{X}}_{\mathcal{W},i}^{\mathrm{T}}, \\
& \widehat{\Sigma}_{\mathcal{W},2} = \frac{1}{n}\sum_{i=1}^n\widehat{F}_{U}\left(\widehat{Z}_{i}\right)\left(1-\widehat{F}_{U}\left(\widehat{Z}_{i}\right)\right)\widehat{\mathfrak{Z}}_{\mathcal{W},i}\widehat{\mathfrak{Z}}_{\mathcal{W},i}^{\mathrm{T}}.
\end{align*}
Then $\Sigma_{\mathcal{W},1} + \Sigma_{\mathcal{W},2}$ is estimated by $\widehat{\Sigma}_{\mathcal{W},1}+\widehat{\Sigma}_{\mathcal{W},2}$. 

\section{Monte Carlo Simulations} \label{mcsimulations}
\setcounter{equation}{0}
This section conducts some simulation experiments to evaluate the performance of the proposed estimators. The data generating process we consider is (\ref{sel_eq}) and (\ref{out_eq}). We generate $\widetilde z_{0,i},x_{0,i}\sim_{\text{i.i.d.}}N(0,1), \widetilde z_{1,i},x_{1,i}\sim_{\text{i.i.d.}}\text{Bernoulli}(0.5), \widetilde z_{2,i},x_{2,i}\sim_{\text{i.i.d.}}\text{Poisson}(2), z_{j,i}, x_{j,i}\sim_{\text{i.i.d.}} \frac{1}{\sqrt{2}}(\chi^2(1) - 1)$ for $j\geq 3$, and $\widetilde  z_{j,i}$ is independent of  $x_{l,i}$ for any $i$, $0\leq j\leq p_Z$ and $0\leq l\leq p_X$. Then, we set $z_{j,i} = \widetilde z_{j,i}$ for $0\leq j\leq 2$, and $z_{j, i} = \widetilde{z}_{j,i} + \sum_{l=3}^{p_X}\varsigma_{l}x_{l,i}$, where $\varsigma_l$ is randomly drawn from $0.4\cdot (U(0,1) - 0.5)$ in each round of . 
$(U_i, V_i)$ is i.i.d. across $i$ and is independent of $\boldsymbol{Z}_{e,i}$ and $\boldsymbol{X}_{e,i}$. We set 
\[\boldsymbol{\delta}_0 = \left(-1, 1, 0.5, -0.5, 1.2, -1.4, -2.2, 1.8, 0.05\cdot \boldsymbol{1}_{10}, -0.05\cdot \boldsymbol{1}_{10}\right)^{\mathrm{T}}\] and  \[\boldsymbol{\beta}_0 = \left(1, -1, 0.5, -0.5, 0.8, -1.0, -1.8, 2.1, 0.05\cdot\boldsymbol{1}_{10}, -0.05\cdot\boldsymbol{1}_{10}\right)^{\mathrm{T}},\] where $\boldsymbol{1}_{10}$ is a $10\times 1$ row vector whose arguments are all 1. In the following we report simulation results under different sample sizes $n$ and  setups of joint distributions of $U_i$ and $V_i$.

\begin{table}[t!]
\small
\centering{}\caption{Bias and RMSE of Second-Step Estimator: $U = \eta_1, V= U+\eta_2, \eta_1, \eta_2\sim Cauchy$}\label{table1}
\begin{tabular}{lllccccccccc}
\hline
                            &                                           & Method & $\beta_1$ & $\beta_2$ & $\beta_3$ & $\beta_4$ & $\beta_5$ & $\beta_6$ & $\beta_7$ & $\beta_8$ & all    \\ \hline
\multirow{8}{*}{$n=25000$}  & \multicolumn{1}{c}{\multirow{4}{*}{Bias}} & 2NLS   & 0.0124     & 0.0139     & 0.0090     & 0.0028     & 0.0121     & 0.0331     & 0.0898     & 0.0331     & 0.2741 \\
                            & \multicolumn{1}{c}{}                      & MLE    & 0.0139     & 0.0313     & 0.0155     & 0.0157     & 0.0188     & 0.0986     & 0.2791     & 0.0284     & 0.5692 \\
                            & \multicolumn{1}{c}{}                      & M-BGD  & 0.0335     & 0.0001     & 0.0071     & 0.0143     & 0.0036     & 0.0068     & 0.0202     & 0.0194     & 0.1713 \\
                            & \multicolumn{1}{c}{}                      & S-BGD  & 0.0123     & 0.0013     & 0.0046     & 0.0117     & 0.0014     & 0.0002     & 0.0096     & 0.0067     & 0.1055 \\ \cline{2-12} 
                            & \multirow{4}{*}{RMSE}                     & 2NLS   & 0.0818     & 0.0539     & 0.0516     & 0.0436     & 0.0566     & 0.0764     & 0.1334     & 0.1138     & 0.2950 \\
                            &                                           & MLE    & 0.0886     & 0.0639     & 0.0599     & 0.0515     & 0.0639     & 0.1221     & 0.2952     & 0.1276     & 0.4258 \\
                            &                                           & M-BGD  & 0.0919     & 0.0612     & 0.0497     & 0.0582     & 0.0618     & 0.0832     & 0.1249     & 0.1172     & 0.3117 \\
                            &                                           & S-BGD  & 0.0843     & 0.0552     & 0.0487     & 0.0541     & 0.0537     & 0.0769     & 0.1067     & 0.1095     & 0.2888 \\ \cline{2-12} 
\multirow{8}{*}{$n=50000$} & \multicolumn{1}{c}{\multirow{4}{*}{Bias}} & 2NLS   & 0.0015     & 0.0104     & 0.0068     & 0.0043     & 0.0194     & 0.0281     & 0.0889     & 0.0370     & 0.2526 \\
                            & \multicolumn{1}{c}{}                      & MLE    & 0.0064     & 0.0260     & 0.0110     & 0.0201     & 0.0259     & 0.0835     & 0.2806     & 0.0132     & 0.5417 \\
                            & \multicolumn{1}{c}{}                      & M-BGD  & 0.0158     & 0.0093     & 0.0017     & 0.0063     & 0.0109     & 0.0188     & 0.0124     & 0.0248     & 0.1506 \\
                            & \multicolumn{1}{c}{}                      & S-BGD  & 0.0036     & 0.0060     & 0.0002     & 0.0043     & 0.0069     & 0.0134     & 0.0025     & 0.0105     & 0.0946 \\ \cline{2-12} 
                            & \multirow{4}{*}{RMSE}                     & 2NLS   & 0.0673     & 0.0349     & 0.0354     & 0.0302     & 0.0395     & 0.0516     & 0.1091     & 0.0786     & 0.2169 \\
                            &                                           & MLE    & 0.0699     & 0.0431     & 0.0390     & 0.0383     & 0.0475     & 0.0949     & 0.2878     & 0.0883     & 0.3646 \\
                            &                                           & M-BGD  & 0.0780     & 0.0395     & 0.0321     & 0.0368     & 0.0391     & 0.0571     & 0.0763     & 0.0827     & 0.2157 \\
                            &                                           & S-BGD  & 0.0675     & 0.0362     & 0.0318     & 0.0344     & 0.0360     & 0.0517     & 0.0777     & 0.0727     & 0.2026 \\ \hline
\end{tabular}

\bigskip{}
\bigskip{}
\bigskip{}

\centering{}\caption{Bias and RMSE of Second-Step Estimator: $U = \eta_1, V= U/2+\eta_2, \eta_1\sim \chi^2(5) - 5, \eta_2\sim \exp(3) - 2$}\label{table2}
\begin{tabular}{lllccccccccc}
\hline
                            &                                           & Method & $\beta_1$ & $\beta_2$ & $\beta_3$ & $\beta_4$ & $\beta_5$ & $\beta_6$ & $\beta_7$ & $\beta_8$ & all    \\ \hline
\multirow{8}{*}{$n=50000$}  & \multicolumn{1}{c}{\multirow{4}{*}{Bias}} & 2NLS   & 0.0193    & 0.0117    & 0.0055    & 0.0069    & 0.0121    & 0.0156    & 0.0307    & 0.0479    & 0.2008 \\
                            & \multicolumn{1}{c}{}                      & MLE    & 0.0207    & 0.0231    & 0.0112    & 0.0139    & 0.0357    & 0.0345    & 0.0636    & 0.1820    & 0.4335 \\
                            & \multicolumn{1}{c}{}                      & M-BGD  & 0.0449    & 0.0173    & 0.0137    & 0.0119    & 0.0159    & 0.0261    & 0.0431    & 0.0493    & 0.2753 \\
                            & \multicolumn{1}{c}{}                      & S-BGD  & 0.0160    & 0.0019    & 0.0027    & 0.0030    & 0.0004    & 0.0045    & 0.0049    & 0.0054    & 0.0868 \\ \cline{2-12} 
                            & \multirow{4}{*}{RMSE}                     & 2NLS   & 0.0765    & 0.0445    & 0.0393    & 0.0398    & 0.0450    & 0.0572    & 0.0875    & 0.0931    & 0.2326 \\
                            &                                           & MLE    & 0.0740    & 0.0487    & 0.0364    & 0.0389    & 0.0548    & 0.0618    & 0.0994    & 0.1942    & 0.2900 \\
                            &                                           & M-BGD  & 0.0858    & 0.0482    & 0.0407    & 0.0410    & 0.0493    & 0.0571    & 0.0939    & 0.1011    & 0.2415 \\
                            &                                           & S-BGD  & 0.0729    & 0.0405    & 0.0365    & 0.0360    & 0.0429    & 0.0486    & 0.0731    & 0.0769    & 0.2088 \\ \cline{2-12} 
\multirow{8}{*}{$n=100000$} & \multicolumn{1}{c}{\multirow{4}{*}{Bias}} & 2NLS   & 0.0003    & 0.0051    & 0.0055    & 0.0037    & 0.0181    & 0.0133    & 0.0172    & 0.0610    & 0.1612 \\
                            & \multicolumn{1}{c}{}                      & MLE    & 0.0008    & 0.0160    & 0.0153    & 0.0092    & 0.0427    & 0.0282    & 0.0508    & 0.1940    & 0.3906 \\
                            & \multicolumn{1}{c}{}                      & M-BGD  & 0.0139    & 0.0057    & 0.0066    & 0.0054    & 0.0050    & 0.0117    & 0.0184    & 0.0229    & 0.1228 \\
                            & \multicolumn{1}{c}{}                      & S-BGD  & 0.0021    & 0.0034    & 0.0002    & 0.0011    & 0.0058    & 0.0001    & 0.0072    & 0.0040    & 0.0538 \\ \cline{2-12} 
                            & \multirow{4}{*}{RMSE}                     & 2NLS   & 0.0561    & 0.0301    & 0.0258    & 0.0280    & 0.0336    & 0.0394    & 0.0566    & 0.0820    & 0.1692 \\
                            &                                           & MLE    & 0.0505    & 0.0327    & 0.0265    & 0.0278    & 0.0505    & 0.0457    & 0.0702    & 0.2005    & 0.2540 \\
                            &                                           & M-BGD  & 0.0565    & 0.0288    & 0.0261    & 0.0276    & 0.0285    & 0.0378    & 0.0517    & 0.0628    & 0.1564 \\
                            &                                           & S-BGD  & 0.0517    & 0.0277    & 0.0228    & 0.0255    & 0.0278    & 0.0347    & 0.0474    & 0.0556    & 0.1459 \\ \hline
\end{tabular}
\end{table}

\begin{table}[t!]
\small
\centering{}\caption{Bias and RMSE of Second-Step Estimator: $U = \eta_1, V= -U+\eta_2, \eta_1\sim Unif(-10, 10), \eta_2 \sim Unif(0,1)$}\label{table3}
\begin{tabular}{lllccccccccc}
\hline
                           &                                           & Method & $\beta_1$ & $\beta_2$ & $\beta_3$ & $\beta_4$ & $\beta_5$ & $\beta_6$ & $\beta_7$ & $\beta_8$ & all    \\ \hline
\multirow{8}{*}{$n=25000$} & \multicolumn{1}{c}{\multirow{4}{*}{Bias}} & 2NLS   & 0.0280    & 0.0003    & 0.0022    & 0.0087    & 0.0083    & 0.0202    & 0.0091    & 0.0839    & 0.2792 \\
                           & \multicolumn{1}{c}{}                      & MLE    & 0.0256    & 0.0028    & 0.0036    & 0.0009    & 0.0007    & 0.0146    & 0.0116    & 0.0619    & 0.2288 \\
                           & \multicolumn{1}{c}{}                      & M-BGD  & 0.0983    & 0.0977    & 0.0518    & 0.0517    & 0.0882    & 0.1352    & 0.2167    & 0.2505    & 1.1660 \\
                           & \multicolumn{1}{c}{}                      & S-BGD  & 0.0244    & 0.0002    & 0.0037    & 0.0029    & 0.0028    & 0.0205    & 0.0092    & 0.0018    & 0.1644 \\ \cline{2-12} 
                           & \multirow{4}{*}{RMSE}                     & 2NLS   & 0.1568    & 0.0896    & 0.0756    & 0.0825    & 0.0799    & 0.1285    & 0.1647    & 0.1867    & 0.4821 \\
                           &                                           & MLE    & 0.1168    & 0.0727    & 0.0654    & 0.0644    & 0.0701    & 0.0960    & 0.1219    & 0.1473    & 0.3832 \\
                           &                                           & M-BGD  & 0.2142    & 0.1549    & 0.1036    & 0.1111    & 0.1348    & 0.2140    & 0.3085    & 0.3383    & 0.7033 \\
                           &                                           & S-BGD  & 0.1532    & 0.0863    & 0.0744    & 0.0819    & 0.0774    & 0.1281    & 0.1617    & 0.1612    & 0.4619 \\ \cline{2-12} 
\multirow{8}{*}{$n=50000$} & \multicolumn{1}{c}{\multirow{4}{*}{Bias}} & 2NLS   & 0.0037    & 0.0048    & 0.0043    & 0.0010    & 0.0198    & 0.0106    & 0.0011    & 0.0792    & 0.2123 \\
                           & \multicolumn{1}{c}{}                      & MLE    & 0.0038    & 0.0050    & 0.0050    & 0.0029    & 0.0107    & 0.0087    & 0.0001    & 0.0591    & 0.1599 \\
                           & \multicolumn{1}{c}{}                      & M-BGD  & 0.0774    & 0.0515    & 0.0203    & 0.0346    & 0.0602    & 0.0551    & 0.1193    & 0.1435    & 0.6607 \\
                           & \multicolumn{1}{c}{}                      & S-BGD  & 0.0021    & 0.0060    & 0.0107    & 0.0016    & 0.0054    & 0.0106    & 0.0029    & 0.0054    & 0.1387 \\ \cline{2-12} 
                           & \multirow{4}{*}{RMSE}                     & 2NLS   & 0.1230    & 0.0528    & 0.0563    & 0.0583    & 0.0672    & 0.0826    & 0.1033    & 0.1269    & 0.3428 \\
                           &                                           & MLE    & 0.0932    & 0.0450    & 0.0480    & 0.0487    & 0.0561    & 0.0678    & 0.0782    & 0.0949    & 0.2783 \\
                           &                                           & M-BGD  & 0.1657    & 0.0864    & 0.0624    & 0.0712    & 0.0955    & 0.1009    & 0.1741    & 0.1958    & 0.4365 \\
                           &                                           & S-BGD  & 0.1180    & 0.0528    & 0.0522    & 0.0541    & 0.0618    & 0.0740    & 0.0996    & 0.0994    & 0.3172 \\ \hline
\end{tabular}

\bigskip{}
\bigskip{}
\bigskip{}

\centering{}\caption{Bias and RMSE of Second-Step Estimator: $U = \eta_1, V= U/3+\eta_2, \eta_1\sim N(0,9), \eta_2\sim N(0,4)$}\label{table4}
\begin{tabular}{lllccccccccc}
\hline
                           &                                           & Method & $\beta_1$ & $\beta_2$ & $\beta_3$ & $\beta_4$ & $\beta_5$ & $\beta_6$ & $\beta_7$ & $\beta_8$ & all    \\ \hline
\multirow{8}{*}{$n=25000$} & \multicolumn{1}{c}{\multirow{4}{*}{Bias}} & 2NLS   & 0.0090    & 0.0045    & 0.0020    & 0.0051    & 0.0044    & 0.0074    & 0.0029    & 0.0066    & 0.0813 \\
                           & \multicolumn{1}{c}{}                      & MLE    & 0.0101    & 0.0048    & 0.0028    & 0.0037    & 0.0045    & 0.0051    & 0.0010    & 0.0049    & 0.0739 \\
                           & \multicolumn{1}{c}{}                      & M-BGD  & 0.0368    & 0.0147    & 0.0115    & 0.0125    & 0.0171    & 0.0248    & 0.0347    & 0.0393    & 0.2380 \\
                           & \multicolumn{1}{c}{}                      & S-BGD  & 0.0114    & 0.0063    & 0.0028    & 0.0049    & 0.0062    & 0.0065    & 0.0030    & 0.0073    & 0.0873 \\ \cline{2-12} 
                           & \multirow{4}{*}{RMSE}                     & 2NLS   & 0.0482    & 0.0273    & 0.0213    & 0.0297    & 0.0271    & 0.0446    & 0.0582    & 0.0528    & 0.1524 \\
                           &                                           & MLE    & 0.0437    & 0.0244    & 0.0195    & 0.0254    & 0.0256    & 0.0388    & 0.0525    & 0.0476    & 0.1365 \\
                           &                                           & M-BGD  & 0.0602    & 0.0306    & 0.0242    & 0.0308    & 0.0312    & 0.0493    & 0.0684    & 0.0651    & 0.1668 \\
                           &                                           & S-BGD  & 0.0457    & 0.0262    & 0.0201    & 0.0272    & 0.0264    & 0.0406    & 0.0553    & 0.0487    & 0.1424 \\ \cline{2-12} 
\multirow{8}{*}{$n=50000$} & \multicolumn{1}{c}{\multirow{4}{*}{Bias}} & 2NLS   & 0.0003    & 0.0006    & 0.0014    & 0.0007    & 0.0007    & 0.0007    & 0.0000    & 0.0005    & 0.0292 \\
                           & \multicolumn{1}{c}{}                      & MLE    & 0.0004    & 0.0003    & 0.0016    & 0.0015    & 0.0005    & 0.0005    & 0.0006    & 0.0006    & 0.0265 \\
                           & \multicolumn{1}{c}{}                      & M-BGD  & 0.0144    & 0.0044    & 0.0029    & 0.0031    & 0.0056    & 0.0101    & 0.0173    & 0.0165    & 0.0965 \\
                           & \multicolumn{1}{c}{}                      & S-BGD  & 0.0010    & 0.0003    & 0.0012    & 0.0012    & 0.0002    & 0.0006    & 0.0004    & 0.0001    & 0.0277 \\ \cline{2-12} 
                           & \multirow{4}{*}{RMSE}                     & 2NLS   & 0.0323    & 0.0190    & 0.0167    & 0.0203    & 0.0174    & 0.0270    & 0.0392    & 0.0381    & 0.1043 \\
                           &                                           & MLE    & 0.0288    & 0.0171    & 0.0145    & 0.0175    & 0.0152    & 0.0244    & 0.0336    & 0.0341    & 0.0922 \\
                           &                                           & M-BGD  & 0.0354    & 0.0196    & 0.0161    & 0.0193    & 0.0180    & 0.0297    & 0.0385    & 0.0418    & 0.1055 \\
                           &                                           & S-BGD  & 0.0296    & 0.0179    & 0.0153    & 0.0184    & 0.0154    & 0.0256    & 0.0346    & 0.0369    & 0.0964 \\ \hline
\end{tabular}
\end{table}

We consider four competing methods, two being parametric and two semiparametric.
The first method is parametric estimation using nonlinear least squares (two-step NLS), which is similar to Heckman's two-step estimation but accounts for the binary response in the second stage. To apply such method, we assume that   $U_i$ and $V_i$ in (\ref{sel_eq}) and (\ref{out_eq}) have zero mean and unit variance, and are jointly normally distributed with covariance $\rho$.
Then we can jointly estimate $\boldsymbol{\delta}_0$, $\boldsymbol{\beta}_0$ together with $\rho$. In particular, let $F_1(\cdot)$ and $F_2(\cdot, \cdot, \rho)$ denote  the CDF's of univariate standard normal distribution and bivariate normal distribution with covariance matrix $[\sigma_{ij}]_{2\times2}$, where $\sigma_{11} = \sigma_{22} = 1$, and $\sigma_{12} = \sigma_{21} = \rho$. Also define $\overline{\boldsymbol{Z}}_{e,i} = \left(1, z_{0,i}, \boldsymbol{Z}_i^{\mathrm{T}}\right)^{\mathrm{T}}$, $\overline{\boldsymbol{X}}_{e,i} = \left(1, x_{0,i}, \boldsymbol{X}_i^{\mathrm{T}}\right)^{\mathrm{T}}$, $\overline{\boldsymbol{\delta}} = \left(c_{\delta, 0}, c_{\delta, 1}, \boldsymbol{\delta}^{\mathrm{T}}\right)^{\mathrm{T}}$, $\overline{\boldsymbol{\beta}} = \left(c_{\beta, 0}, c_{\beta, 1}, \boldsymbol{\beta}^{\mathrm{T}}\right)^{\mathrm{T}}$.  In the first step, we minimize the following loss function 
\[
L_{1,n}(\overline{\boldsymbol{ \delta}}) = \frac{1}{n}\sum_{i=1}^n (D_i - F_1(\overline{\boldsymbol{ Z}}_{e,i}^{\mathrm{T}}\overline{\boldsymbol{ \delta}}))^2
\]
and obtain the minimizer $\widehat{\overline{\boldsymbol{\delta}}}$. Then in the second step, we minimize the following loss function 
\[
L_{2,n}(\overline{\boldsymbol{\beta}}, \rho) = \frac{1}{S_n}\sum_{i=1}^nD_i\left(Y_i - \frac{F_2(\overline{\boldsymbol{Z}}_{e,i}^{\mathrm{T}}\widehat{\overline{\boldsymbol{\delta}}}, \overline{\boldsymbol{X}}_{e,i}^{\mathrm{T}}\overline{\boldsymbol{\beta}}, \rho)}{F_1(\overline{\boldsymbol{Z}}_{e,i}^{\mathrm{T}}\widehat{\overline{\boldsymbol{\delta}}})}\right)^2
\]
and obtain the minimizer $\widehat{\overline{\boldsymbol{\beta}}}$. Then the two-step NLS estimators for $\boldsymbol{\delta}_0$ and $\boldsymbol{\beta}_0$ are given by $\widehat c_{\delta, 1}^{-1}\widehat{\boldsymbol{\delta}}$ and $\widehat c_{\beta, 1}^{-1}\widehat {\boldsymbol{\beta}}$.

 The second method is parametric maximum likelihood estimation (MLE).  Like in the first method, we also assume that  $U_i$ and $V_i$ in (\ref{sel_eq}) and (\ref{out_eq}) are jointly normally distributed, then the log-likelihood function is then given by 
\begin{align*}
    L_{3,n}\left(\overline{\boldsymbol{ \delta}}, \overline{\boldsymbol{\beta}},  \rho \right) = \frac{1}{n}\sum_{i=1}^n& \left((1-D_i)\log(1-F_1(\overline{\boldsymbol{Z}}_{e,i}^{\mathrm{T}} \overline{\boldsymbol{\delta}})) + D_iY_i\log\left(F_2 (\overline{\boldsymbol{Z}}_{e,i}^{\mathrm{T}}\overline{\boldsymbol{\delta}}, \overline{\boldsymbol{X}}_{e,i}^{\mathrm{T}}\overline{\boldsymbol{\beta}}, \rho)\right)\right.\\
& \left.+ D_i(1-Y_i)\log\left(F_1(\overline{\boldsymbol{Z}}_{e,i}^{\mathrm{T}} \overline{\boldsymbol{\delta}})-F_2 (\overline{\boldsymbol{Z}}_{e,i}^{\mathrm{T}}\overline{\boldsymbol{\delta}}, \overline{\boldsymbol{X}}_{e,i}^{\mathrm{T}}\overline{\boldsymbol{\beta}}, \rho)\right)\right).
\end{align*}
Suppose the MLE estimators are given by $\widehat{\overline{\boldsymbol{\delta}} }$ and  $\widehat{\overline{\boldsymbol{\beta}}}$, then the MLE estimators for $\boldsymbol{\delta}_0$ and $\boldsymbol{\beta}_0$ are given by $\widehat c_{\delta, 1}^{-1}\widehat{\boldsymbol{\delta}}$ and $\widehat c_{\beta, 1}^{-1}\widehat{\boldsymbol{\beta}}$.

The third method is semiparametric estimation based on matching. In particular, we first obtain the estimator of $\delta_0$ in the first step, then we conduct Algorithm 1. To improve the computational efficiency, in the first and second step estimation, we use nearest neighbor matching with $m=[(\log(n))^{1.1}]$ and $m=[(\log(S_n))^{1.1}]$, where recall that $S_n = \sum_{i=1}^nD_i$. We update 2000 times for both first- and second-step estimation. 

The fourth method is semiparametric estimation based on series approximation. For the sieve functions, we consider the Legendre polynomials used in \citet{khanetal2021}, and use tensor products of one-variate sieve functions as the sieve functions for bivariate functions. The order of sieves is chosen to be 31 for the first-step estimation and 15 for the second-step estimation. Finally, the stopping rule is the same as in that of \citet{khanetal2021} with the tolerance being $10^{-6}$.  

We report the bias and the root mean squared error (RMSE) of the second-step estimator for  all methods. Let the simulation be repeated for $R$ times, and in the $r$-th round of repetition the estimator of $\beta_{j, 0}$ be $\widehat \beta_j^r$. Then the bias of  the estimator is given by $B_j = |\frac{1}{R}\sum_{r=1}^R\widehat \beta_{j}^r - \beta_{j,0}|$, and the RMSE is given by $RMSE_j = \sqrt{\frac{1}{R}\sum_{r=1}^R(\widehat \beta_{j}^r - \beta_{j,0})^2}$.  We report the bias and RMSE for $\beta_{1,0}$ through  $\beta_{8,0}$. We also report the total bias and RMSE, which are defined as $B_{\text{all}} = \sum_{j=1}^{p_X}B_j$ and $RMSE_{\text{all}} = \sqrt{\sum_{j=1}^{p_X} RMSE_j^2}$. We choose $R=100$. Results are reported in \autoref{table1} to \autoref{table4}.

\autoref{table1} through \autoref{table4} correspond to fours setups of the error terms. The first three setups feature model misspecification, while the last one features correct specification. Moreover, in the first setup  both $U_i$ and $V_i$ are  symmetric and are positively correlated. For the second step, $U_i$ and $V_i$ are also positive correlated but are both heavily skewed. The third setup corresponds to the case in which $U_i$ and $V_i$ are both bounded and are strongly (negatively) correlated with each other.  Several insights can be drawn from the simulation results. First of all,  when the model is misspecified, that is, the error terms $U_i,V_i$ are not jointly normally distributed, parametric estimators always have nonvanishing sizable bias, while the semiparametric sieve-based estimator constantly has minimal bias which is close to zero. This highlights the fact that the key advantage of semiparametric estimation lies in the robustness to model misspecifications. Further comparisons of the RMSE between different methods reveal that even though parametric estimators have $O(1/\sqrt{n})$ standard deviation, the sizable bias significantly contaminates the RMSE so that the sieve-based estimator always outperforms two parametric estimators in terms of RMSE. Notably, when the model is correctly specified (\autoref{table4}), both two-step NLS and joint MLE estimators have trivial bias, and joint MLE estimators have smallest RMSE among all methods. However, even in this case, the sieve-based semiparametric estimator is competitive compared with two-step NLS in terms of RMSE. We finally point out that the performance of matching-based semiparametric estimator is less promising compared with the sieve-based on in terms of both bias and RMSE. This may be explained by limited number of iterations (2000 iterations for both steps) or small neighborhood size $k$ ($k= \log(n)^{1.1}$ for first step and $k = \log(S)^{1.1}$ for second step). Nevertheless, as we mentioned above, the matching-based estimator can be used as a decent initial point for sieve-based estimator.

\section{Empirical Application: Stanford Policing Data}\label{section5}

\begin{table}[!t]

\begin{center}
\caption{Data Description}
    \label{emp_table1}
\begin{tabular}{>{\raggedright}p{5cm}>{\raggedright}p{8cm}>{\raggedright}p{1.5cm}}
\hline 
Variable & Value & Mean\tabularnewline
\hline 
Search & $1$ if the police officer conducts a search, and $0$ otherwise & 0.0423\tabularnewline
Contraband Found & $1$ if the police officer conducts a search and contraband items
are found, and $0$ otherwise & 0.0088\tabularnewline
Age & Ranging from 10 to 99 & 37.073\tabularnewline
Black & $1$ if the stopped individual is black, and 0 otherwise & 0.3859\tabularnewline
White & $1$ if the stopped individual is white, and 0 otherwise & 0.5470\tabularnewline
Gender & $1$ if female and $0$ if male & 0.4095\tabularnewline
Precinct1  & $1$ if the police officer is from precinct1, and 0 otherwise & 0.1138\tabularnewline
Precinct2  & $1$ if the police officer is from precinct 2, and 0 otherwise & 0.1420\tabularnewline
Precinct3  & $1$ if the police officer is from precinct 3, and 0 otherwise & 0.1558\tabularnewline
Precinct4 & $1$ if the police officer is from precinct 4, and 0 otherwise & 0.0873\tabularnewline
Precinct5  & $1$ if the police officer is from precinct 5, and 0 otherwise & 0.1668\tabularnewline
Precinct6  & $1$ if the police officer is from precinct 6, and 0 otherwise & 0.1068\tabularnewline
Precinct7  & $1$ if the police officer is from precinct 7, and 0 otherwise & 0.0729\tabularnewline
Reason1 & $1$ if the reason for stop is ``investigative stop'', and 0 otherwise & 0.0186\tabularnewline
Reason2 & $1$ if the reason for stop is ``moving traffic violation'', and 0
otherwise & 0.4952\tabularnewline
Reason3 & $1$ if the reason for stop is ``registration'', and 0 otherwise & 0.0616\tabularnewline
Reason4 & $1$ if the reason for stop is ``safety violation'', and 0 otherwise & 0.0607\tabularnewline
Reason5 & $1$ if the reason for stop is ``seatbelt violation'', and 0 otherwise & 0.0345\tabularnewline
\hline 
\end{tabular}
\par\end{center}
Note: We leave out data points whose ethnicity is Asian. When Black and White both take  value  0, the stopped individual is Hispanic. 
\end{table}

\begin{table}[!t]

\begin{center}
\caption{Estimation Results of Stanford Policing Data}
    \label{emp_table2}
\begin{tabular}{>{\raggedright}p{3cm}>{\raggedright}p{3cm}>{\raggedright}p{3cm}}
\hline 
  & First-Step Results & Second-Step Results \tabularnewline
\hline 
Age & $-0.0229^{***}$\\$(0.0004)$ & $-0.0541^{***}$\\$(0.0103)$ \tabularnewline
Black & $0.0785^{***}$\\$(0.0073)$ & $2.6126^{***}$\\$(0.2734)$\tabularnewline
White & $-0.3200^{***}$\\$(0.0092)$ & $2.5389^{***}$\\$(0.2422) $\tabularnewline
Gender & $-0.5583^{***}$\\$(0.0106)$ & $-0.3493^{***}$\\$(0.1015)$\tabularnewline
Precinct1  & $0.0028^{***}$\\$(0.0084)$ & \tabularnewline
Precinct2  &  $0.1574^{***}$\\$(0.0077)$& \tabularnewline
Precinct3  & $-0.0966^{***}$\\(0.0076) & \tabularnewline
Precinct4 & $-0.1228^{***}$\\$(0.0094)$ & \tabularnewline
Precinct5  & $0.1537^{***}$\\$(0.0074)$ & \tabularnewline
Precinct6  & $0.0813^{***}$\\$(0.0080)$ & \tabularnewline
Precinct7  & $0.3827^{***}$\\$(0.0106)$ & \tabularnewline
Reason2 & $-0.0701^{***}$\\$(0.0046)$ & $0.0800$\\$(0.1080)$\tabularnewline
Reason3 & $0.0149^{***}$\\$(0.0081)$ & $0.4221^{**}$\\$(0.2013)$\tabularnewline
Reason4 & $-0.0652^{***}$\\$(0.0085)$ & $0.1107$\\$(0.1261)$\tabularnewline
Reason5 & $0.0596^{***}$\\$(0.0100)$ & $0.4439^{**}$\\$(0.2112)$\tabularnewline
\hline 
\end{tabular}\\

Note: $^{***}$ and $^{**}$ indicate significance at 1\% and 5\% level of significance. 
\par\end{center}
\end{table}

In this section we apply our new algorithm to analyze the Stanford Open Policing Project data set \citep{pierson2020large}. This large-scale data set records traffic stops made by police officers and various features related to the stops so can be used to provide insights into the potential  racial disparities during police stops, which has been widely studied in the existing literature \citep{goel2016precinct}. 

In this paper we use the data set from Nashville, which contains a total of 3092351	 raw observations and 2608109 observations after data clearing. To analyze this data set, we define the first-stage binary outcome as, after a stop has been made, whether the police officer further decides to conduct a search for  the person or vehicle.  Conditioned on that   a search has been made, the second stage outcome is  whether the police officer found any contraband items such as drugs or weapons. For the first stage equation, we consider a series of regressors including police precinct, the reason for the stop, as well as the gender, race, and age of the stopped subject. For the second stage equation, we consider all the regressors in the first stage but the precinct so that the precinct is used as the exclusion restriction.  We provide detailed description of the variables in \autoref{emp_table1}.

When conducting SBGD estimation, we set the coefficient of Reason1 to be 1 for both selection and outcome equations, whose reason is detailed as follows. When police officer stops individuals to conduct investigation, it's highly likely that the police officer notice something unusual and hence a search is more likely to follow and contraband items are more likely to be found. This implies that the coefficients of Reason1 in both selection and outcome equations should be positive. We also note that the ratio of maximum and minimum eigenvalues of the covariate matrices for both selection and outcome equation is extremely large, which may lead to poor numerical performance during iterations. For example, in this case small learning rate $\gamma_k$ has to be chosen during the estimation of both stages to guarantee convergence of the algorithms, which leads to extensive rounds of iterations. To solve this issue, we use Gram-Schimidt method to orthogonalize the covariate matrix so that the resulting covariate matrix has unit column variance and zero column correlation, and the first column of the orthogonalized covariate matrix is equal to Reason1. See \autoref{details of empirical applications} for more details of the empirical setups.  The estimation results are reported in \autoref{emp_table2}.

Several insights can be drawn from the results. First of all, age is negatively related to the conditional probability of being searched after a stop is made; it is also negatively associated with the probability that illegal items are found. Second, compared with male individuals, female individuals are less likely to be searched after the stop, and conditioned on that the stopped individual is searched, female individuals are less likely to be found carrying illegal items. Finally, for racial disparities during stops,  we find that the coefficient of Black in the selection equation is significantly positive, while that of white is negative.  This implies that when other conditions are held constant,   individuals are more likely to be searched (compared with Hispanic individuals) if they are black, while white people are less likely to be searched. We further notice that for the outcome equation, the coefficients of Black and White almost coincide with each other, implying that conditioned on being searched, black people are almost as likely as the white people to be found carrying contraband items. Such a result implies that even though black community may confront certain disparities in terms of a search decision, the hit rate (of finding illegal items)  does not differ across races.

\section{
  Selective Labeling with Endogenous Treatment
}\label{section6}

This section extends the previous estimation method to selective labeling models with endogenous treatments. In his seminal work, 
\cite{leebounds} considers  partial identification of the  treatment parameters in treatment effect models with attrition but does not allow for explanatory variables. \citet{SEMENOVA2025106055} extends \citet{leebounds}'s bound and studies the asymptotic properties of the proposed bounds under both fixed and high dimensionality. For more discussion on \citet{leebounds}'s model and its recent development, see \citet{SEMENOVA2025106055} and references therein. 

Extending our selective labeling model to allow for    endogenous treatment status which is  denoted by  $T_i$   can be expressed as 
\begin{align}
&T_i= I\left(r_{0,i} + \boldsymbol{R}_i^{\mathrm{T}}\boldsymbol{\varphi}_0 > W_i\right), \label{treateq}\\
&D_i=I\left(z_{0,i}+\boldsymbol{Z}_i^{\mathrm{T}}\boldsymbol{\delta}_0 +\tau_{1,0} T_i  > U_i\right),\label{seleceq}\\
&Y_i=D_i\cdot I\left(x_{0,i}+\boldsymbol{X}_i^{\mathrm{T}}\boldsymbol{\beta}_0 +\tau_{2,0} T_i> V_i\right), \label{outceq}
\end{align}
where $\boldsymbol{R}_{e,i} = (r_{0,i}, \boldsymbol{R}_{i}^{\mathrm{T}})^{\mathrm{T}}\in R^{p_R+1}$ and $\boldsymbol{\varphi}_{0}\in R^{p_R}$.  In the above system of equations, the observed binary variable $D_i$ indicates
whether or not the $i$-th agent outcome variable is observed in the sample, $Y_i$ denotes the observed outcome variable for the selected sample, with selection governed by $D_i$, and $T_i$ denotes an observed binary variable indicating treatment status, whose coefficient in the outcome equation, $\tau_{2,0}$,  is the main parameter of interest. Such system of equations generalizes \citet{leebounds}'s original model by explicitly modeling the determination of the treatment status. Moreover, since  the unobserved random errors $(W_i, U_i,V_i)$ are potentially mutually correlated,  
 the treatment status is allowed to be endogenous and correlated with both the selection status $D_i$ and the outcome $Y_i$.  We also note  that $\boldsymbol{R}_{e,i}, \boldsymbol{Z}_{e,i},\boldsymbol{X}_{e,i}$ can be large dimensional, which, similar to the selective labeling models, makes the estimation of the above system computationally intensive.  

The above system of selective labeling model with endogenous treatment nests many models in important work in the literature.
In a model where for $D_i=1$, $Y_i$ was linear and there were no regressors besides $T_i$, \cite{leebounds} considered {\em partial} identification of $\tau_{2,0}$. In a model where there was no selection/attrition issue so $D_i$ was identical to 1,  identification and estimation were considered 
in \cite{vytlacilyildiz}, \cite{abhausmankhan}, \cite{shaikhvytlacil}. \cite{vytlacilyildiz} attained point identification under a monotonicity condition as well as support conditions on exogenous covariates effecting $Y_i$. See also \cite{khanmaurelzhang} and \cite{chenkhantang} for point identification results in similar models. However, none of the methods proposed in the above papers are applicable to the model above where there are endogenous treatment, attrition, and selective labeling on the same time, even for low dimensional models. In contrast, the methods we introduced in Section \ref{algorithms}
can be extended to estimate  $\boldsymbol{\varphi}_0 , \boldsymbol{\delta}_0, \boldsymbol{\beta}_0$, and $\tau_{1,0},\tau_{2,0}$ even when the covariates in the system are all large dimensional. 

Before we demonstrate our main algorithm, we first  show point identification for the parameters in the above system under a set of conditions. Denote the marginal CDF of $W_i$, the joint CDF of $W_i$ and $U_i$, and the joint CDF of $W_i$, $U_i$, and $V_i$ as $F_W(w)$,  $F_{W,U}(w,u)$, and $F_{W,U,V}(w,u,v)$, respectively. Moreover,  rearrange the order of the regressors such that $\boldsymbol{R} = (\boldsymbol{R}_c^{\mathrm{T}}, \boldsymbol{R}_d^{\mathrm{T}})^{\mathrm{T}}$, where $\boldsymbol{R}_c$ is continuous and $\boldsymbol{R}_d$ is discrete. Similarly, we write $\boldsymbol{Z} = (\boldsymbol{Z}_c^{\mathrm{T}}, \boldsymbol{Z}_d^{\mathrm{T}})^{\mathrm{T}}$  and $\boldsymbol{X} = (\boldsymbol{X}_c^{\mathrm{T}}, \boldsymbol{X}_d^{\mathrm{T}})^{\mathrm{T}}$.  We impose the following conditions.

\begin{condition}\label{condition10}
 $\{T_i, D_i,Y_i, \boldsymbol{R}_{e,i},\boldsymbol{Z}_{e,i},\boldsymbol{X}_{e,i},W_{i}, U_{i},V_{i}\}$ satisfy (\ref{treateq})--(\ref{outceq}) and are iid over $i$. The random errors $W_i, U_{i}$ and $V_{i}$
	are jointly independent of $\boldsymbol{R}_{e,i}$, $\boldsymbol{Z}_{e,i}$ and $\boldsymbol{X}_{e,i}$. We observe the data set $\mathcal{S}_{n}=\left\{ T_i, D_{i},Y_{i},\boldsymbol{R}_{e,i},\boldsymbol{Z}_{e,i},\boldsymbol{X}_{e,i}\right\} _{i=1}^{n}$.
    \end{condition}
    
    \begin{condition}\label{condition11} (i) $F_W(\cdot)$  has continuous derivative; (ii) $r_0$ is continuous; (iii) $\boldsymbol{R}_{e}\in\mathcal{R}_e\subseteq R^{p_R+1}$, and moreover,    $E(T|\boldsymbol{R}_{e})$  can be estimated consistently uniformly for $\boldsymbol{R}_{e}\in\mathcal{R}_e$; (iv) there exists at least one interior point $\boldsymbol{R}_e^{1}\in \mathcal{R}_e$ such that  $\nabla_{r_0} E(T|\boldsymbol{R}_e^1)  \neq 0 $; (v) There exist  points $\boldsymbol{R}_{e}^2, \boldsymbol{R}_{e}^3, \cdots, \boldsymbol{R}_{e}^{p_R^d+2}\in\mathcal{R}_e$  such that $\boldsymbol{R}_{e}^2$ is interior to $\mathcal{R}_e$,  $\nabla_{r_0} E(T|\boldsymbol{R}_e^2)  \neq 0 $ and $E(T|\boldsymbol{R}_e^2) = E(T|\boldsymbol{R}_e^j)$ for all $3\leq j\leq p_R^{d}+2$, where $p_R^d$ is the dimension of $\boldsymbol{R}_d$, and moreover, $(\boldsymbol{R}_d^3 -\boldsymbol{R}_d^2 , \cdots, \boldsymbol{R}_d^{p_R^{d}+2}-\boldsymbol{R}_d^2)$ has full rank. 
    \end{condition}

    \begin{condition}\label{condition12}(i) $\nabla_{w} F_{W,U}(w,u)$, $\nabla_u F_{W,U}(w,u) $,  and $\nabla_{wu} F_{W,U}(w,u)$  exist and are continuous; (ii) $z_0$ is continuous; (iii) define  $R_0 = r_0 + \boldsymbol{R}^{\mathrm{T}}\boldsymbol{\varphi}_0$, $(R_0, \boldsymbol{Z}_e^{\mathrm{T}})^{\mathrm{T}}\in \overline{\mathcal{Z}}_e\subseteq R^{ 2 + p_Z}$, and  $E(T|R_0)$, $E(D|T=1, R_0, \boldsymbol{Z}_e)$ and $E(D|T=0, R_0, \boldsymbol{Z}_e)$   can be estimated consistently uniformly for any point $(R_0, \boldsymbol{Z}_e^{\mathrm{T}})^{\mathrm{T}}\in \overline{\mathcal{Z}}_e$; (iv)  there exists at least one interior point $(R_0^1, \boldsymbol{Z}_e^{1})\in \overline{\mathcal{Z}}_e$ such that   $\nabla_{z_0}E(D|T=1, R_0, \boldsymbol{Z}_e)   \neq 0 $ or $\nabla_{z_0} E(D|T=0, R_0, \boldsymbol{Z}_e)   \neq 0 $; (v) There exist $R_0^2$ and $\boldsymbol{Z}_{e}^2, \cdots, \boldsymbol{Z}_{e}^{p_Z^d+2}$ such that $(R_0^2, (\boldsymbol{Z}_{e}^{2})^{\mathrm{T}})^{\mathrm{T}}$ is interior to $\overline{\mathcal{Z}}_e$, and for $3\leq j\leq p_Z^d$ there hold $(R_0^2, (\boldsymbol{Z}_{e}^{j})^{\mathrm{T}})^{\mathrm{T}}\in \overline{\mathcal{Z}}_e$, $\nabla_{z_0} E(D|T=\iota, R_0^2,\boldsymbol{Z}_e^2) \neq 0 $, and $E(D|T=\iota, R_0^2,\boldsymbol{Z}_e^2) = E(D|T=\iota, R_0^2, \boldsymbol{Z}_e^j)$, where $\iota = 0$ or 1; moreover, $(\boldsymbol{Z}_d^3 -\boldsymbol{Z}_d^2 , \cdots, \boldsymbol{Z}_d^{p_Z^{d}+2}-\boldsymbol{Z}_d^2)$ has full rank; (vi) Define  $Z_0 = z_0 + \boldsymbol{Z}^{\mathrm{T}}\boldsymbol{\delta}_0$, there exist interior points $(R_0^{\prime}, (\boldsymbol{Z}_0^{\prime})^{\mathrm{T}})^{\mathrm{T}}, (R_0^{\prime}, (\boldsymbol{Z}_0^{\prime\prime})^{\mathrm{T}})^{\mathrm{T}}\in \overline{\mathcal{Z}}_e$ such that $0<E(T|R_0^{\prime})<1$, $\nabla_{R_0, Z_0} [E(D|T =1, {R}_0^{\prime},{Z}_0^{\prime} )E(T|R_0^{\prime})]\neq0 $, and  $-\nabla_{R_0} [E(D|T =0, {R}_0^{\prime},{Z}_0^{\prime} )(1-E(T|R_0^{\prime}))]  = \nabla_{R_0}[E(D|T =1, {R}_0^{\prime\prime},{Z}_0^{\prime\prime} )E(T|R_0^{\prime\prime})]$.

\end{condition}

\begin{condition}\label{condition13}(i) $\nabla_{v} F_{W,U,V}(w,u,v) $, $\nabla_{wu} F_{W,U,V}(w,u,v) $,  and $\nabla_{wuv} F_{W,U,V}(w,u,v)$  exist and are continuous; (ii) $x_0$ is continuous; (iii)  $(R_0, Z_0, \boldsymbol{X}_e^{\mathrm{T}})^{\mathrm{T}}\in \overline{\mathcal{X}}_e\subseteq R^{ 3 + p_X}$, and $P(T = 1, D = 1|R_0, Z_0)$, $P(T = 0, D = 1|R_0, Z_0)$,   $E(Y|T=1, D = 1,  R_0, Z_0, \boldsymbol{X}_e)$ and $E(Y|T=0, D= 1, R_0, Z_0, \boldsymbol{X}_e)$   can be estimated consistently uniformly for any point $(R_0, Z_0, \boldsymbol{X}_e^{\mathrm{T}})^{\mathrm{T}}\in \overline{\mathcal{X}}_e$; (iv)  there exists at least one interior point $(R_0^1, Z_0^1, \boldsymbol{X}_e^{1})\in \overline{\mathcal{X}}_e$ such that   either $\nabla_{x_0} E(Y|T=1, D= 1,  R_0^1, Z_0^1, \boldsymbol{X}_e^1)   \neq 0 $ or $\nabla_{x_0} E(Y|T=0, D= 1, R_0^1, Z_0^1, \boldsymbol{X}_e^1) \neq 0 $; (v) There exist $R_0^2, Z_0^2$ and $\boldsymbol{X}_{e}^2, \cdots, \boldsymbol{X}_{e}^{p_X^d+2}$ such that $(R_0^2, Z_0^2, (\boldsymbol{X}_{e}^{2})^{\mathrm{T}})^{\mathrm{T}}$ is interior to $\overline{\mathcal{X}}_e$, and for $3\leq j\leq p_Z^d$  there hold $(R_0^2, Z_0^2,  (\boldsymbol{X}_{e}^{j})^{\mathrm{T}})^{\mathrm{T}}\in \overline{\mathcal{X}}_e$,  $\nabla_{x_0} E(Y|T=\iota, D= 1,  R_0^2, Z_0^2, \boldsymbol{X}_e^2)   \neq 0 $ and  $E(Y|T=\iota, D = 1, R_0^2, Z_0^2, \boldsymbol{X}_e^2) = E(Y|T=\iota, D= 1, R_0^2, Z_0^j, \boldsymbol{X}_e^j)$,  where $\iota = 0$ or 1; moreover, $(\boldsymbol{X}_d^3 -\boldsymbol{X}_d^2 , \cdots, \boldsymbol{X}_d^{p_Z^{d}+2}-\boldsymbol{X}_d^2)$ has full rank; (vi) Define  $X_0 = x_0 + \boldsymbol{X}^{\mathrm{T}}\boldsymbol{\beta}_0$, there exist interior points $(R_0^{\prime}, Z_0^{\prime},  (\boldsymbol{X}_0^{\prime})^{\mathrm{T}})^{\mathrm{T}}, (R_0^{\prime}, Z_0^{\prime}, (\boldsymbol{X}_0^{\prime\prime})^{\mathrm{T}})^{\mathrm{T}}\in \overline{\mathcal{X}}_e$ such that $P(T = 1, D = 1|R_0^{\prime}, Z_0^{\prime})>0$, $P(T = 0, D = 1|R_0^{\prime}, Z_0^{\prime})>0$, $\nabla_{R_0, Z_0, X_0} [E(Y|T =1, D=1, {R}_0^{\prime},{Z}_0^{\prime}, X_0^{\prime} )P(T=1, D=1|R_0^{\prime}, Z_0^{\prime})] \neq0 $, and  $-\nabla_{R_0, Z_0} [E(Y|T =0, D=1, {R}_0^{\prime},{Z}_0^{\prime}, X_0^{\prime})(1-P(T=0, D=1|R_0^{\prime}), Z_0^{\prime})]  = \nabla_{R_0, Z_0}[E(Y|T =1, D=1, {R}_0^{\prime},{Z}_0^{\prime}, X_0^{\prime\prime})P(T=1, D=1|R_0^{\prime}, Z_0^{\prime})] $.

\end{condition}

\begin{remark}
\autoref{condition10}--\autoref{condition13} are high level but can be easily broken down to mild  conditions  that are commonly used in the identification literature.  \autoref{condition10} specifies the data structure that we observe. Part (ii) of  \autoref{condition11} -- \autoref{condition13} requires that in each equation among (\ref{treateq})--(\ref{outceq}), the covariate whose coefficient is normalized to 1 is continuous. Since identification of each equation generally requires at least one continuous regressor whose coefficient is not zero, we can choose such regressor and normalize its coefficient. Part (iii) of \autoref{condition11} -- \autoref{condition13} requires accessibility of the values of the conditional expectations. In practice, uniform consisteny may only hold for a subset of features spaces. But as long as all the conditions hold for a known truncated feature space, then the identification results are still valid.   Moreover, for \autoref{condition12} (iii) and \autoref{condition13} (iii), the estimability of the conditional expectations may require some exclusion restrictions and support conditions. For example, when $E(D|T=1, R_0, \boldsymbol{Z}_e)$ can be  consistently estimated, it's generally required that at least one continuous argument of $\boldsymbol{R}_e$ is not included in $\boldsymbol{Z}_e$, and such argument has large support. Also note that the estimability of the conditional expectations also imposes requirements  on the rate of divergence of the dimensionality ($p_R, p_Z,$ and $p_X$) of each model under increasing dimensionality. Part (iv) and (v)
 of the above conditions generally require  that the error term of each equation has large enough support.      Finally, part (vi) of \autoref{condition12} and \autoref{condition13} allows us to identify the treatment effects $\tau_{1,0}$ and $\tau_{2,0}$ by matching the partial derivatives of the distribution functions, which is similar to \citet{abhausmankhan} and \citet{khanmaurelzhang}. 
\end{remark}

Based on the above conditions, we have the following identification results. 

 \begin{theorem}\label{theorem4}
If \autoref{condition10} and \autoref{condition11} hold, then $\boldsymbol{\varphi}_0$ is point identified. If \autoref{condition12} additionally holds, then $\boldsymbol{\delta}_0$ and $\tau_{1,0}$ are identified. Finally, if \autoref{condition13} further holds, then  $\boldsymbol{\beta}_0$ and $\tau_{2,0}$ are identified.  
    \end{theorem}

Given the identification results, now  we  can illustrate our estimation method in detail, which involves four steps. In particular, we first sequentially estimate $\boldsymbol{\varphi}_0$, $\boldsymbol{\delta}_0$, and $\boldsymbol{\beta}_0$ in the first three steps. Then with the estimators in hand, we finally estimate $\tau_{1,0}$ and $\tau_{2,0}$. First of all, note that the parameter $\boldsymbol{\varphi}_0$ in  equation (\ref{treateq}) indicating the treatment status  can be readily estimated using the SBGD algorithm proposed in \citet{khanetal2021} because it is a binary choice process. Denote the first-step estimator as $\widehat{\boldsymbol{\varphi}}$. 

In the following, we use $\boldsymbol{\Phi}_q$ to denote generic vectors of sieves functions, whose length depends on $q$ and arguments may have different dimensions depending on the specific functions we would like to approximate. Correspondingly, we use $\boldsymbol{\Pi}_q$ to denote the pseudo true sieve parameters. Now we proceed to the second step. Note that 
\begin{align*}
E\left(\left.D_i\right|T_i = 1, \boldsymbol{R}_{e,i}, \boldsymbol{Z}_{e,i}\right) & = \frac{F_{W,U} \left(R_{0,i},Z_{0,i}+\tau_{1,0}\right)}{F_W(R_{0,i})} \equiv G_1\left(R_{0,i},Z_{0,i}\right), 
\end{align*}
\begin{align*}
E\left(\left.D_i\right|T_i = 0, \boldsymbol{R}_{e,i}, \boldsymbol{Z}_{e,i}\right) & = \frac{F_{U}(Z_{0,i}) - F_{W,U} \left(R_{0,i},Z_{0,i}\right)}{ 1 - F_W(R_{0,i})} \equiv G_2\left(R_{0,i},Z_{0,i}\right), 
\end{align*}
where $G_1(\cdot, \cdot)$ and $G_2(\cdot, \cdot)$ are both monotonically increasing in their second argument. Following the intuition of Algorithm 2 in Section \ref{algorithms}, we have the following estimation procedure for $\boldsymbol{\delta}_0$. 

\fbox{\begin{minipage}\textwidth
    
\textbf{Algorithm 3.1 for Estimating $\boldsymbol{\delta}_0$:}
\begin{enumerate}
\item Start with $k=0$, the first-step estimator $\widehat {\boldsymbol{\varphi}}$, initial guess of $\boldsymbol{\delta}_0$, $\widehat {\boldsymbol{\delta}}^0$, initial guesses of the sieve parameter $\widehat {\boldsymbol{\Pi}}_{1,q}^0$ and $\widehat {\boldsymbol{\Pi}}_{2,q}^0$, and initial guesses of the conditional expectation functions $\widehat G^0_1(w,u)$ and $\widehat G^0_2(w,u)$.  
\item In the $k$-th round,  with $\widehat{\boldsymbol{\delta}}^k$, update $\widehat {\boldsymbol{\Pi}}_{1,q}^{k}$  and $\widehat {\boldsymbol{\Pi}}_{2,q}^{k}$to $\widehat {\boldsymbol{\Pi}}_{1,q}^{k+1}$ and $\widehat {\boldsymbol{\Pi}}_{2,q}^{k+1}$ using 
\begin{align*}
\widehat{\boldsymbol{\Pi}}^{k+1}_{1,q} =  \left[\sum_{i=1}^n T_i\boldsymbol{\Phi}_{q}\left(\widehat{R}_i,  \widehat{Z}_i^k\right)\boldsymbol{\Phi}_q\left(\widehat{R}_i,  \widehat{Z}_i^k\right)^{\mathrm{T}}\right]^{-1}\times
\left[\sum_{i=1}^n T_iD_i\boldsymbol{\Phi}_q\left(\widehat{R}_i,  \widehat{Z}_i^k\right)\right],
\end{align*}
\begin{align*}
\widehat{\boldsymbol{\Pi}}^{k+1}_{2,q} =  \left[\sum_{i=1}^n (1-T_i)\boldsymbol{\Phi}_{q}\left(\widehat{R}_i,  \widehat{Z}_i^k\right)\boldsymbol{\Phi}_q\left(\widehat{R}_i,  \widehat{Z}_i^k\right)^{\mathrm{T}}\right]^{-1}\times
\left[\sum_{i=1}^n (1-T_i)D_i\boldsymbol{\Phi}_q\left(\widehat{R}_i,  \widehat{Z}_i^k\right)\right],
\end{align*}
where $\widehat R_i = r_{0,i} + \boldsymbol{R}_i^{\mathrm{T}}\widehat{\boldsymbol{\varphi}}$, and $\widehat Z_i^k = z_{0,i} + \boldsymbol{Z}_i^{\mathrm{T}}\widehat{\boldsymbol{\delta}}^k$.

\item With $\widehat {\boldsymbol{\Pi}}^{k+1}_{1,q}$ and $\widehat {\boldsymbol{\Pi}}^{k+1}_{2,q}$, update $\widehat G_1^k(w,u)$ and $\widehat G_2^{k}(w,u)$ to $\widehat G_1^{k+1}(w,u)$ to $\widehat G_2^{k+1}(w,u)$ using $\widehat G_1^{k+1} \left(w, 
u\right)= \boldsymbol{\Phi}_{1}( w, u)^{\mathrm{T}}\widehat{\boldsymbol{\Pi}}_{1,q}^{k+1}$ and $\widehat G_2^{k+1} \left(w, 
u\right)= \boldsymbol{\Phi}_q( w, u)^{\mathrm{T}}\widehat{\boldsymbol{\Pi}}_{2,q}^{k+1}$.
\item Update $\widehat {\boldsymbol{\delta}}^k $ to $\widehat {\boldsymbol{\delta}}^{k+1}$ using
\[
\widehat {\boldsymbol{\delta}}^{k+1} = \widehat {\boldsymbol{\delta}}^{k} - \frac{\gamma_k}{n} \sum_{i=1}^n \left( T_i\widehat G_1^{k+1}\left(\widehat{R}_i, \widehat{Z}_{i,k}\right) + (1-T_i) \widehat G_2^{k+1}\left(\widehat{R}_i, \widehat{Z}_{i,k}\right) - D_i\right)\boldsymbol{Z}_i,
\]
where $\gamma_k>0$ is the learning rate.
\item Set $k = k+1$ and go back to Step 2 unless some terminating conditions are satisfied.
\end{enumerate}
\end{minipage}}

In the third step, we estimate $\boldsymbol{\beta}_0$ based on the first-step estimator $\widehat{\boldsymbol{\varphi}}$ and second-step estimator $\widehat{\boldsymbol{\delta}}$. The estimation procedure is also motivated by the following observation 
\begin{align*}
& E\left(\left.
Y_i\right| T_i = 1, D_i = 1, \boldsymbol{R}_{e,i}, \boldsymbol{Z}_{e,i}, \boldsymbol{X}_{e,i}\right)=\frac{F_{W, U, V}(R_{0,i}, Z_{0,i}, X_{0,i} +\tau_{2,0})}{F_{W,U}(R_{0,i},Z_{0,i})}
\equiv G_3\left(R_{0,i}, Z_{0,i}, X_{0,i}\right),
\end{align*}
\begin{align*}
 E\left(\left.
Y_i \right| T_i = 0, D_i = 1, \boldsymbol{R}_{e,i}, \boldsymbol{Z}_{e,i}, \boldsymbol{X}_{e,i}\right) & = \frac{F_{U,V}(Z_{0,i}, X_{0,i}) - F_{W, U, V}(R_{0,i}, Z_{0,i}, X_{0,i} )}{F_U(Z_{0,i}) - F_{W,U}(R_{0,i},Z_{0,i})}\\
&\equiv G_4\left(R_{0,i}, Z_{0,i}, X_{0,i}\right),
\end{align*}
where, similar to the previous step,  $G_3(\cdot, \cdot, \cdot)$ and $G_4(\cdot, \cdot, \cdot)$ are both monotonically increasing in the third argument. Then we get the following estimation procedure for $\boldsymbol{\beta}_0$.

\fbox{\begin{minipage}\textwidth
    
\textbf{Algorithm 3.2 for Estimating $\boldsymbol{\beta}_0$:}
\begin{enumerate}
\item Start with $k=0$, the first-step estimator $\widehat {\boldsymbol{\varphi}}$, the second-step estimator $\widehat {\boldsymbol{\delta}}$, the initial guess of $\boldsymbol{\beta}_0$, $\widehat {\boldsymbol{\beta}}^0$, initial guess of the sieve parameters $\widehat {\boldsymbol{\Pi}}_{1,q}^0$ and $\widehat {\boldsymbol{\Pi}}_{2,q}^0$, and initial guess of the conditional expectation functions $\widehat G^0_3(w,u,v)$ and $\widehat G^0_4(w,u,v)$.  
\item In the $k$-th round,  with $\widehat{\boldsymbol{\beta}}^k$, update $\widehat {\boldsymbol{\Pi}}_{1,q}^{k}$ and $\widehat {\boldsymbol{\Pi}}_{2,q}^{k}$ to $\widehat {\boldsymbol{\Pi}}_{1,q}^{k+1}$  and $\widehat {\boldsymbol{\Pi}}_{2,q}^{k+1}$  using 
\begin{align*}
\widehat{\boldsymbol{\Pi}}^{k+1}_{1,q} =  \left[\sum_{i=1}^n T_iD_i\boldsymbol{\Phi}_q\left(\widehat{R}_i,  \widehat{Z}_i, \widehat{X}_i^k\right)\boldsymbol{\Phi}_q\left(\widehat{R}_i,  \widehat{Z}_i,\widehat{X}_i^k\right)^{\mathrm{T}}\right]^{-1}\times
\left[\sum_{i=1}^n T_iD_iY_i\boldsymbol{\Phi}_q\left(\widehat{R}_i,  \widehat{Z}_i, \widehat{X}_i^k\right)\right],
\end{align*}
\begin{align*}
\widehat{\boldsymbol{\Pi}}^{k+1}_{2,q} =  \left[\sum_{i=1}^n (1-T_i)D_i\boldsymbol{\Phi}_q\left(\widehat{R}_i,  \widehat{Z}_i, \widehat{X}_i^k\right)\boldsymbol{\Phi}_q\left(\widehat{R}_i,  \widehat{Z}_i,\widehat{X}_i^k\right)^{\mathrm{T}}\right]^{-1}\times
\left[\sum_{i=1}^n (1-T_i)D_iY_i\boldsymbol{\Phi}_q\left(\widehat{R}_i,  \widehat{Z}_i, \widehat{X}_i^k\right)\right],
\end{align*}

\item With $\widehat {\boldsymbol{\Pi}}^{k+1}_{1,q}$ and $\widehat {\boldsymbol{\Pi}}^{k+1}_{2,q}$, update $\widehat G_3^k(w,u,v)$ and $\widehat G_4^k(w,u,v)$ to $\widehat G_3^{k+1}(w,u,v)$ and $\widehat G_4^{k+1}(w,u,v)$ using $\widehat G_3^{k+1} \left(w, 
u,v\right)= \boldsymbol{\Phi}_q( w, u,v)^{\mathrm{T}}\widehat{\boldsymbol{\Pi}}_{1,q}^{k+1}$ and $\widehat G_4^{k+1} \left(w, 
u,v\right)= \boldsymbol{\Phi}_q( w, u,v)^{\mathrm{T}}\widehat{\boldsymbol{\Pi}}_{2,q}^{k+1}$.
\item With $\widehat G_3^{k+1}(w,u,v)$ and $\widehat G_4^{k+1}(w,u,v)$, update $\widehat {\boldsymbol{\beta}}^k $ to $\widehat {\boldsymbol{\beta}}^{k+1}$ using
\[
\widehat {\boldsymbol{\beta}}^{k+1} = \widehat {\boldsymbol{\beta}}^{k} - \frac{\gamma_k}{S_n} \sum_{i=1}^n D_i\left(T_i\widehat G_3^{k+1}\left(\widehat{R}_i, \widehat{Z}_{i}, \widehat{X}_i^k\right) + (1-T_i)\widehat G_4^{k+1}\left(\widehat{R}_i, \widehat{Z}_{i}, \widehat{X}_i^k\right) - Y_i\right)\boldsymbol{X}_i,
\]
where $\gamma_k>0$ is the learning rate.
\item Set $k = k+1$ and go back to Step 2 unless some terminating conditions are satisfied.
\end{enumerate}
\end{minipage}}

Denote the third-step estimator as $\widehat{\boldsymbol{\beta}}$, now we proceed to the final step for the estimation of the treatment effects $\tau_{1,0}$ and $\tau_{2,0}$. We first describe the estimator for $\tau_{1,0}$, whose idea can be similarly extended to the estimation of $\tau_{2,0}$.  Note that under \autoref{condition12}, we can identify the following two functions
\begin{align*}
F_{W,U}(w, u+\tau_{1,0}) & = G_1(w, u)E(T_i|R_{0,i} = w), \\
F_U(u) -F_{W,U}(w,u)  & = G_2(w, u)(1-E(T_i|R_{0,i} = w))
\end{align*} 
when $ (R_{0,i}, \boldsymbol{Z}_{e,i}^{\mathrm{T}})^{\mathrm{T}}$ is interior point of $\overline{\mathcal{Z}}_e$, then 
$\nabla_w F_{W,U}(w, u+\tau_{1,0})$ and 
$\nabla_w  F_{W,U}(w, u) = -\nabla_w\left(F_U(u) - F_{W,U}(w, u)\right)$ can also be identified. So we can match the values of the derivatives of the CDF functions and their  arguments to estimate $\tau_{1,0}$. Typically, under \autoref{condition12}, $\tau_{1,0}$ uniquely minimizes the following loss function  
\begin{align}\label{loss1}
\int \left(\nabla_w F_{W,U}\left(w,u+\tau_{1,0}\right) - \nabla_w F_{W,U}\left(w,u+\tau_{1}\right)\right)^2 \omega(w,u)dwdu,
\end{align}
where $\omega_1(w,u)\geq 0$ is a well-chosen weight function. 
The above observation motivates the following  algorithm for estimating $\tau_{1,0}$.

\fbox{\begin{minipage}\textwidth
\textbf{Algorithm 3.3 for Estimating $\boldsymbol{\tau_{1,0}$}}
    \begin{enumerate}
        \item Choose grid points for $w$ and $u$, denoted as $(w_1,u_1), (w_2, u_2), \cdots, (w_J, u_J)$, and a nonnegative weight function $\omega_1(w, u)$.

        \item Estimate $\nabla_wF_{W,U}\left(w,u+\tau_{1,0}\right)$ and $\nabla_w F_{W,U}\left(w,u+\tau_{1}\right)$ at $(w,u) = (w_j, u_j), j= 1, 2, \cdots, J$, denoted as $\nabla_w\widehat F_{W,U}\left(w,u+\tau_{1,0}\right)$ and $\nabla_w \widehat F_{W,U}\left(w,u+\tau_{1}\right)$.

        \item Minimize $\sum_{j=1}^J (\nabla_w\widehat F_{W,U}(w_j,u_j+\tau_{1,0}) - \nabla_w \widehat F_{W,U}(w_j,u_j+\tau_{1}))^2\omega_1(w_j, u_j)$ with respect to $\tau_{1}$. 
    \end{enumerate}
\end{minipage}}

\begin{remark}
    (i)Note that the loss function (\ref{loss1}) is non-convex with respect to $\tau_1$, so gradient-based optimization may fail to work and lead to local optimum. In this case, we can use grid search to find the minimizer of the loss function. The computational burden is acceptable since the  optimization problem is only one-dimensional. (ii) The estimator of $\tau_{1,0}$ (and also for $\tau_{2,0}$) is sensitive to the choice of the derivative estimators of $\nabla_wF_{W,U}\left(w,u+\tau_{1,0}\right)$ and $\nabla_w F_{W,U}\left(w,u+\tau_{1}\right)$. To improve robustness, we suggest using local polynomials for estimation. 
\end{remark}

We finally describe the algorithm for estimating $\tau_{2,0}$. Under \autoref{condition13},   the following two probabilities can be identified 
\begin{align*}
    F_{W,U,V}(w, u+\tau_{1,0}, v+\tau_{2,0})
    & = G_3(w, u, v)P(T_i=1, D_i=1|R_{0,i} = w, Z_{0,i} = u)\\
    F_{U,V}(u, v) - F_{W,U,V}(w,u,v) 
    & = G_4(w, u, v)P(T_i=0, D_i=1|R_{0,i} = w, Z_{0,i} = u)
\end{align*}
when $(R_{0,i}, Z_{0,i}, \boldsymbol{X}_i^{\mathrm{T}})^{\mathrm{T}}$ is interior point of $\overline{\mathcal{X}}_e$. 
So the cross partial derivative $\nabla_{wu}F_{W,U,V}(w, u+\tau_{1,0}, v+\tau_{2,0})$ and $\nabla_{wu}F_{W,U,V}(w, u, v)$  can be identified. Then since $\nabla_{wu}F_{W,U,V}(w, u, v)$   is non-decreasing with respect to $v$, under \autoref{condition13},  $\tau_{2,0}$ uniquely minimizes the following loss function 
\[
\int \left(\nabla_{wu}F_{W,U,V}(w, u+\tau_{1,0}, v+\tau_{2,0}) - \nabla_{wu}F_{W,U,V}(w, u+\tau_{1,0}, v+\tau_{2})\right)^2\omega_2(w,u,v)dwdudv, 
\]
where $\omega_2(w,u,v)$ is some well-chosen weight function. This immediately leads to the following algorithm for estimating $\tau_{2,0}$.

\fbox{\begin{minipage}\textwidth
\textbf{Algorithm 3.4 for Estimating $\boldsymbol{\tau_{2,0}$}}
    \begin{enumerate}
        \item Choose grid points for $w$,  $u$, and $v$, denoted as $(w_1,u_1, v_1), (w_2, u_2,v_2), \cdots, (w_J, u_J, v_J)$, and weight function $\omega_2(w,u,v)$.

        \item Estimate $\nabla_{wu} F_{W,U, V}\left(w,u, v+\tau_{2,0}\right)$ and $\nabla_{wu} F_{W,U,V}\left(w,u+\tau_{1,0}, v+\tau_{2}\right)$ at $(w,u,v) = (w_j, u_j, v_j)$, $j= 1, 2, \cdots, J$, denoted as $\nabla_{wu}\widehat F_{W,U, V}\left(w,u,v+\tau_{2,0}\right)$ and $\nabla_{wu} \widehat F_{W,U,V}\left(w,u,v+\tau_{2}\right)$.

        \item Minimize $\sum_{j=1}^J (\nabla_{wu}\widehat F_{W,U,V}(w_j,u_j +\tau_{1,0}, v_j+\tau_{2,0}) - \nabla_{wu} \widehat F_{W,U,V}(w_j,u_j+\widehat{\tau}_1, v_j+\tau_{2}))^2\omega_2(w,u,v)$ with respect to $\tau_{2}$. 
    \end{enumerate}
\end{minipage}}

\ \ \

\section{Conclusions}\label{section7}
This paper considers estimation and inference for large dimensional semiparametric selective labeling models. Statistically these models have a similar structure to sample selection models with binary, as opposed to linear outcome equations in the second stage. It is this binary/binary structure which makes computation of the model particularly difficult when compared to the standard selection model, especially for large dimensional (i.e many regressor) models.

To address this problem we propose novel algorithmic procedures
which are computationally fast, and derive their asymptotic properties even for the case where the dimension increases with the sample size.  We demonstrate the finite sample properties of our proposed procedures by a simulation study.

Our work here motivates areas for future research. For example to further ease implementation, a bivariate penalization scheme 
would be useful for model selection in this settings, and its asymptotic validity would need to be proven. Furthermore, the usefulness of our methods in other empirical settings in economics, biostatistics and medicine would be worthy of exploration.
\bibliography{overall_1}

@ARTICLE{abhausmankhan,
  title = {Testing for Causal Effects in a Generalized Regression Model with Endogenous Regressors},
author = {J. Abrevaya and J. Hausman and S. Khan},
  journal = {Econometrica},
  year = {2010},
  volume = {78},
  pages={2043-2061}
}

@article{gronau,
  title={Wage comparisons--A selectivity bias},
  author={Gronau, Reuben},
  journal={Journal of political Economy},
  volume={82},
  number={6},
  pages={1119--1143},
  year={1974},
  publisher={The University of Chicago Press}
}

@article{hsiaozhou2024,
author={
Cheng Hsiao and Qiankun Zhou},
title={Statistical inference for the low dimensional parameters of linear regression models in the presence of high-dimensional data: An orthogonal projection approach},
journal={Journal of Econometrics},
year={2024},
volume={105851}
}

@article{xaviermaurel2025,
author={
X. D'Haultfoeuille and C. Gaillac and A. Maurel},
title={Partially Linear Models under Data Combination},
journal={Review of Economic Studies},
year={2025},
volume={92},
issue={1},
pages={238--267}
}

@article{bellonietal2014,
author={Belloni,  A. and  Chernozhukov, V. and  Kato, K.},
year={2014},
title={Uniform post-selection inference for least absolute deviation regression and other Z-estimation problems.},
journal={Biometrika},
volume={102},
number={1},
pages={77--94}
}

@article{vytlacilyildiz,
author={Vytlacil, E. and  Yildiz, N.},
year={2007},
title= {Dummy Endogenous Variables in Weakly Separable Models},
journal= {Econometrica},
volume={75},
pages={757-779}
}

@article{shaikhvytlacil,
author={Shaikh, A. and Vytlacil, E.},
year={2011},
title= {Partial Identification in Triangular Systems of Equations with Binary Dependent Variables},
journal= {Econometrica},
volume={79},
pages={949-955}
}

@article{chenkhantang,
author={Chen, S. and Khan, S. and Tang, X.},
year={2024},
title= {Endogeneity in Weakly Separable Models without Monotonicity},
journal= {Journal of Econometrics},
volume={238},
pages={1-14}
}

@article{khanmaurelzhang,
author={Khan, S. and Maurel, A. and Zhang, Y.},
year={2023},
title= {Informational Content of Factor Structures in Simultaneous Binary Response Models},
journal= {Advances in Econometrics},
volume={45},
pages={385-410}
}

@article{leebounds,
author={Lee, D.},
year={2009},
title= {Training, Wages and Sample Selection: Estimating Sharp Bounds on Treatment Effects},
journal= {Review of Economic Studies},
volume={76},
pages={1071-1102}
}

@artical{chernozhukhovetal2018,
title={Double machine learning for treatment and causal parameters.},
year={2018},
author={Chernozhukov, V. and Chetverikov, D. and Demirer, M. and Duflo, E. and Hansen, C. and Newey, W.},
journal={Econometrics Journal},
volume={21},
pages={C1--C68}
}

@article{robinson1988,
title={Root-N-Consistent Semiparametric Regression},
year={1988},
author={P.M. Robinson},
journal={Econometrica},
volume={56},
pages={931--954}
}

@article{speckman1988,
title={Kernel Smoothing in Partial Linear Models},
author={P. Speckman},
year={1988},
journal={Journal of the Royal Statistical Society, Series B},
volume={50},
pages={413--436}
}

@UNPUBLISHED{khantamerwei2025,
  author = {Shakeeb Khan and Elie Tamer and Shengbin Wei},
  title = {Two Step Estimation of Large Dimensional Partially Identified Models},
  note = {Boston College Working Paper},
  year = {2025}
}

@article{chib1,
    author = {S. Chib and E. Greenberg and I. Jeliazkov},
    title = { Estimation of Semiparametric Models in the Presence of Endogeneity and Sample Selection},
    journal = {Journal of Computational and Graphical Statistics} ,
    year = {2009},
    volume = {18},
    pages = {321-348}
  }

@article{chib2,
    author = {S. Chib and E. Greenberg},
    title = {Semiparametric Modeling and Estimation of Instrumental Variable Models},
    journal = {Journal of Computational and Graphical Statistics} ,
    year = {2007},
    volume = {16},
    pages = {1-29}
  }

@unpublished{lakkarujuetal2017,
title={The Selective Labels Problem: Evaluating Algorithmic Predictions in the Presence of Unobservables},
author={Himabindu Lakkaraju and Jon Kleinberg and Jure Leskovec and Jens Ludwig and Sendhil Mullainathan},
year={2017},
note={KDD Research Paper}
}

@ARTICLE{heckman1974,
  author = {Heckman, James},
  title = {Shadow Prices, Market Wages, and Labor Supply},
  journal = {Econometrica},
  year = {1974},
  volume = {42},
  pages = {679--694},
  number = {4}
}

@article{chen2007large,
  title={Large sample sieve estimation of semi-nonparametric models},
  author={Chen, Xiaohong},
  journal={Handbook of Econometrics},
  volume={6},
  pages={5549--5632},
  year={2007},
  publisher={Elsevier}
}

@article{neweyej,
  title={Two-step Series Estimation of Sample Selection Models},
  author={Newey, Whitney K},
  journal={Econometrics Journal},
  volume={12},
  number={1},
  pages={217--229},
  year={2009},
  }

@article{neweyetal,
  title={Nonparametric Estimation of Sample Selection Models},
  author={Das, M. and Newey, W.K and Vella, F.},
  journal={Review of Economic Studies},
  volume={70},
  number={1},
  pages={33--58},
  year={2003},
  }

@article{kleinbergetal,
  title={Human Decisions and Machine Predictions},
  author={Jon Kleinberg and Himabindu Lakkaraju and Jure Leskovec   and Jens Ludwig and Sendhil Mullainathan},
  journal={Quarterly Journal of Economics},
  volume={133},
  number={1},
  pages={237--293},
  year={2018},
  }

@article{ichimura1993semiparametric,
	title={Semiparametric least squares (SLS) and weighted SLS estimation of single-index models},
	author={Ichimura, Hidehiko},
	journal={Journal of econometrics},
	volume={58},
	number={1-2},
	pages={71--120},
	year={1993},
	publisher={Elsevier}
}

@unpublished{khanetal2021,
	title={Estimating High Dimensional Monotone Index Models By Iterative Convex Optimization },
	author={Khan, Shakeeb and Lan, Xiaoying and Tamer, Eile and Yao, Qingsong},
	year={2024},
	note={forthcoming, \em {Journal of Econometrics}}}

@article{ahn1993,
  title={Semiparametric estimation of censored selection models with a nonparametric selection mechanism},
  author={Ahn, H. and Powell, J.L.},
  journal={Journal of Econometrics},
  volume={58},
  number={1},
  pages={3--29},
  year={1993},
  publisher={Elsevier}
}

@article{asheshsl,
title={Charecterizing Fairness Over the Set of Good Models Under Selective Labels},
author={Coston, A. and Rambachan, A. and Chouldechova, A. },
journal={Proceedings of the 38th International Conference on Machine Learning},
year={2021},
pages={2144-2155},
volume={139}
}

@book{biau2015lectures,
  title={Lectures on the nearest neighbor method},
  author={Biau, G{\'e}rard and Devroye, Luc},
  volume={246},
  year={2015},
  publisher={Springer}
}

@article{pierson2020large,
  title={A large-scale analysis of racial disparities in police stops across the United States},
  author={Pierson, Emma and Simoiu, Camelia and Overgoor, Jan and Corbett-Davies, Sam and Jenson, Daniel and Shoemaker, Amy and Ramachandran, Vignesh and Barghouty, Phoebe and Phillips, Cheryl and Shroff, Ravi and others},
  journal={Nature human behaviour},
  volume={4},
  number={7},
  pages={736--745},
  year={2020},
  publisher={Nature Publishing Group}
}

@article{goel2016precinct,
  title={Precinct or prejudice? Understanding racial disparities in New York city’s stop-and-frisk policy},
  author={Goel, Sharad and Rao, Justin M and Shroff, Ravi},
  journal={Annals of Applied Statistics},
  volume={10},
  number={1},
  pages={365--394},
  year={2016},
  publisher={Institute of Mathematical Statistics}
}

@article{SEMENOVA2025106055,
title = {Generalized Lee bounds},
journal = {Journal of Econometrics},
volume = {251},
pages = {106055},
year = {2025},
issn = {0304-4076},
doi = {https://doi.org/10.1016/j.jeconom.2025.106055},
url = {https://www.sciencedirect.com/science/article/pii/S0304407625001095},
author = {Vira Semenova}
}

\newpage

\appendix
\textbf{\Large Appendix}

\section{Details for Empirical Applications}\label{details of empirical applications}

Denote the covariate matrix of selection and outcome equations as $\mathbf{Z}_n =\left(\boldsymbol{Z}_{0,n}, \cdots, \boldsymbol{Z}_{p_Z,n}\right)$ and $\mathbf{X}_n =\left(\boldsymbol{X}_{0,n}, \cdots, \boldsymbol{X}_{p_X,n}\right)$, where $\boldsymbol{Z}_{j,n} = (Z_{j,1}, \cdots, Z_{j,n})^{\mathrm{T}}$ and $\boldsymbol{X}_{j,n} = (X_{j,1}, \cdots, X_{j,n})^{\mathrm{T}}$. Note that when conducting iteration, the ratios of maximum and minimum of eigenvalues of $\mathbf{Z}_n$ and $\mathbf{X}_n$ will significantly affect the numerical performance of algorithm. In particular, when the ratio of max/min eigenvalues is large, small learning rate has to be selected, which leads to extensive amount of iteration time. To solve this issue, we seek matrix $\boldsymbol{Q}_Z$ and $\boldsymbol{Q}_X$, such that both $\widetilde{\mathbf{Z}}_n = \mathbf{Z}_n\boldsymbol{Q}_Z$ and $\widetilde{\mathbf{X}}_n = \mathbf{X}_n\boldsymbol{Q}_X$ have unit column variance and zero column correlation. Then we set the coefficients of the first column of $\widetilde{\mathbf{Z}}_n$ and $\widetilde{\mathbf{X}}_n$ to be 1, and apply \textbf{Algorithm 0} and \textbf{Algorithm 2} to estimate the parameters. Denote the estimated parameters based on $\widetilde{\mathbf{Z}}_n$ and $\widetilde{\mathbf{X}}_n$ as $\widetilde{\widehat{\boldsymbol{\delta}}}$ and $\widetilde{\widehat{\boldsymbol{\beta}}}$. Denote  $\overline{\widehat{\boldsymbol{\delta}}} = \boldsymbol{Q}_Z\left[1; \widetilde{\widehat{\boldsymbol{\delta}}}\right]\equiv \left[\overline{\widehat{\delta}}_0; \overline{\widehat{\boldsymbol{\delta}}}_1\right]$ and  $\overline{\widehat{\boldsymbol{\beta}}} = \boldsymbol{Q}_X\left[1; \widetilde{\widehat{\boldsymbol{\beta}}}\right]\equiv \left[\overline{\widehat{\beta}}_0; \overline{\widehat{\boldsymbol{\beta}}}_1\right]$. Then the SBGD estimators $\widehat{\boldsymbol{\delta}}$ and $\widehat{\boldsymbol{\beta}}$ are given by $\widehat{\boldsymbol{\delta}} = \overline{\widehat{\boldsymbol{\delta}}}_1/\overline{\widehat{\delta}}_0$ and $\widehat{\boldsymbol{\beta}} = \overline{\widehat{\boldsymbol{\beta}}}_1/\overline{\widehat{\beta}}_0$. 

Note that based on the above-mentioned estimation strategy, the transformation matrices $\boldsymbol{Q}_Z$ and $\boldsymbol{Q}_X$  should be chosen such that the first columns of $\widetilde{\mathbf{Z}}_n$  and $\widetilde{\mathbf{X}}_n$ have positive impacts on the conditional probability. We use a Gram-Schmidt-type transformation. In particular, we consider 
\begin{align*}
    & \widetilde{\boldsymbol{Z}}_{0,n}= \boldsymbol{Z}_{0.n}\\
    & \widetilde{\boldsymbol{Z}}_{1,n} =\boldsymbol{Z}_{1,n} - \frac{cov(\widetilde{\boldsymbol{Z}}_{0,n}, \boldsymbol{Z}_{1,n})}{cov(\widetilde{\boldsymbol{Z}}_{0,n},\widetilde{\boldsymbol{Z}}_{0,n})}\widetilde{\boldsymbol{Z}}_{0,n}\\
    & \cdots
     \\
     &\widetilde{\boldsymbol{Z}}_{p_Z,n} = \boldsymbol{Z}_{p_Z,n} - \sum_{j=1}^{p_Z}\frac{cov(\widetilde{\boldsymbol{Z}}_{j,n}, \boldsymbol{Z}_{p_Z,n})}{cov(\widetilde{\boldsymbol{Z}}_j,\widetilde{\boldsymbol{Z}}_{j,n})}\widetilde{\boldsymbol{Z}}_{n,j}
\end{align*}
and 
\begin{align*}
    & \widetilde{\boldsymbol{X}}_{0,n}= \boldsymbol{X}_{0.n}\\
    & \widetilde{\boldsymbol{X}}_{1,n} =\boldsymbol{X}_{1,n} - \frac{cov(\widetilde{\boldsymbol{X}}_{0,n}, \boldsymbol{X}_{1,n})}{cov(\widetilde{\boldsymbol{X}}_{0,n},\widetilde{\boldsymbol{X}}_{0,n})}\widetilde{\boldsymbol{X}}_{0,n}\\
    & \cdots
     \\
     &\widetilde{\boldsymbol{X}}_{p_X,n} = \boldsymbol{X}_{p_X,n} - \sum_{j=1}^{p_X}\frac{cov(\widetilde{\boldsymbol{X}}_{j,n}, \boldsymbol{X}_{p_X,n})}{cov(\widetilde{\boldsymbol{Z}}_j,\widetilde{\boldsymbol{X}}_{j,n})}\widetilde{\boldsymbol{X}}_{n,j}
\end{align*}

Note that the above transformation  standardizes each column such that the transformed columns have unit variance, and most importantly, it preserves the first column of the original matrix, which guarantees that the first column of the transformed matrix has positive coefficient. The transformation matrices  $\boldsymbol{Q}_Z$ and $\boldsymbol{Q}_X$ can be numerically obtained by $\boldsymbol{Q}_Z = \left(\boldsymbol{Z}_n^{\mathrm{T}}\boldsymbol{Z}_n\right)^{-1}\boldsymbol{Z}_n^{\mathrm{T}}\widetilde{\boldsymbol{Z}}_n$ and $\boldsymbol{Q}_X = \left(\boldsymbol{X}_n^{\mathrm{T}}\boldsymbol{X}_n\right)^{-1}\boldsymbol{X}_n^{\mathrm{T}}\widetilde{\boldsymbol{X}}_n$. Finally, based on $\widehat{ \boldsymbol{\delta}}$ and $\widehat {\boldsymbol{\beta}}$ we can estimate the covariance matrix of the estimator. 

\section{Proofs of Main Theorems}\label{proofs}

In this section, we use $C$ to denote generic positive constants whose value may vary even it appears on the same line. We will also use $a_{n}\lesssim b_{n}$ ($a_{n}\gtrsim b_{n}$)
if there exists some constant $C$ such that $a_{n}\leq b_{n}$ ($a_{n}\geq Cb_{n}$)
for all sufficiently large $n$.

\subsection{Proof of \autoref{theorme3.1-1}}

Define $\boldsymbol{\theta}=\left(\boldsymbol{Z}_{e},\boldsymbol{\delta},\boldsymbol{X}_{e},\boldsymbol{\beta}\right)\in\varTheta\equiv \mathscr{Z}_{e}\times\mathscr{D}\times\mathscr{X}_{e}\times\mathscr{B}\subseteq R^{2\mathfrak{P}+2}$.
Also recall that  $Z\left(\boldsymbol{\delta}\right)=z_{0}+\boldsymbol{Z}^{\mathrm{T}}\boldsymbol{\delta}$,
$Z_{i}\left(\boldsymbol{\delta}\right)=z_{0,i}+\boldsymbol{Z}_{i}^{\mathrm{T}}\boldsymbol{\delta}$,
$X\left(\boldsymbol{\beta}\right)=x_{0}+\boldsymbol{X}^{\mathrm{T}}\boldsymbol{\beta},$
and $X_{i}\left(\boldsymbol{\beta}\right)=x_{0,i}+\boldsymbol{X}_{i}^{\mathrm{T}}\boldsymbol{\beta}$.
For any fixed $\boldsymbol{\delta}$ and $\boldsymbol{\beta}$, 
$
\mathcal{Y}\left(\nu_{Z},\nu_{X},\boldsymbol{\delta},\boldsymbol{\beta}\right)=E\left(\left.Y\right|D=1,Z\left(\boldsymbol{\delta}\right)=\nu_{Z},X\left(\boldsymbol{\beta}\right)=\nu_{X}\right).
$
Then for each $i$, we can decompose $Y_{i}$ as 
$
Y_{i}=\mathcal{Y}\left(Z_{i}\left(\boldsymbol{\delta}\right),X_{i}\left(\boldsymbol{\beta}\right),\boldsymbol{\delta},\boldsymbol{\beta}\right)+\epsilon_{i}\left(\boldsymbol{\delta},\boldsymbol{\beta}\right),
$
where $\epsilon_{i}\left(\boldsymbol{\delta},\boldsymbol{\beta}\right)=Y_{i}-\mathcal{Y}\left(Z_{i}\left(\boldsymbol{\delta}\right),X_{i}\left(\boldsymbol{\beta}\right),\boldsymbol{\delta},\boldsymbol{\beta}\right)$.
So for any fixed $\boldsymbol{\delta}$ and $\boldsymbol{\beta}$,
we have that 
$
E\left(\left.\epsilon_{i}\left(\boldsymbol{\delta},\boldsymbol{\beta}\right)\right|D_{i}=1,Z_{i}\left(\boldsymbol{\delta}\right),X_{i}\left(\boldsymbol{\beta}\right)\right)=0.
$
Now consider the following reordering mechanism. For each fixed $\boldsymbol{\theta}\in\varTheta$,
we reorder $i$ with $D_{i}=1$ as $\varrho_{n,1}^{\boldsymbol{\theta}},\varrho_{n,2}^{\boldsymbol{\theta}},\cdots,\varrho_{n,S_{n}}^{\boldsymbol{\theta}}$
such that 
\begin{align*}
 \left\Vert \left(Z_{\varrho_{n,1}^{\boldsymbol{\theta}}}\left(\boldsymbol{\delta}\right),X_{\varrho_{n,1}^{\boldsymbol{\theta}}}\left(\boldsymbol{\beta}\right)\right)-\left(Z\left(\boldsymbol{\delta}\right),X\left(\boldsymbol{\beta}\right)\right)\right\Vert 
 & \leq\left\Vert \left(Z_{\varrho_{n,2}^{\boldsymbol{\theta}}}\left(\boldsymbol{\delta}\right),X_{\varrho_{n,2}^{\boldsymbol{\theta}}}\left(\boldsymbol{\beta}\right)\right)-\left(Z\left(\boldsymbol{\delta}\right),X\left(\boldsymbol{\beta}\right)\right)\right\Vert  \leq\cdots\\
 &\leq\left\Vert \left(Z_{\varrho_{n,S_{n}}^{\boldsymbol{\theta}}}\left(\boldsymbol{\delta}\right),X_{\varrho_{n,S_{n}}^{\boldsymbol{\theta}}}\left(\boldsymbol{\beta}\right)\right)-\left(Z\left(\boldsymbol{\delta}\right),X\left(\boldsymbol{\beta}\right)\right)\right\Vert .
\end{align*}
We now provide a probability bound for $\sup_{\boldsymbol{\theta}\in\varTheta}\left|\frac{1}{m}\sum_{j=1}^{m}Y_{\varrho_{n,j}^{\boldsymbol{\theta}}}-\mathcal{Y}\left(Z\left(\boldsymbol{\delta}\right),X\left(\boldsymbol{\beta}\right),\boldsymbol{\delta},\boldsymbol{\beta}\right)\right|$. 
Note that 
\begin{align*}
 & \left|\frac{1}{m}\sum_{j=1}^{m}Y_{\varrho_{n,j}^{\boldsymbol{\theta}}}-\mathcal{Y}\left(Z\left(\boldsymbol{\delta}\right),X\left(\boldsymbol{\beta}\right),\boldsymbol{\delta},\boldsymbol{\beta}\right)\right|\\
 & \leq\left|\frac{1}{m}\sum_{j=1}^{m}\mathcal{Y}\left(Z_{\varrho_{n,j}^{\boldsymbol{\theta}}}\left(\boldsymbol{\delta}\right),X_{\varrho_{n,j}^{\boldsymbol{\theta}}}\left(\boldsymbol{\beta}\right),\boldsymbol{\delta},\boldsymbol{\beta}\right)-\mathcal{Y}\left(Z\left(\boldsymbol{\delta}\right),X\left(\boldsymbol{\beta}\right),\boldsymbol{\delta},\boldsymbol{\beta}\right)\right|+\left|\frac{1}{m}\sum_{j=1}^{m}\epsilon_{\varrho_{n,j}^{\boldsymbol{\theta}}}\left(\boldsymbol{\delta},\boldsymbol{\beta}\right)\right|
\end{align*}
Next we provide probability bounds for the two terms on the RHS of
the above inequality. 
\begin{lemma}\label{lemma1}
Suppose that \autoref{condition1}, \autoref{condition3}, \autoref{condition3.1-1} and \autoref{condition3.1-2} hold.  If further $m/n\rightarrow0$, we have that 
\begin{align*}
\sup_{\boldsymbol{\theta}\in\varTheta} \left|\frac{1}{m}\sum_{j=1}^{m}\mathcal{Y}\left(Z_{\varrho_{n,j}^{\boldsymbol{\theta}}}\left(\boldsymbol{\delta}\right),X_{\varrho_{n,j}^{\boldsymbol{\theta}}}\left(\boldsymbol{\beta}\right),\boldsymbol{\delta},\boldsymbol{\beta}\right)-\mathcal{Y}\left(Z\left(\boldsymbol{\delta}\right),X\left(\boldsymbol{\beta}\right),\boldsymbol{\delta},\boldsymbol{\beta}\right)\right|=O_{P}\left(\sqrt{\frac{\mathfrak{P}(\mathfrak{P}+m)\log\left(n\right)}{n\alpha_n}}\right).
\end{align*}
\end{lemma}
\begin{proof}
Note that according to \autoref{condition3.1-2}, we have 
\begin{align*}
\left|\mathcal{Y}\left(Z_{\varrho_{n,j}^{\boldsymbol{\theta}}}\left(\boldsymbol{\delta}\right),X_{\varrho_{n,j}^{\boldsymbol{\theta}}}\left(\boldsymbol{\beta}\right),\boldsymbol{\delta},\boldsymbol{\beta}\right)-\mathcal{Y}\left(Z\left(\boldsymbol{\delta}\right),X\left(\boldsymbol{\beta}\right),\boldsymbol{\delta},\boldsymbol{\beta}\right)\right|\leq C\sqrt{\mathfrak{P}}\cdot \left\Vert \left(Z_{\varrho_{n,j}^{\boldsymbol{\theta}}}\left(\boldsymbol{\delta}\right),X_{\varrho_{n,j}^{\boldsymbol{\theta}}}\left(\boldsymbol{\beta}\right)\right)-\left(Z\left(\boldsymbol{\delta}\right),X\left(\boldsymbol{\beta}\right)\right)\right\Vert .
\end{align*}
So we only need to look at the distance
$\sup_{\boldsymbol{\theta}\in\varTheta} \left\Vert \left(Z_{\varrho_{n,m}^{\boldsymbol{\theta}}}\left(\boldsymbol{\delta}\right),X_{\varrho_{n,m}^{\boldsymbol{\theta}}}\left(\boldsymbol{\beta}\right)\right)-\left(Z\left(\boldsymbol{\delta}\right),X\left(\boldsymbol{\beta}\right)\right)\right\Vert$. 
Note that since $\varTheta$ has dimension $2\mathfrak{P}+2$ and
is uniformly bounded in each argument, for any positive integer $L$, we can find a sequence
of disjoint sets $\varTheta^{1},\varTheta^{2},\cdots,\varTheta^{L}$
such that $\bigcup_{l=1}^{L}\varTheta^{l}=\varTheta$ and for any
$\boldsymbol{\theta},\boldsymbol{\theta}^{\prime}\in\varTheta^{l}$,
$\left\Vert \boldsymbol{\theta}^{\prime}-\boldsymbol{\theta}\right\Vert \leq C \sqrt{\mathfrak{P}}L^{-\frac{1}{2\mathfrak{P}+2}}$\footnote{To see this more clearly, for each argument of $\boldsymbol{Z}_e, \boldsymbol{\delta}, \boldsymbol{X}_e$, and $\boldsymbol{\beta}$, we evenly divide its support into $[L^{\frac{1}{2\mathfrak{P}+2}}]$ segments. Then the Cartesian products of these segments form a total of no more than $T$ disjoint subsets of  $\varTheta$, and their union is $\varTheta$. For each subset, the maximum variation along each argument is bounded by $C\cdot L^{-\frac{1}{2\mathfrak{P}+2}}$, where $C$ is a positive constant that does not depend on $\mathfrak{P}$ or $L$. So for any
$\boldsymbol{\theta},\boldsymbol{\theta}^{\prime}$ within the same subset,
$\left\Vert \boldsymbol{\theta}^{\prime}-\boldsymbol{\theta}\right\Vert \leq C \sqrt{\mathfrak{P}}L^{-\frac{1}{2\mathfrak{P}+2}}. $ },
where $C$ is a positive constant that depends on $\varTheta$ only.
For each $l,$ pick and fix one $\boldsymbol{\theta}^{l}=\left(\boldsymbol{Z}_{e}^{l},\boldsymbol{X}_{e}^{l},\boldsymbol{\delta}^{l},\boldsymbol{\beta}^{l}\right)\in\varTheta^{l}$,
and for each $\boldsymbol{\theta}$, with slight abuse of notation define $l\left(\boldsymbol{\theta}\right)$
such that $\boldsymbol{\theta}\in\varTheta^{l\left(\boldsymbol{\theta}\right)}$.
Then for each $1\leq i\leq n$ and $\boldsymbol{\theta}\in\varTheta$, we have that 
\begin{align*}
\left\Vert \left(Z_{i}\left(\boldsymbol{\delta}\right),X_{i}\left(\boldsymbol{\beta}\right)\right)-\left(Z\left(\boldsymbol{\delta}\right),X\left(\boldsymbol{\beta}\right)\right)\right\Vert &\leq \left\Vert\left(Z_{i}\left(\boldsymbol{\delta}\right),X_{i}\left(\boldsymbol{\beta}\right)\right) - \left(Z_{i}\left(\boldsymbol{\delta}^{l(\boldsymbol{\theta})}\right),X_{i}\left(\boldsymbol{\beta}^{l(\boldsymbol{\theta})}\right)\right)\right\Vert\\
& + \left\Vert \left(Z_{i}\left(\boldsymbol{\delta}^{l(\boldsymbol{\theta})}\right),X_{i}\left(\boldsymbol{\beta}^{l(\boldsymbol{\theta})}\right)\right)-\left(Z^{l(\boldsymbol{\theta})}\left(\boldsymbol{\delta}^{l(\boldsymbol{\theta})}\right),X^{l(\boldsymbol{\theta})}\left(\boldsymbol{\beta}^{l(\boldsymbol{\theta})}\right)\right)\right\Vert\\
& + \left\Vert \left(Z^{l(\boldsymbol{\theta})}\left(\boldsymbol{\delta}^{l(\boldsymbol{\theta})}\right),X^{l(\boldsymbol{\theta})}\left(\boldsymbol{\beta}^{l(\boldsymbol{\theta})}\right)\right) - \left(Z\left(\boldsymbol{\delta}\right),X\left(\boldsymbol{\beta}\right)\right)\right\Vert\\
&  \leq \left\Vert \left(Z_{i}\left(\boldsymbol{\delta}^{t(\boldsymbol{\theta})}\right),X_{i}\left(\boldsymbol{\beta}^{l(\boldsymbol{\theta})}\right)\right)-\left(Z^{l(\boldsymbol{\theta})}\left(\boldsymbol{\delta}^{l(\boldsymbol{\theta})}\right),X^{l(\boldsymbol{\theta})}\left(\boldsymbol{\beta}^{l(\boldsymbol{\theta})}\right)\right)\right\Vert\\
& + C \mathfrak{P}L^{-\frac{1}{2\mathfrak{P}+2}},
\end{align*}
where $Z^{l\left(\boldsymbol{\theta}\right)}\left(\boldsymbol{\delta}^{l\left(\boldsymbol{\theta}\right)}\right)=z_{0}^{l\left(\boldsymbol{\theta}\right)}+\left(Z^{l\left(\boldsymbol{\theta}\right)}\right)^{\mathrm{T}}\boldsymbol{\delta}^{l\left(\boldsymbol{\theta}\right)}$
and $X^{l\left(\boldsymbol{\theta}\right)}\left(\boldsymbol{\beta}^{l\left(\boldsymbol{\theta}\right)}\right)=x_{0}^{l\left(\boldsymbol{\theta}\right)}+\left(X^{l\left(\boldsymbol{\theta}\right)}\right)^{\mathrm{T}}\boldsymbol{\beta}^{l\left(\boldsymbol{\theta}\right)}$. The second term on RHS of the second inequality comes from Cauchy-Schwarz inequality. This implies that for each $\boldsymbol{\theta}\in\varTheta$, we can find $m$ indices $\varrho_{n,1}^{\boldsymbol{\theta}^{l(\boldsymbol{\theta})}}, \varrho_{n,2}^{\boldsymbol{\theta}^{l(\boldsymbol{\theta})}},\cdots, \varrho_{n,m}^{\boldsymbol{\theta}^{l(\boldsymbol{\theta})}}$, such that each index $i\in\{\varrho_{n,1}^{\boldsymbol{\theta}^{l(\boldsymbol{\theta})}}, \varrho_{n,2}^{\boldsymbol{\theta}^{l(\boldsymbol{\theta})}},\cdots,\varrho_{n,m}^{\boldsymbol{\theta}^{l(\boldsymbol{\theta})}}\}$ satisfies that 
\begin{align*}
& \left\Vert \left(Z_{i}\left(\boldsymbol{\delta}\right),X_{i}\left(\boldsymbol{\beta}\right)\right)-\left(Z\left(\boldsymbol{\delta}\right),X\left(\boldsymbol{\beta}\right)\right)\right\Vert \\
&\leq \left\Vert \left(Z_{i}\left(\boldsymbol{\delta}^{l(\boldsymbol{\theta})}\right),X_{i}\left(\boldsymbol{\beta}^{l(\boldsymbol{\theta})}\right)\right)-\left(Z^{l(\boldsymbol{\theta})}\left(\boldsymbol{\delta}^{l(\boldsymbol{\theta})}\right),X^{l(\boldsymbol{\theta})}\left(\boldsymbol{\beta}^{l(\boldsymbol{\theta})}\right)\right)\right\Vert+ C \mathfrak{P}L^{-\frac{1}{2\mathfrak{P}+2}}\\
&\leq \left\Vert \left(Z_{\varrho_{n,m}^{\boldsymbol{\theta}^{l(\boldsymbol{\theta})}}}\left(\boldsymbol{\delta}^{l\left(\boldsymbol{\theta}\right)}\right),X_{\varrho_{n,m}^{\boldsymbol{\theta}^{l(\boldsymbol{\theta})}}}\left(\boldsymbol{\beta}^{l\left(\boldsymbol{\theta}\right)}\right)\right)-\left(Z^{l\left(\boldsymbol{\theta}\right)}\left(\boldsymbol{\delta}^{l\left(\boldsymbol{\theta}\right)}\right),X^{l\left(\boldsymbol{\theta}\right)}\left(\boldsymbol{\beta}^{l\left(\boldsymbol{\theta}\right)}\right)\right)\right\Vert + C \mathfrak{P}L^{-\frac{1}{2\mathfrak{P}+2}}
\end{align*}
The above implies that 
\begin{align*}
&\left\Vert \left(Z_{\varrho_{n,m}^{\boldsymbol{\theta}}}\left(\boldsymbol{\delta}\right),X_{\varrho_{n,m}^{\boldsymbol{\theta}}}\left(\boldsymbol{\beta}\right)\right)-\left(Z\left(\boldsymbol{\delta}\right),X\left(\boldsymbol{\beta}\right)\right)\right\Vert \\
& \leq \left\Vert \left(Z_{\varrho_{n,m}^{\boldsymbol{\theta}^{l(\boldsymbol{\theta})}}}\left(\boldsymbol{\delta}^{l\left(\boldsymbol{\theta}\right)}\right),X_{\varrho_{n,m}^{\boldsymbol{\theta}^{l(\boldsymbol{\theta})}}}\left(\boldsymbol{\beta}^{l\left(\boldsymbol{\theta}\right)}\right)\right)-\left(Z^{l\left(\boldsymbol{\theta}\right)}\left(\boldsymbol{\delta}^{l\left(\boldsymbol{\theta}\right)}\right),X^{l\left(\boldsymbol{\theta}\right)}\left(\boldsymbol{\beta}^{l\left(\boldsymbol{\theta}\right)}\right)\right)\right\Vert + C \mathfrak{P}L^{-\frac{1}{2\mathfrak{P}+2}},
\end{align*}
and hence
\begin{align*}
 & \sup_{\boldsymbol{\theta}\in\varTheta}\left\Vert \left(Z_{\varrho_{n,m}^{\boldsymbol{\theta}}}\left(\boldsymbol{\delta}\right),X_{\varrho_{n,m}^{\boldsymbol{\theta}}}\left(\boldsymbol{\beta}\right)\right)-\left(Z\left(\boldsymbol{\delta}\right),X\left(\boldsymbol{\beta}\right)\right)\right\Vert \\
&\leq \max_{1\leq l\leq L}\left\Vert \left(Z_{\varrho_{n,m}^{\boldsymbol{\theta}^l}}\left(\boldsymbol{\delta}^{l}\right),X_{\varrho_{n,m}^{\boldsymbol{\theta}^l}}\left(\boldsymbol{\beta}^{l}\right)\right)-\left(Z^{l}\left(\boldsymbol{\delta}^{l}\right),X^{l}\left(\boldsymbol{\beta}^{l}\right)\right)\right\Vert  + C \mathfrak{P}L^{-\frac{1}{2\mathfrak{P}+2}}.
\end{align*}
We will look at the first term on the RHS of the inequality. Note that for any fixed $\boldsymbol{\theta}$,
we have that 
\begin{align*}
 & P\left[\left\Vert \left(Z_{\varrho_{n,m}^{\boldsymbol{\theta}}}\left(\boldsymbol{\delta}\right),X_{\varrho_{n,m}^{\boldsymbol{\theta}}}\left(\boldsymbol{\beta}\right)\right)-\left(Z\left(\boldsymbol{\delta}\right),X\left(\boldsymbol{\beta}\right)\right)\right\Vert >\varsigma\right]\\
 & =P\left[\left.\left\Vert \left(Z_{\varrho_{n,m}^{\boldsymbol{\theta}}}\left(\boldsymbol{\delta}\right),X_{\varrho_{n,m}^{\boldsymbol{\theta}}}\left(\boldsymbol{\beta}\right)\right)-\left(Z\left(\boldsymbol{\delta}\right),X\left(\boldsymbol{\beta}\right)\right)\right\Vert >\varsigma\right|S_{n}\leq\frac{P_{D}n}{2}\right]\times P\left[S_{n}\leq\frac{C_{D}n}{2}\right]\\
 & +P\left[\left.\left\Vert \left(Z_{\varrho_{n,m}^{\boldsymbol{\theta}}}\left(\boldsymbol{\delta}\right),X_{\varrho_{n,m}^{\boldsymbol{\theta}}}\left(\boldsymbol{\beta}\right)\right)-\left(Z\left(\boldsymbol{\delta}\right),X\left(\boldsymbol{\beta}\right)\right)\right\Vert >\varsigma\right|S_{n}>\frac{C_{D}n}{2}\right]\times P\left[S_{n}>\frac{C_{D}n}{2}\right].
\end{align*}
Define $P(\boldsymbol{Z}_e,\boldsymbol{\delta},\boldsymbol{X}_e,\boldsymbol{\beta}, \varsigma) = P\left(\left.\left\Vert \left(Z_{i}\left(\boldsymbol{\delta}\right),X_{i}\left(\boldsymbol{\beta}\right)\right)-\left(Z\left(\boldsymbol{\delta}\right),X\left(\boldsymbol{\beta}\right)\right)\right\Vert \leq \varsigma\right|D_{i}=1\right)$. Under \autoref{condition3.1-1},  we obviously have that $\inf_{\boldsymbol{\theta}\in\varTheta}P(\boldsymbol{Z}_e,\boldsymbol{\delta},\boldsymbol{X}_e,\boldsymbol{\beta}, \varsigma)\geq C\alpha_n\varsigma^2$ for some $C>0$. Since $m/n\rightarrow0$, for $n$ sufficiently large  we have that
\begin{align*}
 & P\left[\left.\left\Vert \left(Z_{\varrho_{n,m}^{\boldsymbol{\theta}}}\left(\boldsymbol{\delta}\right),X_{\varrho_{n,m}^{\boldsymbol{\theta}}}\left(\boldsymbol{\beta}\right)\right)-\left(Z\left(\boldsymbol{\delta}\right),X\left(\boldsymbol{\beta}\right)\right)\right\Vert >\varsigma\right|S_{n}>\frac{C_{D}n}{2}\right]\\
 &=P\left[ \left.\sum_{j=0}^{m-1}\frac{S_n!}{j!(S_n - j)!}P(\boldsymbol{Z}_e,\boldsymbol{\delta},\boldsymbol{X}_e,\boldsymbol{\beta}, \varsigma)^{j}(1-P(\boldsymbol{Z}_e,\boldsymbol{\delta},\boldsymbol{X}_e,\boldsymbol{\beta}, \varsigma))^{S_n - j}\right|S_n > \frac{C_Dn}{2}\right]\\
 & \leq P\left[\left.Cmn^{m}\left(1-C\alpha_n\varsigma^{2}\right)^{S_{n}-m}\right|S_{n}>\frac{C_{D}n}{2}\right]\leq\exp\left(Cm\log\left(n\right)-\left(Cn-Cm\right)\alpha_n\varsigma^{2}\right).
\end{align*}
Then since $P\left[S_{n}\leq\frac{C_{D}n}{2}\right]\leq P\left[\left|S_{n}-ES_{n}\right|\geq\frac{C_{D}n}{2}\right]\leq1/C_{D}^{2}n$, we have that 
\begin{align*}
& P\left[\max_{1\leq l\leq L}\left\Vert \left(Z_{\varrho_{n,m}^{\boldsymbol{\theta}^l}}\left(\boldsymbol{\delta}^{l}\right),X_{\varrho_{n,m}^{\boldsymbol{\theta}^l}}\left(\boldsymbol{\beta}^{l}\right)\right)-\left(Z^{l}\left(\boldsymbol{\delta}^{l}\right),X^{l}\left(\boldsymbol{\beta}^{l}\right)\right)\right\Vert >\varsigma\right]\\
& =P\left[\left.\max_{1\leq l\leq L}\left\Vert \left(Z_{\varrho_{n,m}^{\boldsymbol{\theta}^l}}\left(\boldsymbol{\delta}^{l}\right),X_{\varrho_{n,m}^{\boldsymbol{\theta}^l}}\left(\boldsymbol{\beta}^{l}\right)\right)-\left(Z^{l}\left(\boldsymbol{\delta}^{l}\right),X^{l}\left(\boldsymbol{\beta}^{l}\right)\right)\right\Vert >\varsigma\right|S_{n}\leq\frac{C_{D}n}{2}\right]\times P\left[S_{n}\leq\frac{C_{D}n}{2}\right]\\
 & +P\left[\left.\max_{1\leq l\leq L}\left\Vert \left(Z_{\varrho_{n,m}^{\boldsymbol{\theta}^l}}\left(\boldsymbol{\delta}^{l}\right),X_{\varrho_{n,m}^{\boldsymbol{\theta}^l}}\left(\boldsymbol{\beta}^{l}\right)\right)-\left(Z^{l}\left(\boldsymbol{\delta}^{l}\right),X^{l}\left(\boldsymbol{\beta}^{l}\right)\right)\right\Vert>\varsigma\right|S_{n}>\frac{C_{D}n}{2}\right]\times P\left[S_{n}>\frac{C_{D}n}{2}\right]\\
 &\leq \sum_{l=1}^LP\left[\left.\left\Vert \left(Z_{\varrho_{n,m}^{\boldsymbol{\theta}^l}}\left(\boldsymbol{\delta}^{l}\right),X_{\varrho_{n,m}^{\boldsymbol{\theta}^l}}\left(\boldsymbol{\beta}^{l}\right)\right)-\left(Z^{l}\left(\boldsymbol{\delta}^{l}\right),X^{l}\left(\boldsymbol{\beta}^{l}\right)\right)\right\Vert >\varsigma\right|S_{n}>\frac{C_{D}n}{2}\right] + P\left[S_n\leq \frac{C_Dn}{2}\right]\\
 &\leq \exp\left(\log(T) +Cm\log(n) - (Cn - Cm)\alpha_n\varsigma^2\right) + 1/C_D^2n.
\end{align*}
This implies that 
\[
\max_{1\leq l\leq L}\left\Vert \left(Z_{\varrho_{n,m}^{\boldsymbol{\theta}^l}}\left(\boldsymbol{\delta}^{l}\right),X_{\varrho_{n,m}^{\boldsymbol{\theta}^l}}\left(\boldsymbol{\beta}^{l}\right)\right)-\left(Z^{l}\left(\boldsymbol{\delta}^{l}\right),X^{l}\left(\boldsymbol{\beta}^{l}\right)\right)\right\Vert =O_{P}\left(\sqrt{\frac{m\log\left(n\right)+\log\left(L\right)}{n\alpha_n}}\right).
\]
So 
\[
\sup_{\boldsymbol{\theta}\in\varTheta}\left\Vert \left(Z_{\varrho_{n,m}^{\boldsymbol{\theta}}}\left(\boldsymbol{\delta}\right),X_{\varrho_{n,m}^{\boldsymbol{\theta}}}\left(\boldsymbol{\beta}\right)\right)-\left(Z\left(\boldsymbol{\delta}\right),X\left(\boldsymbol{\beta}\right)\right)\right\Vert = O_P\left(\sqrt{\frac{m\log\left(n\right)+\log\left(L\right)}{n\alpha_n}} + \mathfrak{P}L^{-\frac{1}{2\mathfrak{P}+2}}\right)
\]
It remains to choose $L=\left[\mathfrak{P}^2n\alpha_n/m\log\left(n\right)\right]^{\mathfrak{P}+1}$
to show that 
\[
\sup_{\boldsymbol{\theta}\in\varTheta}\left\Vert \left(Z_{\varrho_{n,m}^{\boldsymbol{\theta}}}\left(\boldsymbol{\delta}\right),X_{\varrho_{n,m}^{\boldsymbol{\theta}}}\left(\boldsymbol{\beta}\right)\right)-\left(Z\left(\boldsymbol{\delta}\right),X\left(\boldsymbol{\beta}\right)\right)\right\Vert = O_P\left(\sqrt{\frac{(\mathfrak{P}+m)\log\left(n\right)}{n\alpha_n}}\right),
\]
and finally conclude the proof of the lemma.
\end{proof}
We next look at $|\frac{1}{m}\sum_{j=1}^{m}\epsilon_{\varrho_{n,j}^{\boldsymbol{\theta}}}\left(\boldsymbol{\delta},\boldsymbol{\beta}\right)|$.
We provide the following lemma.\\

\begin{lemma}\label{lemma2}
If $m/\log\left(n\right)\rightarrow\infty$ and
$m/n\rightarrow0$, we have that 
\[
\sup_{\boldsymbol{\theta}\in\varTheta}\left|\frac{1}{m}\sum_{j=1}^{m}\epsilon_{\varrho_{n,j}^{\boldsymbol{\theta}}}\left(\boldsymbol{\delta},\boldsymbol{\beta}\right)\right|=O_{P}\left(\sqrt{\frac{\log\left(n\right)}{m}}\right).
\]
\end{lemma}
\begin{proof}
Consider one realization of data set $\mathcal{S}_{n}=\left\{ \left(\boldsymbol{Z}_{e,1},D_{1},\boldsymbol{X}_{e,1},Y_{1}\right),\cdots,\left(\boldsymbol{Z}_{e,n},D_{n},\boldsymbol{X}_{e,n},Y_{n}\right)\right\} $.
Corresponding to such realization of $\mathcal{S}_{n}$, there is
one realization of $S_{n}$ and $i_{1},i_{2},\cdots,i_{S_{n}}$ such
that $i_{1}<i_{2}\cdots<i_{S_{n}}$ and for each $1\leq j\leq S_{n}$,
$D_{i_{j}}=1$. Conditioned on $\mathcal{S}_{n}$, according
to \citet{biau2015lectures}, for any $\boldsymbol{\theta}\in\varTheta$,
\[
\left(\boldsymbol{Z}_{e,\varrho_{n,1}^{\boldsymbol{\theta}}},\boldsymbol{X}_{e,\varrho_{n,1}^{\boldsymbol{\theta}}},Y_{e,\varrho_{n,1}^{\boldsymbol{\theta}}}\right),\left(\boldsymbol{Z}_{e,\varrho_{n,2}^{\boldsymbol{\theta}}},\boldsymbol{X}_{e,\varrho_{n,2}^{\boldsymbol{\theta}}},Y_{e,\varrho_{n,2}^{\boldsymbol{\theta}}}\right),\cdots,\left(\boldsymbol{Z}_{e,\varrho_{n,S_n}^{\boldsymbol{\theta}}},\boldsymbol{X}_{e,\varrho_{n,S_n}^{\boldsymbol{\theta}}},Y_{e,\varrho_{n,S_n}^{\boldsymbol{\theta}}}\right)
\]
are independent to each other, and satisfy that 
\[
E\left(\left.\epsilon_{\varrho_{n,j}^{\boldsymbol{\theta}}}\left(\boldsymbol{\delta},\boldsymbol{\beta}\right)\right|\mathcal{S}_{n}\right)=0, j = 1, 2, \cdots, S_n.
\]
Moreover, conditional on $\mathcal{S}_{n}$, according to \citet{biau2015lectures}, there are at most $\left(Cn^2/\mathfrak{P}\right)^{2\mathfrak{P}+2}$
different realizations of $(\varrho_{n,1}^{\boldsymbol{\theta}},\cdots,\varrho_{n,S_{n}}^{\boldsymbol{\theta}})$, where $C$ is a fixed constant that does not depend on $n$, $\mathfrak{P}$, or $\mathcal{S}_n$.  
For each $\mathcal{S}_{n}$, denote all potential realizations as
$\{ (\varrho_{n,1}^{l},\cdots,\varrho_{n,S_{n}}^{l})\} _{l=1}^{N_{n}}$
(which are $\mathcal{S}_{n}$-dependent). Furthermore, define $\widetilde{\varTheta}^{l}$
as the collection of $\boldsymbol{\theta}$'s that lead to the order
$(\varrho_{n,1}^{l},\cdots,\varrho_{n,S_{n}}^{l})$, so
$\{ \widetilde{\varTheta}^{l}\} _{l=1}^{N_{n}}$ are disjoint
and $\bigcup_{l=1}^{N_{n}}\widetilde{\varTheta}^{l}=\varTheta$. Define 
$\varTheta^{l,l^{\prime}}=\varTheta^{l}\bigcap\widetilde{\varTheta}^{l^{\prime}}$,
and we pick an arbitrary point $\boldsymbol{\theta}^{l,l^{\prime}}\in\varTheta$. Then conditioned
on $\mathcal{S}_{n}$, we have that 
\begin{align*}
  \sup_{\boldsymbol{\theta}\in\varTheta}\left|\frac{1}{m}\sum_{j=1}^{m}\epsilon_{\varrho_{n,j}^{\boldsymbol{\theta}}}\left(\boldsymbol{\delta},\boldsymbol{\beta}\right)\right|&=\max_{1\leq l\leq L,1\leq l^{\prime}\leq N_{n}}\sup_{\boldsymbol{\theta}\in\varTheta^{tl}}\left|\frac{1}{m}\sum_{j=1}^{m}\epsilon_{\varrho_{n,j}^{\boldsymbol{\theta}}}\left(\boldsymbol{\delta},\boldsymbol{\beta}\right)\right|\\
 & \leq\max_{1\leq l\leq L,1\leq l^{\prime}\leq N_{n}}\left|\frac{1}{m}\sum_{j=1}^{m}\epsilon_{\varrho_{n,j}^{l}}\left(\boldsymbol{\delta}^{l,l^{\prime}},\boldsymbol{\beta}^{l,l^{\prime}}\right)\right|+C\mathfrak{P}L^{-\frac{1}{2\mathfrak{P}+2}}
\end{align*}
Now we look at the first term on the right-hand side of the last inequality.
We have that 
\begin{align*}
 & P\left[\left.\max_{1\leq l\leq L,1\leq l^{\prime}\leq N_{n}}\left|\frac{1}{m}\sum_{j=1}^{m}\epsilon_{\varrho_{n,j}^{l^{\prime}}}\left(\boldsymbol{\delta}^{l,l^{\prime}},\boldsymbol{\beta}^{l,l^{\prime}}\right)\right|>\varsigma\right|\mathcal{S}_{n},S_{n}>C_{D}n/2\right]\\
 & \leq\sum_{1\leq l\leq L,1\leq l^{\prime}\leq N_{n}}P\left[\left.\left|\frac{1}{m}\sum_{j=1}^{m}\epsilon_{\varrho_{n,j}^{l^{\prime}}}\left(\boldsymbol{\delta}^{l,l^{\prime}},\boldsymbol{\beta}^{l,l^{\prime}}\right)\right|>\varsigma\right|\mathcal{S}_{n},S_{n}>C_{D}n/2\right]\\
 & \leq\sum_{1\leq l\leq L,1\leq l^{\prime}\leq N_{n}}C\exp\left(-Cm\varsigma^{2}\right)\leq C\exp\left(C\left(\mathfrak{P}\left(\log\left(L\right)+\log\left(n\right)\right)-m\varsigma^{2}\right)\right).
\end{align*}
where the second last inequality comes from the fact that for any fixed $\boldsymbol{\delta}$ and $\boldsymbol{\beta}$, $\{\epsilon_{\varrho_{n,j}^{l}}\}_{j=1}^{S_n}$ are independent with each other conditioned on the data set $\mathcal{S}_n$ and that $\epsilon_{\varrho_{n,j}^l}$ is bounded. This implies that 
\begin{align*}
 & P\left[\max_{1\leq l\leq L,1\leq l^{\prime}\leq N_{n}}\left|\frac{1}{m}\sum_{j=1}^{m}\epsilon_{\varrho_{n,j}^{l^{\prime}}}\left(\boldsymbol{\delta}^{tl},\boldsymbol{\beta}^{tl}\right)\right|>\varsigma\right]\\
 & \leq P\left[\left.\max_{1\leq l\leq L,1\leq l^{\prime}\leq N_{n}}\left|\frac{1}{m}\sum_{j=1}^{m}\epsilon_{\varrho_{n,j}^{l^{\prime}}}\left(\boldsymbol{\delta}^{l,l^{\prime}},\boldsymbol{\beta}^{l,l^{\prime}}\right)\right|>\varsigma\right|S_{n}>C_{D}n/2\right]\times P\left[S_{n}>C_{D}n/2\right]+P\left[S_{n}\leq C_{D}n/2\right]\\
 & =E\left[\left.P\left[\left.\max_{1\leq l\leq L,1\leq l^{\prime}\leq N_{n}}\left|\frac{1}{m}\sum_{j=1}^{m}\epsilon_{\varrho_{n,j}^{l^{\prime}}}\left(\boldsymbol{\delta}^{l,l^{\prime}},\boldsymbol{\beta}^{l,l^{\prime}}\right)\right|>\varsigma\right|\mathcal{S}_{n},S_{n}>C_{D}n/2\right]\right|S_{n}>C_{D}n/2\right]\\
 & \times P\left[S_{n}>C_{D}n/2\right]+P\left[S_{n}\leq C_{D}n/2\right] \leq C\exp\left(C\left(\mathfrak{P}\left(\log\left(L\right) + \log\left(n\right)\right)-m\varsigma^{2}\right)\right)+1/C_{D}^{2}n.
\end{align*}
This implies that 
\[
\max_{1\leq l\leq L,1\leq l^{\prime}\leq N_{n}}\left|\frac{1}{m}\sum_{j=1}^{m}\epsilon_{\varrho_{n,j}^{l^{\prime}}}\left(\boldsymbol{\delta}^{l,l^{\prime}},\boldsymbol{\beta}^{l,l^{\prime}}\right)\right|=O_{P}\left(\sqrt{\frac{\mathfrak{P}\log\left(Ln\right)}{m}}\right),
\]
and 
\[
\sup_{\boldsymbol{\theta}\in\varTheta}\left|\frac{1}{m}\sum_{j=1}^{m}\epsilon_{\varrho_{n,j}^{\boldsymbol{\theta}}}\left(\boldsymbol{\delta},\boldsymbol{\beta}\right)\right|=O_{P}\left(\sqrt{\frac{\mathfrak{P}\log\left(Ln\right)}{m}}+ \mathfrak{P}L^{-\frac{1}{2\mathfrak{P}+2}}\right).
\]
Then if we choose $L=[m\sqrt{\mathfrak{P}}/\log(n)]^{\mathfrak{P}+1}$, we have that 
\[
\sup_{\boldsymbol{\theta}\in\varTheta}\left|\frac{1}{m}\sum_{j=1}^{m}\epsilon_{\varrho_{n,j}^{\boldsymbol{\theta}}}\left(\boldsymbol{\delta},\boldsymbol{\beta}\right)\right|=O_{P}\left(\sqrt{\frac{\mathfrak{P}^2\log (n)}{m}}\right).
\]
\end{proof}
Now we prove \autoref{theorme3.1-1}. Combine \autoref{lemma1} and \autoref{lemma2}, we obviously have that 
\[
\sup_{\boldsymbol{\theta}\in\varTheta}\left|\frac{1}{m}\sum_{j=1}^{m}Y_{\varrho_{n,j}^{\boldsymbol{\theta}}}-\mathcal{Y}\left(Z\left(\boldsymbol{\delta}\right),X\left(\boldsymbol{\beta}\right),\boldsymbol{\delta},\boldsymbol{\beta}\right)\right|=O_{P}\left(\sqrt{\frac{\mathfrak{P}(\mathfrak{P}+m) \log\left(n\right)}{n\alpha_n} + \frac{\mathfrak{P}^2\log (n)}{m}} \right).
\]
Then we have that 
\begin{align*}
 &\sup_{k\ge1} \left\Vert \frac{1}{S_{n}}\sum_{i=1}^{n}\sum_{j=1}^{n}w_{ij}D_{i}D_{j}Y_{j}-\frac{1}{nP_{D}}\sum_{i=1}^{n}D_{i}\mathcal{Y}\left(Z_i\left(\widehat{\boldsymbol{\delta}}\right),X_i\left(\widehat{\boldsymbol{\beta}}^{k}\right),\widehat{\boldsymbol{\delta}},\widehat{\boldsymbol{\beta}}^{k}\right)\boldsymbol{X}_{i}\right\Vert \\
 & =O_{P}\left(\sqrt{\frac{\mathfrak{P}(\mathfrak{P}+m)\cdot \log\left(n\right)}{n\alpha_n} + \frac{\mathfrak{P}^2\log (n)}{m}} \right).
\end{align*}
Since $\left\Vert\widehat{\boldsymbol{\delta}}-\boldsymbol{\delta}_{0}\right\Vert=O_{p}\left(\sqrt{p_Z/n}\right)$,
we have that 
\begin{align*}
& \sup_{k\ge1} \left\Vert \frac{1}{S_{n}}\sum_{i=1}^{n}\sum_{j=1}^{n}w_{ij}D_{i}D_{j}Y_{j}-\frac{1}{nP_{D}}\sum_{i=1}^{n}D_{i}\mathcal{Y}\left(Z_{0,i},X_i\left(\widehat{\boldsymbol{\beta}}^{k}\right),\boldsymbol{\delta}_{0},\widehat{\boldsymbol{\beta}}^{k}\right)\boldsymbol{X}_{i}\right\Vert \\
 & =O_{P}\left(\sqrt{\frac{\mathfrak{P}(\mathfrak{P}+m)\cdot \log\left(n\right)}{n\alpha_n} + \frac{\mathfrak{P}^2\log (n)}{m}} \right),
\end{align*}
where recall that $Z_{0,i}=z_{0,i}+\boldsymbol{Z}_i^{\mathrm{T}}\boldsymbol{\delta}_{0}$.
Moreover, given all the conditions, it's not difficult to prove that 
\begin{align*}
\sup_{\boldsymbol{\beta}\in\mathscr{B}} \left\Vert \frac{1}{nC_{D}}\sum_{i=1}^{n}D_{i}\mathcal{Y}\left(Z_{0},X\left(\boldsymbol{\beta}\right),\boldsymbol{\delta}_{0},\boldsymbol{\beta}\right)\boldsymbol{X}_{i}-E\left(\left.\mathcal{Y}\left(Z_{0},X\left(\boldsymbol{\beta}\right),\boldsymbol{\delta}_{0},\boldsymbol{\beta}\right)\boldsymbol{X}_{i}\right|D_{i}=1\right)\right\Vert =O_{P}\left(\sqrt{\frac{\mathfrak{P}\log\left(n\right)}{n}}\right).
\end{align*}
This leads to that 
\begin{align*}
&\sup_{k\ge1} \left\Vert \frac{1}{S_{n}}\sum_{i=1}^{n}\sum_{j=1}^{n}w_{ij}D_{i}D_{j}Y_{j}-E\left(\left.\mathcal{Y}\left(Z_{0},X\left(\widehat{\boldsymbol{\beta}}^{k}\right),\boldsymbol{\delta}_{0},\widehat{\boldsymbol{\beta}}^{k}\right)\boldsymbol{X}\right|D=1\right)\right\Vert \\
& =O_{P}\left(\sqrt{\frac{\mathfrak{P}(\mathfrak{P}+m)\cdot \log\left(n\right)}{n\alpha_n} + \frac{\mathfrak{P}^2\log (n)}{m}} \right).
\end{align*}

For any $\boldsymbol{\beta}$, we have that 
\begin{align*}
 & E\left(\left.\mathcal{Y}\left(Z_{0},X\left(\boldsymbol{\beta}\right),\boldsymbol{\delta}_{0},\boldsymbol{\beta}\right)\boldsymbol{X}\right|D=1\right)-E\left(\left.Y\boldsymbol{X}\right|D=1\right)\\
= & E\left(\left.\left(E_{\widetilde{\boldsymbol{Z}}_{e},\widetilde{\boldsymbol{X}}_{e},\widetilde{D}}\left(\left.G\left(\widetilde{Z}_{0},\widetilde{X}_{0}\right)\right|\widetilde{Z}_{0}=Z_{0},\widetilde{X}\left(\boldsymbol{\beta}\right)=X\left(\boldsymbol{\beta}\right),\widetilde{D}=1\right)-G\left(Z_{0},X_{0}\right)\right)\boldsymbol{X}\right|D=1\right)\\
= & E\left(\left.\left(E_{\widetilde{\boldsymbol{Z}}_{e},\widetilde{\boldsymbol{X}}_{e},\widetilde{D}}\left(\left.G\left(Z_{0},X\left(\boldsymbol{\beta}\right)-\widetilde{\boldsymbol{X}}^{\mathrm{T}}\Delta\boldsymbol{\beta}\right)-G\left(Z_{0},X_{0}\right)\right|\widetilde{Z}_{0}=Z_{0},\widetilde{X}\left(\boldsymbol{\beta}\right)=X\left(\boldsymbol{\beta}\right),\widetilde{D}=1\right)\right)\boldsymbol{X}\right|D=1\right)
\end{align*}
Note that 
\begin{align*}
&G\left(Z_{0},X\left(\boldsymbol{\beta}\right)-\widetilde{\boldsymbol{X}}^{\mathrm{T}}\Delta\boldsymbol{\beta}\right)-G\left(Z_{0},X_{0}\right) \\
& =\int_{0}^1\nabla_{v}G\left(Z_{0},X_0 +\varsigma\left(\boldsymbol{X}-\widetilde{\boldsymbol{X}}\right)^{\mathrm{T}}\Delta\boldsymbol{\beta}\right)\left(\boldsymbol{X}-\widetilde{\boldsymbol{X}}\right)^{\mathrm{T}}\Delta\boldsymbol{\beta}d\varsigma,
\end{align*}
this implies that 
\[
E\left(\left.\mathcal{Y}\left(Z_{0},X\left(\boldsymbol{\beta}\right),\boldsymbol{\delta}_{0},\boldsymbol{\beta}\right)\boldsymbol{X}\right|D=1\right)-E\left(\left.Y\boldsymbol{X}\right|D=1\right)
 = \boldsymbol{\Psi}_{M}(\boldsymbol{\beta})\Delta\boldsymbol{\beta}
\]
The above analysis implies that
\[
\sup_{k\geq1}\left\Vert \Delta\widehat{\boldsymbol{\beta}}_{k+1}-\left(\boldsymbol{I}_{p_{X}}-\gamma\boldsymbol{\Psi}_{M}\left(\widehat{\boldsymbol{\beta}}_{k}\right)\right)\Delta\widehat{\boldsymbol{\beta}}_{k}\right\Vert =O_{P}\left(\sqrt{\frac{\mathfrak{P}(\mathfrak{P}+m)\cdot \log\left(n\right)}{n\alpha_n} + \frac{\mathfrak{P}^2\log (n)}{m}} \right).
\]
Since each argument of $\boldsymbol{\Psi}_M\left(\widehat{\boldsymbol{\beta}}^k\right)^{\mathrm{T}}\boldsymbol{\Psi}_M\left(\widehat{\boldsymbol{\beta}}^k\right)$ is bounded by $C_G^2p_X$, we have that 
\begin{align*}
& \left|\overline{\lambda}\left(I-\gamma \left(\boldsymbol{\Psi}_M\left(\widehat{\boldsymbol{\beta}}^k\right) + \boldsymbol{\Psi}_M\left(\widehat{\boldsymbol{\beta}}^k\right)^{\mathrm{T}}\right) + \gamma^2 \boldsymbol{\Psi}_M\left(\widehat{\boldsymbol{\beta}}^k\right)^{\mathrm{T}}\boldsymbol{\Psi}_M\left(\widehat{\boldsymbol{\beta}}^k\right)\right)\right.\\
&\left.- \overline{\lambda}\left(I-\gamma \left(\boldsymbol{\Psi}_M\left(\widehat{\boldsymbol{\beta}}^k\right) + \boldsymbol{\Psi}_M\left(\widehat{\boldsymbol{\beta}}^k\right)^{\mathrm{T}}\right)\right)\right|\leq \gamma^2 \left\Vert  \boldsymbol{\Psi}_M\left(\widehat{\boldsymbol{\beta}}^k\right)^{\mathrm{T}}\boldsymbol{\Psi}_M\left(\widehat{\boldsymbol{\beta}}^k\right)\right\Vert \leq \gamma^2 C_G^2p_{X}^2.
\end{align*}
Since \autoref{condition3.1-3} holds, we have that when   $\gamma < \min\{\overline{\lambda}_{\boldsymbol{\Psi}_M}^{-1}, \underline{\lambda}_{\boldsymbol{\Psi}_M}/2p_{X}^2C_G^2\}$, there holds 
\[
\overline{\lambda}\left(I-\gamma \left(\boldsymbol{\Psi}_M\left(\widehat{\boldsymbol{\beta}}^k\right) + \boldsymbol{\Psi}_M\left(\widehat{\boldsymbol{\beta}}^k\right)^{\mathrm{T}}\right) + \gamma^2 \boldsymbol{\Psi}_M\left(\widehat{\boldsymbol{\beta}}^k\right)^{\mathrm{T}}\boldsymbol{\Psi}_M\left(\widehat{\boldsymbol{\beta}}^k\right)\right)\leq 1 - \frac{\underline{\lambda}_{\boldsymbol{\Psi}_M}\gamma}{2}
\]
This implies that $\overline{\sigma}(I-\gamma\boldsymbol{\Psi}_M(\widehat{\boldsymbol{\beta}}^k))\leq \sqrt{1-\underline{\lambda}_{\boldsymbol{\Psi}_M}\gamma/2}\leq 1-\underline{\lambda}_{\boldsymbol{\Psi}_M}\gamma/4$ uniformly for all $k$. Using the method in \citet{khanetal2021}, we finish the proof. 

\subsection{Proof of \autoref{theorem3.2-1}}

Recall that $\widehat{Z}_{i}=z_{0,i}+\boldsymbol{Z}_{i}^{\mathrm{T}}\widehat{\boldsymbol{\delta}}$,
$Z_{0,i}=z_{0,i}+\boldsymbol{Z}_{i}^{\mathrm{T}}\boldsymbol{\delta}_{0}$, $\widehat{X}_{i}^{k}=x_{0,i}+\boldsymbol{X}_{i}^{\mathrm{T}}\widehat{\boldsymbol{\beta}}^{k}$,
and $X_{0,i}=x_{0,i}+\boldsymbol{X}_{i}^{\mathrm{T}}\boldsymbol{\beta}_{0}$. Also denote
$\varepsilon_{i}=Y_{i}-G\left(Z_{0,i},X_{0,i}\right)$.
We can rewite the dynamics of $\widehat{\boldsymbol{\beta}}^k$ as follows
\begin{align*}
\widehat{\boldsymbol{\beta}}^{k+1}=\widehat{\boldsymbol{\beta}}^{k} & -\frac{\gamma}{S_{n}}\sum_{i=1}^{n}D_{i}\left(\boldsymbol{\Phi}_{q}\left(\widehat{Z}_{i},\widehat X_{i}^{k}\right)^{\mathrm{T}}\widehat{\boldsymbol{\Pi}}_{q}^{k}-Y_{i}\right)\boldsymbol{X}_{i}\\
=\widehat{\boldsymbol{\beta}}^{k} & -\frac{\gamma}{S_{n}}\sum_{i=1}^{n}D_{i}\boldsymbol{X}_{i}\boldsymbol{\Phi}_{q}\left(\widehat{Z}_{i},\widehat X_{i}^{k}\right)^{\mathrm{T}}\left(\boldsymbol{\widehat{\Pi}}_{q}^{k}-\boldsymbol{\Pi}_{q}\right)-\frac{\gamma}{S_{n}}\sum_{i=1}^{n}D_{i}\boldsymbol{X}_{i}\left(\boldsymbol{\Phi}_{q}\left(\widehat{Z}_{i},\widehat X_{i}^{k}\right)^{\mathrm{T}}\boldsymbol{\Pi}_{q}-G\left(\widehat Z_{i},\widehat X_{i}^{k}\right)\right)\\
 & -\frac{\gamma}{S_{n}}\sum_{i=1}^{n}D_{i}\boldsymbol{X}_{i}\left(G\left(\widehat{Z}_{i},\widehat X_{i}^{k}\right)-G\left(Z_{0,i},\widehat X_{i}^{k}\right)\right)-\frac{\gamma}{S_{n}}\sum_{i=1}^{n}D_{i}\boldsymbol{X}_{i}\left(G\left(Z_{0,i},\widehat X_{i}^{k}\right)-G\left(Z_{0,i},X_{0,i}\right)\right)\\
 & -\frac{\gamma}{S_{n}}\sum_{i=1}^{n}D_{i}\boldsymbol{X}_{i}\varepsilon_{i}.
\end{align*}
Now we further derive the expression for $\widehat{\boldsymbol{\Pi}}_{q}^{k}-\boldsymbol{\Pi}_{q}$.
Define
\[
\widehat{\boldsymbol{\Gamma}}_{n,q}\left(\boldsymbol{\delta},\boldsymbol{\beta}\right)=\frac{1}{S_n}\sum_{i=1}^{n}D_{i}\boldsymbol{\Phi}_{q}\left(Z_i(\boldsymbol{\delta}),X_i(\boldsymbol{\beta})\right)\boldsymbol{\Phi}_{q}\left(Z_i(\boldsymbol{\delta}),X_i(\boldsymbol{\beta})\right)^{\mathrm{T}}.
\]
Then 
\begin{align*}
\widehat{\boldsymbol{\Pi}}_{q}^{k}-\boldsymbol{\Pi}_{q} & =\widehat{\boldsymbol{\Gamma}}_{n,q}^{-1}\left(\widehat{\boldsymbol{\delta}},\widehat{\boldsymbol{\beta}}^{k}\right)\frac{1}{S_n}\sum_{j=1}^{n}D_{j}\boldsymbol{\Phi}_{q}\left(\widehat{Z}_{j},\widehat X_{j}^{k}\right)Y_{j}-\boldsymbol{\Pi}_{q}\\
 & =\widehat{\boldsymbol{\Gamma}}_{n,q}^{-1}\left(\widehat{\boldsymbol{\delta}},\widehat{\boldsymbol{\beta}}^{k}\right)\frac{1}{S_n}\sum_{j=1}^{n}D_{j}\boldsymbol{\Phi}_{q}\left(\widehat{Z}_{j},\widehat{X}_{j}^{k}\right)\left(G\left(Z_{0,j},X_{0,j}\right)-G\left(Z_{0,j},\widehat{X}_{j}^{k}\right)\right)\\
 & +\widehat{\boldsymbol{\Gamma}}_{n,q}^{-1}\left(\widehat{\boldsymbol{\delta}},\widehat{\boldsymbol{\beta}}^{k}\right)\frac{1}{S_n}\sum_{j=1}^{n}D_{j}\boldsymbol{\Phi}_{q}\left(\widehat{Z}_{j},\widehat{X}_{j}^{k}\right)\left(G\left(Z_{0,j},\widehat{X}_{j}^{k}\right)-G\left(\widehat{Z}_{j},\widehat{X}_{j}^{k}\right)\right)\\
 & +\widehat{\boldsymbol{\Gamma}}_{n,q}^{-1}\left(\widehat{\boldsymbol{\delta}},\widehat{\boldsymbol{\beta}}^{k}\right)\frac{1}{S_n}\sum_{j=1}^{n}D_{j}\boldsymbol{\Phi}_{q}\left(\widehat{Z}_{j},\widehat{X}_{j}^{k}\right)\left(G\left(\widehat{Z}_{j},\widehat{X}_{j}^{k}\right)-\boldsymbol{\Phi}_{q}\left(\widehat{Z}_{j},\widehat{X}_{j}^{k}\right)^{\mathrm{T}}\boldsymbol{\Pi}_{q}\right)\\
 & +\widehat{\boldsymbol{\Gamma}}_{n,q}^{-1}\left(\widehat{\boldsymbol{\delta}},\widehat{\boldsymbol{\beta}}^{k}\right)\frac{1}{S_n}\sum_{j=1}^{n}D_{j}\boldsymbol{\Phi}_{q}\left(\widehat{Z}_{j},\widehat{X}_{j}^{k}\right)\varepsilon_{j}.
\end{align*}
Define \[\widehat{\mathcal{X}}_n\left(\nu_{Z},\nu_{X},\boldsymbol{\beta}\right)=\frac{1}{S_n}\sum_{i=1}^{n}D_{i}\boldsymbol{X}_{i}\boldsymbol{\Phi}_{q}\left(\widehat{Z}_{0,i},x_{0,i}+\boldsymbol{X}_{i}^{\mathrm{T}}\boldsymbol{\beta}\right)^{\mathrm{T}}\widehat{\boldsymbol{\Gamma}}_{n,q}^{-1}\left(\widehat{\boldsymbol{\delta}},\boldsymbol{\beta}\right)\boldsymbol{\Phi}_{q}\left(\nu_{Z},\nu_{X}\right).\]
Taking the expression of $\widehat{\boldsymbol{\Pi}}_{q}^{k}-\boldsymbol{\Pi}_{q}$
into the expression of $\widehat{\boldsymbol{\beta}}^{k+1}$, we have that 

\begin{align}
\widehat{\boldsymbol{\beta}}^{k+1}=\widehat{\boldsymbol{\beta}}^{k} & -\frac{\gamma}{S_{n}}\sum_{i=1}^{n}D_{i}\left(\boldsymbol{X}_{i}-\widehat{\mathcal{X}}_n\left(\widehat{Z}_{i},\widehat{X}_{i}^{k},\boldsymbol{\beta}^k\right)\right)\left(\boldsymbol{\Phi}_{q}\left(\widehat{Z}_{i},\widehat{X}_{i}^{k}\right)^{\mathrm{T}}\boldsymbol{\Pi}_{q}-G\left(\widehat{Z}_{i},\widehat{X}_{i}^{k}\right)\right)\nonumber\\
 & -\frac{\gamma}{S_{n}}\sum_{i=1}^{n}D_{i}\left(\boldsymbol{X}_{i}-\widehat{\mathcal{X}}_{n} \left(\widehat{Z}_{i},\widehat{X}_{i}^{k},\boldsymbol{\beta}^k\right)\right)\left(G\left(\widehat{Z}_{i},\widehat{X}_{i}^{k}\right)-G\left(Z_{0,i},\widehat{X}_{i}^{k}\right)\right)\nonumber\\
 & -\frac{\gamma}{S_{n}}\sum_{i=1}^{n}D_{i}\left(\boldsymbol{X}_{i}-\widehat{\mathcal{X}}_{n}\left(\widehat{Z}_{i},\widehat{X}_{i}^{k},\boldsymbol{\beta}^k\right)\right)\left(G\left(Z_{0,i},\widehat{X}_{i}^{k}\right)-G\left(Z_{0,i},X_{0,i}\right)\right)\nonumber\\
 & -\frac{\gamma}{S_{n}}\sum_{i=1}^{n}D_{i}\left(\boldsymbol{X}_{i}-\widehat{\mathcal{X}}_{n}\left(\widehat{Z}_{i},\widehat{X}_{i}^{k},\boldsymbol{\beta}^k\right)\right)\varepsilon_{i}.\label{dynamics_of_beta}
\end{align}
We next introduce some lemmas that will be useful in analyzing the above dynamics.

\begin{lemma}\label{lemma3}
Let \autoref{condition1}--\autoref{condition3}, \autoref{condition4} and \autoref{condition5} hold, then 
\[
\sup_{\boldsymbol{\beta}\in\mathscr{B}}\left\Vert \widehat{\boldsymbol{\Gamma}}_{n,q}\left(\widehat{\boldsymbol{\delta}},\boldsymbol{\beta}\right)-\boldsymbol{\Gamma}_{q}\left(\boldsymbol{\beta}\right)\right\Vert =O_P\left(\frac{q^2C(0,q)\left(p_{Z}C(1,q) + \sqrt{\log(n)p_{X}}C(0,q)\right)}{\sqrt{n}}\right).
\]
If further $\Xi_{1,n}\rightarrow0$, then \[P\left(\sup_{\boldsymbol{\beta}\in\mathscr{B}}\overline{\lambda}\left(\widehat{\boldsymbol{\Gamma}}_{n,q}^{-1}\left(\widehat{\boldsymbol{\delta}},\boldsymbol{\beta}\right)\right)\leq1.5\underline{\lambda}_{\boldsymbol{\Phi}}^{-1}\right)\rightarrow1.\] 
\end{lemma}
\begin{proof}
Recall that $\boldsymbol{\Gamma}_{q}\left(\boldsymbol{\beta}\right)=E\left[\boldsymbol{\Phi}_{q}\left(Z_{0,i},X_i(\boldsymbol{\beta})\right)\boldsymbol{\Phi}_{q}\left(Z_{0,i},X_i(\boldsymbol{\beta})\right)^{\mathrm{T}}|D_i=1\right]$. To show the first result, we first note that
\begin{align*}
\sup_{\boldsymbol{\beta}\in\mathscr{B}}\left\Vert \widehat{\boldsymbol{\Gamma}}_{n,q}\left(\widehat{\boldsymbol{\delta}},\boldsymbol{\beta}\right)-\widehat{\boldsymbol{\Gamma}}_{n,q}\left(\boldsymbol{\delta}_{0},\boldsymbol{\beta}\right)\right\Vert =\sup_{\boldsymbol{\beta}\in\mathscr{B}} & \left\Vert S_{n}^{-1}\sum_{i=1}^{n}D_{i}\boldsymbol{\Phi}_{q}\left(\widehat{Z}_{i},X_i(\boldsymbol{\beta})\right)\boldsymbol{\Phi}_{q}\left(\widehat{Z}_{i},X_i(\boldsymbol{\beta})\right)^{\mathrm{T}}\right.\\& \left.-S_{n}^{-1}\sum_{i=1}^{n}D_{i}\boldsymbol{\Phi}_{q}\left(Z_{0,i},X_i(\boldsymbol{\beta})\right)\boldsymbol{\Phi}_{q}\left(Z_{0,i},X_i(\boldsymbol{\beta})\right)^{\mathrm{T}}\right\Vert \\
\leq & Cq^{2}C(0,q)C(1,q)\max_{1\leq i\leq n}\left|\widehat{Z}_{i}-Z_{0,i}\right|.
\end{align*}
According to \autoref{condition2} and \autoref{condition3}, $\max_{1\leq i\leq n}\left|\widehat{Z}_{i}-Z_{0,i}\right|\leq  C\sqrt{p_{Z}}\left\Vert \widehat{\boldsymbol{\delta}}-\boldsymbol{\delta}_{0}\right\Vert =O_{P}\left(p_{Z}/\sqrt{n}\right),$
so 
\[
\sup_{\boldsymbol{\beta}\in\mathscr{B}}\left\Vert \widehat{\boldsymbol{\Gamma}}_{n,q}\left(\widehat{\boldsymbol{\delta}},\boldsymbol{\beta}\right)-\widehat{\boldsymbol{\Gamma}}_{n,q}\left(\boldsymbol{\delta}_{0},\boldsymbol{\beta}\right)\right\Vert =O_{P}\left(\frac{p_{Z}q^{2}C(0,q)C(1,q)}{\sqrt{n}}\right).
\]
Also note that 
$\Vert
\widehat{\boldsymbol{\Gamma}}_{n,q}\left(\boldsymbol{\delta}_{0},\boldsymbol{\beta}\right) - \frac{1}{nP_D}\sum_{i=1}^{n}D_{i}\boldsymbol{\Phi}_{q}\left(Z_{0,i},X_i(\boldsymbol{\beta})\right)\boldsymbol{\Phi}_{q}\left(Z_{0,i},X_i(\boldsymbol{\beta})\right)^{\mathrm{T}}\Vert$ is uniformly of order $O_P(q^2C(0,q)^2/\sqrt{n})$, 
it then remains
to bound the following distance
\begin{align*}
\sup_{\boldsymbol{\beta}\in\mathscr{B}}&\left\Vert \frac{1}{n}\sum_{i=1}^{n}D_{i}\boldsymbol{\Phi}_{q}\left(Z_{0,i},X_i(\boldsymbol{\beta})\right)\Phi_{q}\left(Z_{0,i},X_i(\boldsymbol{\beta})\right)^{\mathrm{T}}-E\left(D_{i}\boldsymbol{\Phi}_{q}\left(Z_{0,i},X_i(\boldsymbol{\beta})\right)\boldsymbol{\Phi}_{q}\left(Z_{0,i},X_i(\boldsymbol{\beta})\right)^{\mathrm{T}}\right)\right\Vert .
\end{align*}
Note that each argument
of $D_{i}\boldsymbol{\Phi}_{q}\left(Z_{0,i},X_i(\boldsymbol{\beta})\right)\boldsymbol{\Phi}_{q}\left(Z_{0,i},X_i(\boldsymbol{\beta})\right)^{\mathrm{T}}$
is bounded by $C(0,q)^{2}$, and each argument of $\boldsymbol{\Phi}_{q}\left(Z_{0,i},X_i(\boldsymbol{\beta})\right)\boldsymbol{\Phi}_{q}\left(Z_{0,i},X_i(\boldsymbol{\beta})\right)^{\mathrm{T}}$ has partial derivative with respect to $\boldsymbol{\beta}$
 that is bounded (in norm) by $ C(0,q)C(1,q)\sqrt{p_{X}}$.
So using Lemma A1 of \citet{khanetal2021}, we have that 
\begin{align*}
\sup_{\boldsymbol{\beta}\in\mathscr{B}}&\left\Vert \frac{1}{n}\sum_{i=1}^{n}D_{i}\boldsymbol{\Phi}_{q}\left(Z_{0,i},X_i(\boldsymbol{\beta})\right)\boldsymbol{\Phi}_{q}\left(Z_{0,i},X_i(\boldsymbol{\beta})\right)^{\mathrm{T}}
 -E\left(D_{i}\boldsymbol{\Phi}_{q}\left(Z_{0,i},X_i(\boldsymbol{\beta})\right)\boldsymbol{\Phi}_{q}\left(Z_{0,i},X_i(\boldsymbol{\beta})\right)^{\mathrm{T}}\right)\right\Vert\\
 =O_{P}&\left(\frac{\sqrt{\log\left(n\right)p_{X}}q^{2}C(0,q)^2}{\sqrt{n}}\right).
\end{align*}

This shows the first result.
Since $\Xi_{1,n}\rightarrow0$, we have that $\sup_{\boldsymbol{\beta}\in\mathscr{B}}\left\Vert \widehat{\boldsymbol{\Gamma}}_{n,q}\left(\widehat{\boldsymbol{\delta}},\boldsymbol{\beta}\right)-\boldsymbol{\Gamma}_{q}\left(\boldsymbol{\beta}\right)\right\Vert \rightarrow_{p}0.$
Using the fact that 
\[
\sup_{\boldsymbol{\beta}\in\mathscr{B}}\left|\underline{\lambda}\left(\widehat{\boldsymbol{\Gamma}}_{n,q}\left(\widehat{\boldsymbol{\delta}},\boldsymbol{\beta}\right)\right)-\underline{\lambda}\left(\boldsymbol{\Gamma}_{q}\left(\boldsymbol{\beta}\right)\right)\right|\leq\sup_{\boldsymbol{\beta}\in\mathscr{B}}\left\Vert \widehat{\boldsymbol{\Gamma}}_{n,q}\left(\widehat{\boldsymbol{\delta}},\boldsymbol{\beta}\right)-\boldsymbol{\Gamma}_{q}\left(\boldsymbol{\beta}\right)\right\Vert \rightarrow_{p}0,
\]
we have that $\inf_{\boldsymbol{\beta}\in\mathscr{B}}\underline{\lambda}\left(\widehat{\boldsymbol{\Gamma}}_{n,q}\left(\widehat{\boldsymbol{\delta}},\boldsymbol{\beta}\right)\right)\geq2/3\inf_{\boldsymbol{\beta}\in\mathscr{B}}\underline{\lambda}\left(\boldsymbol{\Gamma}_{q}\left(\boldsymbol{\beta}\right)\right)\geq2/3\underline{\lambda}_{\boldsymbol{\Phi}}$
with probability going to 1. Then $\sup_{\boldsymbol{\beta}\in\mathscr{B}}\overline{\lambda}\left(\widehat{\boldsymbol{\Gamma}}_{n,q}^{-1}\left(\widehat{\boldsymbol{\delta}},\boldsymbol{\beta}\right)\right)\leq1.5\underline{\lambda}_{\boldsymbol{\Phi}}^{-1}$  with probability going to 1. This proves
the result.
\end{proof}

\begin{lemma}\label{lemma4}
Let \autoref{condition1}--\autoref{condition3}, \autoref{condition4}, and \autoref{condition5}  hold and $\Xi_{1,n}\rightarrow0$, then 
\begin{align*}
&\sup_{\nu_Z, \nu_X, \boldsymbol{\beta}\in\mathscr{B}} \left\Vert \widehat{\mathcal{X}}_{n}\left(\nu_Z, \nu_X,\boldsymbol{\beta}\right)-\mathcal{X}\left(\nu_Z, \nu_X,\boldsymbol{\beta}\right)\right\Vert \\
& =O_{p}\left(q^2C(0,q)^2\mathscr{R}_X(q)+\frac{\sqrt{p_{X}}q^4C(0,q)^3\left(p_{Z}C(1,q) + C(0,q)\sqrt{\log(n)p_{X}}\right)}{\sqrt{n}}\right).
\end{align*}
\end{lemma}
\begin{proof}
Recall that according to  \autoref{condition5}(ii), we have that 
\[
\sup_{\nu_Z,\nu_X, \boldsymbol{\beta}\in\mathscr{B}}\left\Vert \mathcal{X}\left(\nu_Z,\nu_X, \boldsymbol{\beta}\right)-\boldsymbol{\Pi}_{q}^{\boldsymbol{X}}\left(\boldsymbol{\beta}\right)\boldsymbol{\Phi}_q\left(\nu_Z, \nu_X\right)\right\Vert \leq\mathscr{R}_X(q),
\]
that 
\begin{align*}
\sup_{\nu_Z,\nu_X, \boldsymbol{\beta}\in\mathscr{B}} & \left\Vert \frac{1}{S_n}\sum_{i=1}^{n}D_{i}\boldsymbol{X}_{i}\boldsymbol{\Phi}_{q}\left(\widehat{Z}_{i},X_i(\boldsymbol{\beta})\right)^{\mathrm{T}}\widehat{\boldsymbol{\Gamma}}_{n,q}^{-1}\left(\widehat{\boldsymbol{\delta}},\boldsymbol{\beta}\right){\boldsymbol{\Phi}}_{q}\left(\nu_Z, \nu_X\right)-\right.\\
 & \left.\frac{1}{nP_D}\sum_{i=1}^{n}D_{i}\boldsymbol{X}_i\boldsymbol{\Phi}_{q}\left(\widehat{Z}_{i},X_i(\boldsymbol{\beta})\right)^{\mathrm{T}}\widehat{\boldsymbol{\Gamma}}_{n,q}^{-1}\left(\widehat{\boldsymbol{\delta}},\boldsymbol{\beta}\right)\boldsymbol{\Phi}_{q}\left(\nu_Z, \nu_X\right)\right\Vert = O_p\left(q^2C(0,q)^2\sqrt{\frac{p_{X}}{n}}\right),
\end{align*}
and that 
\begin{align*}
& \left\Vert E\left(\left.\boldsymbol{X}_i\boldsymbol{\Phi}_{q}\left(Z_{0,i},X_i(\boldsymbol{\beta})\right)^{\mathrm{T}}\boldsymbol{\Gamma}_{q}^{-1}\left(\boldsymbol{\beta}\right)\right|D_i = 1\right)\boldsymbol{\Phi}_{q}\left(\nu_Z, \nu_X\right)-\boldsymbol{\Pi}_q^{\boldsymbol{X}}(\boldsymbol{\beta})\boldsymbol{\Phi}_{q}\left(\nu_Z, \nu_X\right)\right\Vert\\
& \leq \left\Vert E\left(\left.\boldsymbol{\Pi}_q^{\boldsymbol{X}}(\boldsymbol{\beta})\boldsymbol{\Phi}_{q}\left(Z_{0,i},X_i(\boldsymbol{\beta})\right)\boldsymbol{\Phi}_{q}\left(Z_{0,i},X_i(\boldsymbol{\beta})\right)^{\mathrm{T}}\boldsymbol{\Gamma}_{q}^{-1}\left(\boldsymbol{\beta}\right)\right|D_i = 1\right)\boldsymbol{\Phi}_{q}\left(\nu_Z, \nu_X\right)-\boldsymbol{\Pi}_q^{\boldsymbol{X}}(\boldsymbol{\beta})\boldsymbol{\Phi}_{q}\left(\nu_Z, \nu_X\right)\right\Vert\\
&+\left\Vert E\left(\left.\left(\mathcal{X}(Z_{0,i},X_i(\boldsymbol{\beta}), \boldsymbol{\beta}) - \boldsymbol{\Pi}_q^{\boldsymbol{X}}(\boldsymbol{\beta})\boldsymbol{\Phi}_{q}\left(Z_{0,i},X_i(\boldsymbol{\beta})\right)\right)\boldsymbol{\Phi}_{q}\left(Z_{0,i},X_i(\boldsymbol{\beta})\right)^{\mathrm{T}}\boldsymbol{\Gamma}_{q}^{-1}\left(\boldsymbol{\beta}\right)\right|D_i = 1\right)\boldsymbol{\Phi}_{q}\left(\nu_Z, \nu_X\right)\right\Vert\\
&\leq q^2C(0,q)^2\mathscr{R}_X(q),
\end{align*}
so we only need to bound the following
\begin{align*}
\sup_{\nu_Z,\nu_X, \boldsymbol{\beta}\in\mathscr{B}} & \left\Vert \frac{1}{n}\sum_{i=1}^{n}D_{i}\boldsymbol{X}_{i}\boldsymbol{\Phi}_{q}\left(\widehat{Z}_{i},X_i(\boldsymbol{\beta})\right)^{\mathrm{T}}\widehat{\boldsymbol{\Gamma}}_{n,q}^{-1}\left(\widehat{\boldsymbol{\delta}},\boldsymbol{\beta}\right)\boldsymbol{\Phi}_{q}\left(\nu_Z,\nu_X\right)\right.\\
 & \left.-E\left(D_{i}\boldsymbol{X}_{i}\boldsymbol{\Phi}_{q}\left(Z_{0,i},X_i(\boldsymbol{\beta})\right)^{\mathrm{T}}\boldsymbol{\Gamma}_{q}^{-1}\left(\boldsymbol{\beta}\right)\boldsymbol{\Phi}_{q}\left(\nu_Z,\nu_X\right)\right)\right\Vert. 
\end{align*}
Note that the above is bounded by 
\begin{align*}
 &\sup_{\nu_Z,\nu_X,\boldsymbol{\beta}} \left\Vert \frac{1}{n}\sum_{i=1}^{n}D_{i}\boldsymbol{X}_{i}\left(\boldsymbol{\Phi}_{q}\left(\widehat{Z}_{i},X_i(\boldsymbol{\beta})\right)-\boldsymbol{\Phi}_{q}\left(Z_{0,i},X_i(\boldsymbol{\beta})\right)\right)^{\mathrm{T}}\widehat{\boldsymbol{\Gamma}}_{n,q}^{-1}\left(\widehat{\boldsymbol{\delta}},\boldsymbol{\beta}\right)\boldsymbol{\Phi}_{q}\left(\nu_Z,\nu_X)\right)\right\Vert (i)\\
+ & \sup_{\nu_Z,\nu_X,\boldsymbol{\beta}}\left\Vert \frac{1}{n}\sum_{i=1}^{n}D_{i}\boldsymbol{X}_{i}\boldsymbol{\Phi}_{q}\left(Z_{0,i},X_i(\boldsymbol{\beta})\right)^{\mathrm{T}}\left(\widehat{\boldsymbol{\Gamma}}_{n,q}^{-1}\left(\widehat{\boldsymbol{\delta}},\boldsymbol{\beta}\right)-\boldsymbol{\Gamma}_{q}^{-1}\left(\boldsymbol{\beta}\right)\right)\boldsymbol{\Phi}_{q}\left(\nu_Z,\nu_X\right)\right\Vert (ii)\\
+ & \sup_{\nu_Z,\nu_X,\boldsymbol{\beta}}\left\Vert \frac{1}{n}\sum_{i=1}^{n}D_{i}\boldsymbol{X}_{i}\boldsymbol{\Phi}_{q}\left(Z_{0,i},X_i(\boldsymbol{\beta})\right)^{\mathrm{T}}\boldsymbol{\Gamma}_{q}^{-1}\left(\boldsymbol{\beta}\right)\boldsymbol{\Phi}_{q}\left(\nu_Z,\nu_X\right)\right.\\
 & \left.-E\left(D_{i}\boldsymbol{X}_{i}\boldsymbol{\Phi}_{q}\left(Z_{0,i},X_i(\boldsymbol{\beta})\right)^{\mathrm{T}}\boldsymbol{\Gamma}_{q}^{-1}\left(\boldsymbol{\beta}\right)\boldsymbol{\Phi}_{q}\left(\nu_Z,\nu_X\right)\right)\right\Vert (iii)
\end{align*}
According to the proof of \autoref{lemma3}, we know that  $\sup_{\boldsymbol{\beta}\in\mathscr{B}}\overline{\lambda}(\widehat{\Gamma}_{n,q}^{-1}(\widehat{\boldsymbol{\delta}},\boldsymbol{\beta}))\leq1.5\underline{\lambda}_{\boldsymbol{\Phi}}^{-1}$ with probability going
to 1. 
So 
\begin{align*}
\Vert(i)\Vert & \leq C\cdot \sqrt{p_X}q C(0,q)\sup_{1\leq i\leq n, \boldsymbol{\beta}\in\mathscr{B}}\left\Vert\left(\boldsymbol{\Phi}_{q}\left(\widehat{Z}_{i},X_i(\boldsymbol{\beta})\right)-\boldsymbol{\Phi}_{q}\left(Z_{0,i},X_i(\boldsymbol{\beta})\right)\right)\right\Vert\\
& =O_P\left(\frac{p_Z\sqrt{p_X}q^2 C(0,q)C(1,q)}{\sqrt{n}}\right),
\end{align*}
and according to the proof of \autoref{lemma3}, we have that 
\begin{align*}
   (ii) & \leq \sqrt{p_{X}}q^{2}C(0,q)^{2}\sup_{\boldsymbol{\beta}\in\mathscr{B}}\Vert \widehat{\boldsymbol{\Gamma}}_{n,q}^{-1}\left(\widehat{\boldsymbol{\delta}},\boldsymbol{\beta}\right)-\boldsymbol{\Gamma}_{q}^{-1}\left(\boldsymbol{\beta}\right)\Vert\\
   & = O_{P}\left(\frac{\sqrt{p_{X}}q^4C(0,q)^3\left(p_{Z}C(1,q) + C(0,q)\sqrt{\log(n)p_{X}}\right)}{\sqrt{n}}\right).
\end{align*}
For the last term (iii), we know that each argument of $D_{i}\boldsymbol{X}_{i}\boldsymbol{\Phi}_{q}\left(Z_{0,i},X_i(\boldsymbol{\beta})\right)^{\mathrm{T}}$
is bounded by $C(0,q)$ and each argument of $D_{i}\boldsymbol{X}_{i}\boldsymbol{\Phi}_{q}\left(Z_{0,i},X_i(\boldsymbol{\beta}_1)\right)^{\mathrm{T}}-D_{i}\boldsymbol{X}_{i}\boldsymbol{\Phi}_{q}\left(Z_{0,i},X_i(\boldsymbol{\beta}_2)\right)^{\mathrm{T}}$
is bounded by $C(1,q)\sqrt{p_X}\left\Vert \boldsymbol{\beta}_{1}-\boldsymbol{\beta}_{2}\right\Vert $.
Using Lemma A1 of Khan et al (2024), we have that 
\begin{align*}
(iii) & \leq CqC(0,q)\sup_{\boldsymbol{\beta}\in\mathscr{B}}\left\Vert \frac{1}{n}\sum_{i=1}^{n}D_{i}\boldsymbol{X}_{i}\boldsymbol{\Phi}_{q}\left(Z_{0,i},X_i(\boldsymbol{\beta})\right)^{\mathrm{T}}-E\left(D_{i}\boldsymbol{X}_{i}\boldsymbol{\Phi}_{q}\left(Z_{0,i},X_i(\boldsymbol{\beta})\right)^{\mathrm{T}}\right)\right\Vert \\
 & =O_{P}\left(\frac{\sqrt{\log(n)}p_{X}q^2C^2(0,q)}{\sqrt{n}}\right).
\end{align*}
This proves the results.
\end{proof}
\begin{lemma}\label{lemma5}
Let \autoref{condition1}--\autoref{condition3}, \autoref{condition4}, \autoref{condition5} hold and $\Xi_{1,n}\rightarrow0$, then 
\[
\sup_{k\geq1}\left\Vert \frac{1}{S_{n}}\sum_{i=1}^{n}D_{i}\left(\boldsymbol{X}_{i}-\widehat{\mathcal{X}}_{n}\left(\widehat{Z}_{i},\widehat{X}_{i}^{k},\widehat{\boldsymbol{\beta}}^{k}\right)\right)\left(\boldsymbol{\Phi}_{q}\left(\widehat{Z}_{i},\widehat{X}_{i}^{k}\right)^{\mathrm{T}}\boldsymbol{\Pi}_{q}-G\left(\widehat{Z}_{i},\widehat{X}_{i}^{k}\right)\right)\right\Vert =O_{P}\left(\sqrt{p_{X}}\mathscr{R}(q)\right),
\]
\[
\sup_{k\geq1}\left\Vert \frac{1}{S_{n}}\sum_{i=1}^{n}D_{i}\left(\boldsymbol{X}_{i}-\widehat{\mathcal{X}}_{n} \left(\widehat{Z}_{i},\widehat{X}_{i}^{k},\widehat{\boldsymbol{\beta}}^{k}\right)\right)\left(G\left(\widehat{Z}_{i},\widehat{X}_{i}^{k}\right)-G\left(Z_{0,i},\widehat{X}_{i}^{k}\right)\right)\right\Vert =O_{P}\left(p_{Z}\sqrt{p_{X}/n}\right),
\]
\begin{align*}
\sup_{k\geq1} & \left\Vert \frac{1}{S_n}\sum_{i=1}^{n}D_{i}\left(\boldsymbol{X}_{i}-\widehat{\mathcal{X}}_{n}\left(\widehat{Z}_{i},\widehat{X}_{i}^{k},\widehat{\boldsymbol{\beta}}^{k}\right)\right)\left(G\left(Z_{0,i},\widehat{X}_{i}^{k}\right)-G\left(Z_{0,i},X_{0,i}\right)\right) -\boldsymbol{\Psi}_S(\widehat{\boldsymbol{\beta}}^k)\Delta\boldsymbol{\beta}^{k}\right\Vert  \\
& =O_{P}\left(q^2C(0,q)^2\mathscr{R}_X(q)+\frac{\sqrt{p_{X}}q^4C(0,q)^3\left(p_{Z}C(1,q) + C(0,q)\sqrt{\log(n)p_{X}}\right)}{\sqrt{n}}\right),
\end{align*}
and
\begin{align*}
& \sup_{k\geq 1}\left\Vert\frac{1}{S_{n}}\sum_{i=1}^{n}D_{i}\left(\boldsymbol{X}_{i}-\widehat{\mathcal{X}}_{n}\left(\widehat{Z}_{i},\widehat{X}_{i}^{k},\boldsymbol{\beta}^k\right)\right)\varepsilon_{i}\right\Vert \\
& =O_P\left(q^2C(0,q)^2\mathscr{R}_X(q)+\frac{\sqrt{p_{X}}q^4C(0,q)^3\left(p_{Z}C(1,q) + C(0,q)\sqrt{\log(n)p_{X}}\right)}{\sqrt{n}}\right).
\end{align*}
\end{lemma}
\begin{proof}
The first two results are obvious following  that \autoref{lemma2} holds and the fact that $\mathcal{X}\left(\nu_Z,\nu_X,\boldsymbol{\beta}\right)$ is 
uniformly bounded. Next we look at the third term. Note that
\begin{align*}
\sup_{k\geq1} & \left\Vert \frac{1}{S_{n}}\sum_{i=1}^{n}D_{i}\left(\boldsymbol{X}_{i}-\widehat{\mathcal{X}}_{n}\left(\widehat{Z}_{i},\widehat{X}_{i}^{k},\widehat{\boldsymbol{\beta}}^{k}\right)\right)\left(G\left(Z_{0,i},\widehat{X}_{i}^{k}\right)-G\left(Z_{0,i},X_{0,i}\right)\right)\right.\\
 & \left.-\frac{1}{nP_D}\sum_{i=1}^{n}D_{i}\left(\boldsymbol{X}_{i}-\mathcal{X}\left(Z_{0,i},\widehat{X}_{i}^{k},\widehat{\boldsymbol{\beta}}^{k}\right)\right)\left(G\left(Z_{0,i},\widehat{X}_{i}^{k}\right)-G\left(Z_{0,i},X_{0,i}\right)\right)\right\Vert \\
 & =O_{P}\left(q^2C(0,q)^2\mathscr{R}_X(q)+\frac{\sqrt{p_{X}}q^4C(0,q)^3\left(p_{Z}C(1,q) + C(0,q)\sqrt{\log(n)p_{X}}\right)}{\sqrt{n}}\right).
\end{align*}
Then it remains to bound 
\begin{align*}
\sup_{\boldsymbol{\beta}\in\mathscr{B}} & \left\Vert \frac{1}{n}\sum_{i=1}^{n}D_{i}\left(\boldsymbol{X}_{i}-\mathcal{X}\left(Z_{0,i},X_{i}(\boldsymbol{\beta}),\boldsymbol{\beta}\right)\right)\left(G\left(Z_{0,i},X_i(\boldsymbol{\beta})\right)-G\left(Z_{0,i},X_{0,i}\right)\right)\right.\\
 & \left. - E\left(D_{i}\left(\boldsymbol{X}_{i}-\mathcal{X}\left(Z_{0,i},X_i(\boldsymbol{\beta}),\boldsymbol{\beta}\right)\right)\left(G\left(Z_{0,i},X_i(\boldsymbol{\beta})\right)-G\left(Z_{0,i},X_{0,i}\right)\right)\right)\right\Vert .
\end{align*}
Note that each argument of $D_{i}\left(\boldsymbol{X}_{i}-\mathcal{X}\left(Z_{0,i},X_i(\boldsymbol{\beta}),\boldsymbol{\beta}\right)\right)\left(G\left(Z_{0,i},X_i(\boldsymbol{\beta})\right)-G\left(Z_{0,i},X_{0,i}\right)\right)$
is upper bounded, and moreover, the gradient of each argument with
respect to $\boldsymbol{\beta}$ is bounded by $\sqrt{p_{X}}$ up to some scale
in norm. So using \citet{khanetal2021}'s Lemma A1, we have that the above term is of order $O_P(p_{X}\sqrt{\log(n)/n})$. 
We finish the proof by noting that, using Fubini's theorem, 
\begin{align*}
 & P_D^{-1}E\left(D_{i}\left(\boldsymbol{X}_{i}-\mathcal{X}\left(Z_{0,i},X_i(\boldsymbol{\beta}),\boldsymbol{\beta}\right)\right)\left(G\left(Z_{0,i},X_i(\boldsymbol{\beta})\right)-G\left(Z_{0,i},X_{0,i}\right)\right)\right)\\
 & =E\left(\left.\left(\boldsymbol{X}_{i}-\mathcal{X}\left(Z_{0,i},X_i(\boldsymbol{\beta}),\boldsymbol{\beta}\right)\right)\int_{0}^{1}\nabla G_v\left(Z_{0,i},X_{0,i}+\varsigma\boldsymbol{X}_{i}^{\mathrm{T}}\Delta\boldsymbol{\beta}\right)\boldsymbol{X}_{i}^{\mathrm{T}}\Delta\boldsymbol{\beta}d\varsigma\right|D_i = 1\right)\\
 & =\int_{0}^{1}E\left(\left.\nabla_v G\left(Z_{0,i},X_{0,i}+\varsigma\boldsymbol{X}_{i}^{\mathrm{T}}\Delta\boldsymbol{\beta}\right)\left(\boldsymbol{X}_{i}-\mathcal{X}\left(Z_{0,i},X_i(\boldsymbol{\beta}),\boldsymbol{\beta}\right)\right)\boldsymbol{X}_{i}^{\mathrm{T}}\right|D_i=1\right)d\varsigma\Delta\boldsymbol{\beta}.
\end{align*}

We finally prove the last result. Note that 
\begin{align*}
 \left\Vert\frac{1}{S_{n}}\sum_{i=1}^{n}D_{i}\left(\boldsymbol{X}_{i}-\widehat{\mathcal{X}}_{n}\left(\widehat{Z}_{i},\widehat{X}_{i}^{k},\boldsymbol{\beta}^k\right)\right)\varepsilon_{i} \right\Vert
  & \leq \left\Vert\frac{1}{S_{n}}\sum_{i=1}^{n}D_{i}\left(\boldsymbol{X}_{i}-\mathcal{X}\left(Z_{0,i},\widehat{X}_{i}^{k},\boldsymbol{\beta}^k\right)\right)\varepsilon_{i}\right\Vert \\
   & + \left\Vert \frac{1}{S_{n}}\sum_{i=1}^{n}D_{i}\left(\mathcal{X}\left(\widehat{Z}_{i},\widehat{X}_{i}^{k},\boldsymbol{\beta}^k\right) - \mathcal{X}\left(Z_{0,i},\widehat{X}_{i}^{k},\boldsymbol{\beta}^k\right) \right)\varepsilon_{i} \right\Vert\\
   & + \left\Vert\frac{1}{S_{n}}\sum_{i=1}^{n}D_{i}\left(\widehat{\mathcal{X}}_{n}\left(\widehat{Z}_{i},\widehat{X}_{i}^{k},\boldsymbol{\beta}^k\right) - \mathcal{X}\left(\widehat{Z}_{i},\widehat{X}_{i}^{k},\boldsymbol{\beta}^k\right) \right)\varepsilon_{i}\right\Vert,
\end{align*}
we obviously have that the second term on RHS of the inequality is of order $O_P(p_Z\sqrt{p_X/n})$, and the last term is of the same order as $\sup_{\nu_Z, \nu_X, \boldsymbol{\beta}\in\mathscr{B}} \left\Vert \widehat{\mathcal{X}}_{n}\left(\nu_Z, \nu_X,\boldsymbol{\beta}\right)-\mathcal{X}\left(\nu_Z, \nu_X,\boldsymbol{\beta}\right)\right\Vert$. So we only need to look at the first term. Note that for any fixed $\boldsymbol{\beta}$, $E\left(D_{i}\left(\boldsymbol{X}_{i}-\mathcal{X}\left(Z_{0,i},X_{i}(\boldsymbol{\beta}),\boldsymbol{\beta}\right)\right)\varepsilon_{i}\right) =0$, then further note that each argument of $D_{i}\left(\boldsymbol{X}_{i}-\mathcal{X}\left(Z_{0,i},X_{i}(\boldsymbol{\beta}),\boldsymbol{\beta}\right)\right)$  is upper bounded, and each argument's derivative with respect to $\boldsymbol{\beta}$ is bounded in norm by $C\sqrt{p_X}$, we have that the first term is of order $O_P(p_X\sqrt{\log(n)/n})$. This finishes the proof. 
\end{proof}

Based on \autoref{lemma3}--\autoref{lemma5}, we know that  
\begin{align}\label{simplied_dynamics_beta}
\sup_{k\geq 1} \left\Vert\Delta\widehat{\boldsymbol{\beta}}^{k+1} -\left( \boldsymbol{I}_{p_X} - \gamma \boldsymbol{\Psi}_S\left(\widehat{\boldsymbol{\beta}}^k\right)\right)\Delta\widehat{\boldsymbol{\beta}}^{k} \right\Vert = O_P\left(\Xi_{1,n}\right).
\end{align}
Following the proof of \autoref{theorem3.2-1}, we conclude the proof. 

\subsection{Proof of \autoref{theorem3.2-2}}
\begin{lemma}\label{lemma6}
Let \autoref{condition1}--\autoref{condition3} and \autoref{condition4}--\autoref{condition6} hold and $\Xi_{1,n}\rightarrow 0$, we have that 
\begin{align*}
&\sup_{k\geq k_S\left(n,\gamma\right)}  \left\Vert \frac{1}{S_{n}}\sum_{i=1}^{n}D_i\left(\boldsymbol{X}_{i}-\widehat{\mathcal{X}}_{n}\left(\widehat{Z}_{i},\widehat{X}_{i}^{k},\widehat{\boldsymbol{\beta}}^{k}\right)\right)\left(G\left(\widehat{Z}_{i},\widehat{X}_{i}^{k}\right)-G\left(Z_{0,i},\widehat{X}_{i}^{k}\right)\right)-\Sigma_{X,Z}\Delta\widehat{\boldsymbol{\delta}}\right\Vert \\
&=O_{p}\left(\frac{p_{Z}p_{X}\Xi_{1,n}}{\sqrt{n}}\right),
\end{align*}
where $\Sigma_{X,Z}=E\left(\nabla_{u}G\left(Z_{0,i},X_{0,i}\right)\left(\boldsymbol{X}_{i}-\mathcal{X}\left(Z_{0,i},X_{0,i},\boldsymbol{\beta}_{0}\right)\right)\boldsymbol{Z}_{i}^{\mathrm{T}}|D_i=1\right)$. 
\end{lemma}
\begin{proof}
Note that 
\begin{align*}
 & \sup_{k\geq k_S\left(n,\gamma\right)}\left\Vert \frac{1}{S_{n}}\sum_{i=1}^{n}D_i\left(\boldsymbol{X}_{i}-\widehat{\mathcal{X}}_{n}\left(\widehat{Z}_{i},\widehat{X}_{i}^{k},\widehat{\boldsymbol{\beta}}^{k}\right)\right)\left(G\left(\widehat{Z}_{i},\widehat{X}_{i}^{k}\right)-G\left(Z_{0,i},\widehat{X}_{i}^{k}\right)\right)-\Sigma_{X,Z}\Delta\widehat{\boldsymbol{\delta}}\right\Vert \\
 & \leq\left|\frac{n}{S_{n}}-P_{D}^{-1}\right|\sup_{k\geq k_S\left(n,\gamma\right)}\left\Vert \frac{1}{n}\sum_{i=1}^{n}D_i\left(\boldsymbol{X}_{i}-\widehat{\mathcal{X}}_{n}\left(\widehat{Z}_{i},\widehat{X}_{i}^{k},\widehat{\boldsymbol{\beta}}^{k}\right)\right)\left(G\left(\widehat{Z}_{i},\widehat{X}_{i}^{k}\right)-G\left(Z_{0,i},\widehat{X}_{i}^{k}\right)\right)\right\Vert (i)\\
 & +\sup_{k\geq k_S\left(n,\gamma\right)}\left\Vert \frac{1}{nP_D}\sum_{i=1}^{n}D_i\left(\widehat{\mathcal{X}}_{n} \left(\widehat{Z}_{i},\widehat{X}_{i}^{k},\widehat{\boldsymbol{\beta}}^{k}\right)-\mathcal{X}\left(Z_{0,i},X_{0,i},\boldsymbol{\beta}_{0}\right)\right)\left(G\left(\widehat{Z}_{i},\widehat{X}_{i}^{k}\right)-G\left(Z_{0,i},\widehat{X}_{i}^{k}\right)\right)\right\Vert (ii)\\
 & +\sup_{k\geq k_S\left(n,\gamma\right)}\left\Vert \frac{1}{nP_D}\sum_{i=1}^{n}D_i\left(\boldsymbol{X}_{i}-\mathcal{X}\left(Z_{0,i},X_{0,i},\boldsymbol{\beta}_{0}\right)\right)\left(\nabla_{u}G\left(\widetilde{Z}_i,\widehat{X}_{i}^{k}\right)-\nabla_{u}G\left(Z_{0,i},X_{0,i}\right)\right)\boldsymbol{Z}_{i}^{\mathrm{T}}\Delta\widehat{\boldsymbol{\delta}}\right\Vert (iii)\\
 & +\frac{1}{P_D}\left\Vert \frac{1}{n}\sum_{i=1}^{n}D_i\nabla_{u}G\left(Z_{0,i},X_{0,i}\right)\left(\boldsymbol{X}_{i}-\mathcal{X}\left(Z_{0,i},X_{0,i},\boldsymbol{\beta}_{0}\right)\right)\boldsymbol{Z}_{i}^{\mathrm{T}}\right.\\
 &\left. -E\left(D_i\nabla_{u}G\left(Z_{0,i},X_{0,i}\right)\left(\boldsymbol{X}_{i}-\mathcal{X}\left(Z_{0,i},X_{0,i},\boldsymbol{\beta}_{0}\right)\right)\boldsymbol{Z}_{i}^{\mathrm{T}}\right)\right\Vert\cdot \left\Vert\Delta\widehat{\boldsymbol{\delta}}\right\Vert (iv),
\end{align*}
where according to Taylor expansion, $\widetilde{Z}_i$ lies between $Z_{0,i}$ and $\widehat{Z}_{0,i}$. Obviously, with probability going to 1, $(i)\leq C\left|n/S_{n}-P_{D}^{-1}\right|\sqrt{p_X}\sup_{i}\left|\widehat{Z}_{i}-Z_{0,i}\right|=O_{p}(p_{Z}\sqrt{p_{X}}/n)$.
For $(ii)$, we have that 
\[
(ii)\leq C\sup_{k\geq k_S\left(n,\gamma\right),i}\left\Vert \widehat{\mathcal{X}}_{n} \left(\widehat{Z}_{i},\widehat{X}_{i}^{k},\boldsymbol{\beta}^{k}\right)-\mathcal{X} \left(Z_{0,i},X_{0,i},\boldsymbol{\beta}_{0}\right)\right\Vert \sup_{1\leq i\leq n}\left|\widehat{Z}_{i}-Z_{0,i}\right|.
\]
Obviously, 
\begin{align*}
  \sup_{k\geq k_S\left(n,\gamma\right),i}\left\Vert \widehat{\mathcal{X}}_{n} \left(\widehat{Z}_{i},\widehat{X}_{i}^{k},\boldsymbol{\beta}^{k}\right)-\mathcal{X} \left(Z_{0,i},X_{0,i},\boldsymbol{\beta}_{0}\right)\right\Vert
  & \leq\sup_{k\geq k_S\left(n,\gamma\right),i}\left\Vert \widehat{\mathcal{X}}_{n}\left(\widehat{Z}_{i},\widehat{X}_{i}^{k},\widehat{\boldsymbol{\beta}}^{k}\right)-\mathcal{X}\left(\widehat{Z}_i,\widehat{X}_{i}^{k},\widehat{\boldsymbol{\beta}}^{k}\right)\right\Vert \\
 & +\sup_{k\geq k_S\left(n,\gamma\right),i}\left\Vert \mathcal{X}\left(\widehat{Z}_i,\widehat{X}_{i}^{k},\widehat{\boldsymbol{\beta}}^{k}\right)-\mathcal{X}\left(Z_{0,i},\widehat{X}_{i}^{k},\widehat{\boldsymbol{\beta}}^{k}\right)\right\Vert \\
 & +\sup_{k\geq k_S\left(n,\gamma\right),i}\left\Vert \mathcal{X}\left(Z_{0,i},\widehat{X}_{i}^{k},\widehat{\boldsymbol{\beta}}^{k}\right)-\mathcal{X}\left(Z_{0,i},X_{0,i},\widehat{\boldsymbol{\beta}}^{k}\right)\right\Vert\\
 & +\sup_{k\geq k_S\left(n,\gamma\right),i}\left\Vert \mathcal{X}\left(Z_{0,i},X_{0,i},\widehat{\boldsymbol{\beta}}^{k}\right)-\mathcal{X}\left(Z_{0,i},X_{0,i},\boldsymbol{\beta}_0\right)\right\Vert
\end{align*}
which is obviously $O_P(\sqrt{p_X}\Xi_{1,n})$. So $(ii)$ is of order $O_{P}(p_{Z}\sqrt{p_{X}/n}\Xi_{1,n})$.
$(iii)$ is, up to some constant, bounded by $\sqrt{p_Xp_Z}\max_{i}|\nabla_{u}G(\widehat{Z}_i,\widehat{X}_{i}^{k})-\nabla_{u}G\left(Z_{0,i},X_{0,i}\right)|\Vert \Delta\widehat{\boldsymbol{\delta}}\Vert $.
This implies that $(iii)$ is of order $O_{P}\left(p_{Z}p_{X}\Xi_{1,n}/\sqrt{n}\right)$.
Finally, the last term $(iv)$ is obviously of order $O_{p}(p_{Z}\sqrt{p_Zp_{X}}/n)$.
Together we have shown the result. 
\end{proof}
\begin{lemma}\label{lemma7}
Let \autoref{condition1}--\autoref{condition3} and \autoref{condition4}--\autoref{condition6} hold, and $\Xi_{1,n}\rightarrow 0$, we have that 
\begin{align*}
 &\sup_{k\geq k_S\left(n,\gamma\right)} \left\Vert \frac{1}{S_{n}}\sum_{i=1}^{n}D_i\left(\boldsymbol{X}_{i}-\widehat{\mathcal{X}}_{n}\left(\widehat{Z}_{i},\widehat{X}_{i}^{k},\widehat{\boldsymbol{\beta}}^{k}\right)\right)\left(G\left(Z_{0,i},\widehat{X}_{i}^{k}\right)-G\left(Z_{0,i},X_{0,i}\right)\right)-\boldsymbol{\Psi}_S\left(\boldsymbol{\beta}_{0}\right)\Delta\widehat{\boldsymbol{\beta}}^{k}\right\Vert \\
&=O_{P}\left(p_{X}\sqrt{p_X}\Xi_{1,n}^{2}\right).
\end{align*}
\end{lemma}
\begin{proof}
Note that similar to the proof of the above lemma, we have that 
\begin{align*}
 & \sup_{k\geq k_S\left(n,\gamma\right)}\left\Vert \frac{1}{S_{n}}\sum_{i=1}^{n}D_i\left(\boldsymbol{X}_{i}-\widehat{\mathcal{X}}_{n}\left(\widehat{Z}_{i},\widehat{X}_{i}^{k},\beta^{k}\right)\right)\left(G\left(Z_{0,i},\widehat{X}_{i}^{k}\right)-G\left(Z_{0,i},X_{0,i}\right)\right)-\boldsymbol{\Psi}_S\left(\boldsymbol{\beta}_{0}\right)\Delta\widehat{\boldsymbol{\beta}}^{k}\right\Vert \\
 & \leq \left|\frac{n}{S_{n}}-P_{D}^{-1}\right|\sup_{k\geq k_S\left(n,\gamma\right)}\left\Vert \frac{1}{n}\sum_{i=1}^{n}D_i\left(\boldsymbol{X}_{i}-\widehat{\mathcal{X}}_{n}\left(\widehat{Z}_{i},\widehat{X}_{i}^{k},\widehat{\boldsymbol{\beta}}^{k}\right)\right)\left(G\left(Z_{0,i},\widehat{X}_{i}^{k}\right)-G\left(Z_{0,i},X_{0,i}\right)\right)\right\Vert (i)\\
 & +\sup_{k\geq k_S\left(n,\gamma\right)}\left\Vert \frac{1}{nP_D}\sum_{i=1}^{n}D_i\left(\widehat{\mathcal{X}}_{n}\left(\widehat{Z}_{i},\widehat{X}_{i}^{k},\widehat{\boldsymbol{\beta}}^{k}\right)-\mathcal{X}\left(Z_{0,i},X_{0,i},\boldsymbol{\beta}_{0}\right)\right)\left(G\left(Z_{0,i},\widehat{X}_{i}^{k}\right)-G\left(Z_{0,i},X_{0,i}\right)\right)\right\Vert (ii)\\
 & +\sup_{k\geq k_S\left(n,\gamma\right)}\left\Vert \frac{1}{nP_D}\sum_{i=1}^{n}D_i\left(\boldsymbol{X}_{i}-\mathcal{X}\left(Z_{0,i},X_{0,i},\boldsymbol{\beta}_{0}\right)\right)\left(\nabla_{v}G\left(Z_{0,i},\widetilde{X}_{i}^k\right)-\nabla_{v}G\left(Z_{0,i},X_{0,i}\right)\right)\boldsymbol{X}_{i}^{\mathrm{T}}\Delta\widehat{\boldsymbol{\beta}}^{k}\right\Vert (iii)\\
 & +\frac{1}{P_D}\left\Vert \left(\frac{1}{n}\sum_{i=1}^n D_i\nabla_{v}G\left(Z_{0,i},X_{0,i}\right)\left(\boldsymbol{X}_{i}-\mathcal{X}\left(Z_{0,i},X_{0,i},\boldsymbol{\beta}_{0}\right)\right)\boldsymbol{X}_{i}^{\mathrm{T}}\right.\right.\\
 & \left. \left. - E\left(D_i\left(\boldsymbol{X}_{i}-\mathcal{X}\left(Z_{0,i},X_{0,i},\boldsymbol{\beta}_{0}\right)\right)\nabla_{v}G\left(Z_{0,i},X_{0,i}\right)\boldsymbol{X}_{i}^{\mathrm{T}}\right)\right)\Delta\widehat{\boldsymbol{\beta}}^{k}\right\Vert (iv),
\end{align*}
where $\widetilde{X}_i^k$ is between $\widehat{X}_i^k$ and $X_{0,i}$. Obviously, term $(i)$ is of order $O_{P}(p_{X}\Xi_{1,n}/\sqrt{n})$,
term $(ii)$ is of order $O_{P}(p_{X}\Xi_{1,n}^{2})$,
term $(iii)$ is of order $O_{P}(p_X\sqrt{p_X}\Xi_{1,n}^{2})$,
and term $(iv)$ is of order $O_{P}(p_{X}\Xi_{1,n}/\sqrt{n})$. It remains to note that $\Delta\boldsymbol{\beta}_0 = 0$, 
this proves the result. 
\end{proof}
\begin{lemma}\label{lemma8}
Let \autoref{condition1}--\autoref{condition3}, \autoref{condition4}--\autoref{condition6} hold and $\Xi_{1,n}\rightarrow 0$, we have that 
\begin{align*}
 & \sup_{k\geq k_S\left(n,\gamma\right)}\left\Vert \frac{1}{S_{n}}\sum_{i=1}^{n}D_{i}\left(X_{i}-\widehat{\mathcal{X}}_{n} \left(\widehat{Z}_{i},\widehat{X}_{i}^{k},\widehat{\boldsymbol{\beta}}^{k}\right)\right)\varepsilon_{i}-\frac{1}{nP_{D}}\sum_{i=1}^{n}D_{i}\left(X_{i}-\mathcal{X}\left(Z_{0,i},X_{0,i},\boldsymbol{\beta}_{0}\right)\right)\varepsilon_{i}\right\Vert \\
 & =O_P\left(q^2C(0,q)^2\mathscr{R}_X(q) + \sqrt{p_{X}^3}q^{6}C(0,q)^4C_{\Phi,1,q}^2\Xi_{1,n}^{2}+\sqrt{p_{X}^{5}}q^{6}C(0,q)^{2}(C_{\boldsymbol{\Phi},2,q}C_{\boldsymbol{\Phi}, q}+C(1,q)^2)^2\Xi_{1,n}^{4}\right).
\end{align*}
\end{lemma}
\begin{proof}
We only need to bound the following  
\[
\sup_{k\geq k_S\left(n,\gamma\right)}\left\Vert \frac{1}{n}\sum_{i=1}^{n}D_i\left(\widehat{\mathcal{X}}_{n} \left(\widehat{Z}_{i},\widehat{X}_{i}^{k},\widehat{\boldsymbol{\beta}}^{k}\right)-\mathcal{X}\left(Z_{0,i},X_{0,i},\boldsymbol{\beta}_{0}\right)\right)\varepsilon_{i}\right\Vert. 
\]
Using element-wise Taylor expansion, we have that 
\[
\boldsymbol{\Phi}_{q}\left(\widehat{Z}_{j},\widehat{X}_{j}^{k}\right)=\boldsymbol{\Phi}_{q}\left(Z_{0,j},X_{0,j}\right)+\underset{\boldsymbol{\Phi}\left(1,j,n\right)}{\underbrace{\nabla_{u}\boldsymbol{\Phi}_{q}\left(Z_{0,j},X_{0,j}\right)\boldsymbol{Z}_{j}^{\mathrm{T}}\Delta\widehat{\boldsymbol{\delta}}}}+\underset{\boldsymbol{\Phi}\left(2,j,n,k\right)}{\underbrace{\nabla_{v}\boldsymbol{\Phi}_{q}\left(Z_{0,j},X_{0,j}\right)\boldsymbol{X}_{j}^{\mathrm{T}}\Delta\widehat{\boldsymbol{\beta}}^{k}}}+\mathcal{R}_{\boldsymbol{\Phi},q,j}^{k},
\]
where $\sup_j\Vert\boldsymbol{\Phi}(1,j,n)\Vert = O_P(p_ZqC(1,q)/\sqrt{n})$,   $\sup_{j,k\geq k_S(n,\gamma)}\Vert\boldsymbol{\Phi}(2,j,n,k)\Vert = O_P(\sqrt{p_X}qC(1,q)\Xi_{1,n})$,  and  $\sup_{j,k\geq k_S(n,\gamma)}\left\Vert \mathcal{R}_{\boldsymbol{\Phi},q,j}^{k}\right\Vert =O_{P}\left(p_{X}qC(2,q)\Xi_{1,n}^{2}\right)$. Note that all the above terms are $o_P(1)$ when $\Xi_{2,n}\rightarrow 0$.  We also have that
and that 
\begin{align*}
\widehat{\boldsymbol{\Gamma}}_{n,q}\left(\widehat{\boldsymbol{\delta}},\boldsymbol{\widehat{\beta}}^{k}\right) & =\boldsymbol{\Gamma}_{q}\left(\boldsymbol{\beta}_{0}\right) + \underset{\boldsymbol{\Gamma}\left(1,n\right)}{\underbrace{\boldsymbol{\Gamma}_{q}\left(\boldsymbol{\beta}_{0}\right) - S_n^{-1}\sum_{j=1}^nD_i\boldsymbol{\Phi}_{q}\left(Z_{0,j},X_{0,j}\right)\boldsymbol{\Phi}_{q}\left(Z_{0,j},X_{0,j}\right)^{\mathrm{T}} }}\\
& +\underset{\boldsymbol{\Gamma}\left(2,n\right)}{\underbrace{S_{n}^{-1}\sum_{j=1}^{n}D_{j}\nabla_{u}\left[\boldsymbol{\Phi}_{q}\left(Z_{0,j},X_{0,j}\right)\boldsymbol{\Phi}_{q}\left(Z_{0,j},X_{0,j}\right)^{\mathrm{T}}\right]\boldsymbol{Z}_{j}^{\mathrm{T}}\Delta\widehat{\boldsymbol{\delta}}}}\\
 & +\underset{\boldsymbol{\Gamma}\left(3,n,k\right)}{\underbrace{S_{n}^{-1}\sum_{i=1}^{n}D_{i}\nabla_{v}\left[\boldsymbol{\Phi}_{q}\left(Z_{0,j},X_{0,j}\right)\boldsymbol{\Phi}_{q}\left(Z_{0,j},X_{0,j}\right)^{\mathrm{T}}\right]\boldsymbol{X}_{j}^{\mathrm{T}}\Delta\widehat{\boldsymbol{\beta}}^{k}}}+\boldsymbol{\Gamma}(4,n,k),
\end{align*}
where $\left\Vert \boldsymbol{\Gamma}\left(1,n\right)\right\Vert =O_{P}\left(q^2C(0,q)^{2}/\sqrt{n}\right)$, $\left\Vert \boldsymbol{\Gamma}\left(2,n\right)\right\Vert =O_{P}\left(p_{Z}q^{2}C(0,q)C(1,q)/\sqrt{n}\right)$,
$\sup_{k\geq k_S(n,\gamma)}\left\Vert \boldsymbol{\Gamma}\left(3,n,k\right)\right\Vert =O_{P}\left(\sqrt{p_{X}}q^{2}C(0,q)C(1,q)\Xi_{1,n}\right)$,
and $\sup_{k\geq k_S(n,\gamma)}\left\Vert \boldsymbol{\Gamma}(4,n,k) \right\Vert=O_P\left(p_Xq^2(C(2,q)C(0,q) + C(1,q)^2)\Xi_{1,n}^2\right)$.
Since $\Xi_{2,n}\rightarrow0$, all of the above terms are of $o_{P}\left(1\right)$.
Then note that 
\begin{align*}
\widehat{\boldsymbol{\Gamma}}_{n,q}^{-1}\left(\widehat{\boldsymbol{\delta}},\widehat{\boldsymbol{\beta}}^{k}\right)-\boldsymbol{\Gamma}_{q}^{-1}\left(\boldsymbol{\beta}_{0}\right) & =\boldsymbol{\Gamma}_{q}^{-1}\left(\boldsymbol{\beta}_{0}\right)\left(\boldsymbol{\Gamma}_{q}\left(\boldsymbol{\beta}_{0}\right)-\widehat{\boldsymbol{\Gamma}}_{n,q}\left(\boldsymbol{\widehat{\delta}},\widehat{\boldsymbol{\beta}}^{k}\right)\right)\left(\widehat{\boldsymbol{\Gamma}}_{n,q}^{-1}\left(\widehat{\boldsymbol{\delta}},\boldsymbol{\widehat{\beta}}^{k}\right)-\boldsymbol{\Gamma}_{q}^{-1}\left(\boldsymbol{\beta}_{0}\right)\right)\\
 & +\boldsymbol{\Gamma}_{q}^{-1}\left(\boldsymbol{\beta}_{0}\right)\left(\boldsymbol{\Gamma}_{q}\left(\boldsymbol{\beta}_{0}\right)-\widehat{\boldsymbol{\Gamma}}_{n,q}\left(\widehat{\boldsymbol{\delta}},\boldsymbol{\beta}^{k}\right)\right)\boldsymbol{\Gamma}_{q}^{-1}\left(\boldsymbol{\beta}_{0}\right).
\end{align*}
Define 
\[
\mathcal{R}^k_{\boldsymbol{\Gamma},q,n}=\widehat{\boldsymbol{\Gamma}}_{n,q}^{-1}\left(\widehat{\boldsymbol{\delta}},\widehat{\boldsymbol{\beta}}^{k}\right)-\boldsymbol{\Gamma}_{q}^{-1}\left(\boldsymbol{\beta}_{0}\right)+\boldsymbol{\Gamma}_{q}^{-1}\left(\boldsymbol{\beta}_{0}\right)\left(\boldsymbol{\Gamma}\left(1,n\right)+\boldsymbol{\Gamma}\left(2,n\right)+\boldsymbol{\Gamma}\left(3,n,k\right)\right)\boldsymbol{\Gamma}_{q}^{-1}\left(\boldsymbol{\beta}_{0}\right)
\]
we have that with probability going to 1,  \begin{align*}
\sup_{k\geq k_S(n,\gamma)}\left\Vert \mathcal{R}^k_{\boldsymbol{\Gamma},q,n}\right\Vert & \leq C \left\Vert \boldsymbol{\Gamma}_{q}\left(\boldsymbol{\beta}_{0}\right)-\widehat{\boldsymbol{\Gamma}}_{n,q}\left(\widehat{\boldsymbol{\delta}},\boldsymbol{\beta}^{k}\right)\right\Vert^2 + C\left\Vert\boldsymbol{\Gamma}(4,n,k)\right\Vert\\
& = O_{P}\left(p_{X}q^{4}C(0,q)^2C(1,q)^{2}\Xi_{1,n}^{2}+p_Xq^2(C(2,q)C(0,q) + C(1,q)^2)\Xi_{1,n}^2\right).\end{align*}
Then using the expansion of $\boldsymbol{\Phi}_{q}\left(\widehat{Z}_{j},\widehat{X}_{j}^{k}\right)$
and $\widehat{\boldsymbol{\Gamma}}_{n,q}^{-1}\left(\widehat{\boldsymbol{\delta}},\boldsymbol{\widehat{\beta}}^{k}\right)$,
we have that 
\begin{align*}
&\widehat{\mathcal{X}}_{n}\left(\widehat{Z}_{i},\widehat{X}_{i}^{k},\widehat{\boldsymbol{\beta}}^{k}\right)\\ 
&=\left(S_{n}^{-1}\sum_{j=1}^{n} D_{j}\boldsymbol{X}_{j}\left(\boldsymbol{\Phi}_{q}\left(Z_{0,j},X_{0,j}\right)+\boldsymbol{\Phi}\left(1,j,n\right)+\boldsymbol{\Phi}\left(2,j,n,k\right)+\mathcal{R}_{\boldsymbol{\Phi},n,j}^k\right)^{\mathrm{T}}\right)\\
 & \times\left( \boldsymbol{\Gamma}_{ q}^{-1}\left(\boldsymbol{\beta}_{0}\right)-\boldsymbol{\Gamma}_{q}^{-1}\left(\boldsymbol{\beta}_{0}\right)\left(\boldsymbol{\Gamma}\left(1,n\right)+\boldsymbol{\Gamma}\left(2,n\right)+\boldsymbol{\Gamma}\left(3,n,k\right)\right)\boldsymbol{\Gamma}_{q}^{-1}\left(\boldsymbol{\beta}_{0}\right)+\mathcal{R}_{\boldsymbol{\Gamma},q,n}^k\right)\\
 & \times\left(\boldsymbol{\Phi}_{q}\left(Z_{0,i},X_{0,i}\right)+\boldsymbol{\Phi}\left(1,i,n\right)+\boldsymbol{\Phi}\left(2,i,n,k\right)+\mathcal{R}_{\boldsymbol{\Phi},n,i}^k\right)\\
& =S_{n}^{-1}\sum_{j=1}^{n} D_{j}\boldsymbol{X}_{j}\boldsymbol{\Phi}_{q}\left(Z_{0,j},X_{0,j}\right)^{\mathrm{T}}\boldsymbol{\Gamma}_{q}^{-1}\left(\boldsymbol{\beta}_{0}\right)\boldsymbol{\Phi}_{q}\left(Z_{0,i},X_{0,i}\right)\\
& +S_{n}^{-1}\sum_{j=1}^{n} D_{j}\boldsymbol{X}_{j}\left(\boldsymbol{\Phi}\left(1,j,n\right)+\boldsymbol{\Phi}\left(2,j,n,k\right)\right)^{\mathrm{T}}\boldsymbol{\Gamma}_{q}^{-1}\left(\boldsymbol{\beta}_{0}\right)\boldsymbol{\Phi}_{q}\left(Z_{0,i},X_{0,i}\right)\\
& -S_{n}^{-1}\sum_{j=1}^{n} D_{j}\boldsymbol{X}_{j}\boldsymbol{\Phi}_{q}\left(Z_{0,j},X_{0,j}\right)^{\mathrm{T}}\boldsymbol{\Gamma}_{q}^{-1}\left(\boldsymbol{\beta}_{0}\right)\left(\boldsymbol{\Gamma}\left(1,n\right)+\boldsymbol{\Gamma}\left(2,n\right)+\boldsymbol{\Gamma}\left(3,n,k\right)\right)\boldsymbol{\Gamma}_{q}^{-1}\left(\boldsymbol{\beta}_{0}\right)\boldsymbol{\Phi}_{q}\left(Z_{0,i},X_{0,i}\right)\\
 & +S_{n}^{-1}\sum_{j=1}^{n}D_{j}\boldsymbol{X}_{j}\boldsymbol{\Phi}_{q}\left(Z_{0,j},X_{0,j}\right)^{\mathrm{T}}\boldsymbol{\Gamma}_{q}^{-1}\left(\boldsymbol{\beta}_{0}\right)\left(\boldsymbol{\Phi}\left(1,i,n\right)+\boldsymbol{\Phi}\left(2,i,n,k\right)\right) +\mathcal{R}_{\mathcal{X},n,i}^k,
\end{align*}
where 
\[
\sup_{k\geq k_S(n,\gamma),i}\left\Vert \mathcal{R}_{\mathcal{X},n,i}^k\right\Vert =O_{P}\left(p_X\sqrt{p_{X}}q^4C(0,q)^{3}\Xi_{1,n}^{2}\left(q^{2}C(0,q)C(1,q)^2+C(2,q)\right)\right).
\]
Also note that 
\begin{align*}
 & S_{n}^{-1}\sum_{j=1}^{n}D_{j}\boldsymbol{X}_{j}\boldsymbol{\Phi}_{q}\left(Z_{0,j},X_{0,j}\right)^{\mathrm{T}}\boldsymbol{\Gamma}_{q}^{-1}\left(\boldsymbol{\beta}_{0}\right)\boldsymbol{\Phi}_{q}\left(Z_{0,i},X_{0,i}\right)-\mathcal{X}\left(Z_{0,i},X_{0,i},\boldsymbol{\beta}_{0}\right)\\
 & =\left(S_{n}^{-1}\sum_{j=1}^{n}D_{j}\boldsymbol{X}_{j}\boldsymbol{\Phi}_{q}\left(Z_{0,j},X_{0,j}\right)^{\mathrm{T}}\boldsymbol{\Gamma}_{q}^{-1}\left(\boldsymbol{\beta}_{0}\right)-\boldsymbol{\Pi}_{q}^{\boldsymbol{X}}(\boldsymbol{\beta}_0)\right)\boldsymbol{\Phi}_{q}\left(Z_{0,i},X_{0,i}\right)\\
&+\left(\mathcal{X}\left(Z_{0,i},X_{0,i},\boldsymbol{\beta}_{0}\right)-\boldsymbol{\Pi}_{q}^{\boldsymbol{X}}(\boldsymbol{\beta}_0)\boldsymbol{\Phi}_{q}\left(Z_{0,i},X_{0,i}\right)\right)\\
 & =\left(S_{n}^{-1}\sum_{j=1}^{n}D_{j}\boldsymbol{X}_{j}\boldsymbol{\Phi}_{q}\left(Z_{0,j},X_{0,j}\right)^{\mathrm{T}}-\frac{1}{P_{D}}E\left(D_{j}\boldsymbol{X}_{j}\boldsymbol{\Phi}_{q}\left(Z_{0,j},X_{0,j}\right)^{\mathrm{T}}\right)\right)\boldsymbol{\Gamma}_{q}^{-1}\left(\boldsymbol{\beta}_{0}\right)\boldsymbol{\Phi}_{q}\left(Z_{0,i},X_{0,i}\right)\\
 & +\left(E\left(\left.\boldsymbol{X}_{j}\boldsymbol{\Phi}_{q}\left(Z_{0,j},X_{0,j}\right)^{\mathrm{T}}\right|D_{j}=1\right)\boldsymbol{\Gamma}_{q}^{-1}\left(\boldsymbol{\beta}_{0}\right)-\boldsymbol{\Pi}_{q}^{\boldsymbol{X}}(\boldsymbol{\beta}_0)\right)\boldsymbol{\Phi}_{q}\left(Z_{0,i},X_{0,i}\right)\\
 & +\left(\mathcal{X}\left(Z_{0,i},X_{0,i},\boldsymbol{\beta}_{0}\right)-\boldsymbol{\Pi}_{q}^{\boldsymbol{X}}(\boldsymbol{\beta}_0)\boldsymbol{\Phi}_{q}\left(Z_{0,i},X_{0,i}\right)\right).
\end{align*}
Obviously, 
\[
\left\Vert \left(E\left(\left.\boldsymbol{X}_{j}\boldsymbol{\Phi}_{q}\left(Z_{0,j},X_{0,j}\right)^{\mathrm{T}}\right|D_{j}=1\right)\boldsymbol{\Gamma}_{q}^{-1}\left(\boldsymbol{\beta}_{0}\right)-\boldsymbol{\Pi}_{q}^{\boldsymbol{X}}(\boldsymbol{\beta}_0)\right)\boldsymbol{\Phi}_{q}\left(Z_{0,i},X_{0,i}\right)\right\Vert \leq q^2C(0,q)^2\mathscr{R}_X(q),
\]
\[
\left\Vert \mathcal{X}\left(Z_{0,i},X_{0,i},\boldsymbol{\beta}_{0}\right)-\boldsymbol{\Pi}_{q}^{\boldsymbol{X}}(\boldsymbol{\beta}_0)\boldsymbol{\Phi}_{q}\left(Z_{0,i},X_{0,i}\right)\right\Vert \leq\mathscr{R}_X(q).
\]
Moreover,
\begin{align*}
 & S_{n}^{-1}\sum_{j=1}^{n}D_{j}\boldsymbol{X}_{j}\boldsymbol{\Phi}_{q}\left(Z_{0,j},X_{0,j}\right)^{\mathrm{T}}-\frac{1}{P_{D}}E\left(D_{j}\boldsymbol{X}_{j}\boldsymbol{\Phi}_{q}\left(Z_{0,j},X_{0,j}\right)^{\mathrm{T}}\right)\\
 & =\left(\frac{1}{S_{n}/n}-\frac{1}{P_{D}}\right)\frac{1}{n}\sum_{j=1}^{n}D_{j}\boldsymbol{X}_{j}\boldsymbol{\Phi}_{q}\left(Z_{0,j},X_{0,j}\right)^{\mathrm{T}}\\
 & +\frac{1}{P_{D}}\left(\frac{1}{n}\sum_{j=1}^{n}D_{j}\boldsymbol{X}_{j}\boldsymbol{\Phi}_{q}\left(Z_{0,j},X_{0,j}\right)^{\mathrm{T}}-E\left(D_{j}X_{j}\boldsymbol{\Phi}_{q}\left(Z_{0,j},X_{0,j}\right)^{\mathrm{T}}\right)\right).
\end{align*}
Then 
\begin{align*}
&\left\Vert \left(\frac{1}{S_{n}/n}-\frac{1}{P_{D}}\right)\frac{1}{n}\sum_{j=1}^{n}D_{j}\boldsymbol{X}_{j}\boldsymbol{\Phi}_{q}\left(Z_{0,j},X_{0,j}\right)^{\mathrm{T}}\boldsymbol{\Gamma}_{q}^{-1}\left(\boldsymbol{\beta}_{0}\right)\frac{1}{n}\sum_{i=1}^{n}\boldsymbol{\Phi}_{q}\left(Z_{0,i},X_{0,i}\right)\varepsilon_{i}\right\Vert\\
&=O_{P}\left(\sqrt{p_X}q^2C(0,q)^2/n\right),
\end{align*}
and 
\begin{align*}
 & \left\Vert \left(\frac{1}{n}\sum_{j=1}^{n}D_{j}\boldsymbol{X}_{j}\boldsymbol{\Phi}_{q}\left(Z_{0,j},X_{0,j}\right)^{\mathrm{T}}-E\left(D_{j}\boldsymbol{X}_{j}\boldsymbol{\Phi}_{q}\left(Z_{0,j},X_{0,j}\right)^{\mathrm{T}}\right)\right)\boldsymbol{\Gamma}_{q}^{-1}\left(\boldsymbol{\beta}_{0}\right)\frac{1}{n}\sum_{i=1}^{n}\boldsymbol{\Phi}_{q}\left(Z_{0,i},X_{0,i}\right)\varepsilon_{i}\right\Vert \\
 & =O_{P}\left(\sqrt{p_{X}}q^{2}C(0,q)^{2}/n\right).
\end{align*}
The above implies that 
\begin{align*}
&\left\Vert\sum_{i=1}^n\left(\frac{1}{S_n}\sum_{j=1}^{n}D_{j}\boldsymbol{X}_{j}\boldsymbol{\Phi}_{q}\left(Z_{0,j},X_{0,j}\right)^{\mathrm{T}}\boldsymbol{\Gamma}_{q}^{-1}\left(\boldsymbol{\beta}_{0}\right)\boldsymbol{\Phi}_{q}\left(Z_{0,i},X_{0,i}\right)-\mathcal{X}\left(Z_{0,i},X_{0,i},\boldsymbol{\beta}_{0}\right)\right)\varepsilon_i\right\Vert\\
& = O_P\left(\frac{q^2C(0,q)^2\mathscr{R}_X(q)}{\sqrt{n}} + \frac{\sqrt{p_X}q^2C(0,q)^2}{n}\right).
\end{align*}
Now we are ready to derive the result. Note that 
\begin{align*}
 & \left\Vert \frac{1}{S_{n}}\sum_{i=1}^{n}D_{i}\left(\widehat{\mathcal{X}}_{n}\left(\widehat{Z}_{i},\widehat{X}_{i}^{k},\widehat{\boldsymbol{\beta}}^{k}\right)-\mathcal{X}\left(Z_{0,i},X_{0,i},\boldsymbol{\beta}_{0}\right)\right)\varepsilon_{i}\right\Vert \\
 & \leq\left\Vert \frac{1}{S_{n}}\sum_{i=1}^{n}D_{i}\left(\frac{1}{S_{n}}\sum_{i=1}^{n}D_{j}\boldsymbol{X}_{j}\boldsymbol{\Phi}_{q}\left(Z_{0,j},X_{0,j}\right)^{\mathrm{T}}\boldsymbol{\Gamma}_{q}^{-1}\left(\boldsymbol{\beta}_{0}\right)\boldsymbol{\Phi}_{q}\left(Z_{0,i},X_{0,i}\right)-\mathcal{X}\left(Z_{0,i},X_{0,i},\boldsymbol{\beta}_{0}\right)\right)\varepsilon_{i}\right\Vert (i) \\
 & +\left\Vert \frac{1}{S_{n}}\sum_{i=1}^{n}D_{j}\boldsymbol{X}_{j}\left(\boldsymbol{\Phi}\left(1,j,n\right)+\boldsymbol{\Phi}\left(2,j,n,k\right)\right)^{\mathrm{T}}\boldsymbol{\Gamma}_{q}^{-1}\left(\boldsymbol{\beta}_{0}\right)\frac{1}{S_{n}}\sum_{i=1}^{n}D_{i}\boldsymbol{\Phi}_{q}\left(Z_{0,i},X_{0,i}\right)\varepsilon_{i}\right\Vert (ii) \\
 & +\left\Vert \frac{1}{S_{n}}\sum_{i=1}^{n}D_{j}\boldsymbol{X}_{j}\boldsymbol{\Phi}_{q}\left(Z_{0,j},X_{0,j}\right)^{\mathrm{T}}\left(\boldsymbol{\Gamma}_{q}^{-1}\left(\boldsymbol{\beta}_{0}\right)\left(\boldsymbol{\Gamma}\left(1,n\right)+\boldsymbol{\Gamma}\left(2,n\right)+\boldsymbol{\Gamma}\left(3,n,k\right)\right)\right)\boldsymbol{\Gamma}_{q}^{-1}\left(\boldsymbol{\beta}_{0}\right)\frac{1}{S_{n}}\sum_{i=1}^{n}D_{i}\boldsymbol{\Phi}_{q}\left(Z_{0,i},X_{0,i}\right)\varepsilon_{i}\right\Vert (iii) \\
 & +\left\Vert \frac{1}{S_{n}}\sum_{j=1}^{n}D_{j}\boldsymbol{X}_{j}\boldsymbol{\Phi}_{q}\left(Z_{0,j},X_{0,j}\right)^{\mathrm{T}}\boldsymbol{\Gamma}_{q}^{-1}\left(\boldsymbol{\beta}_{0}\right)\frac{1}{S_{n}}\sum_{i=1}^{n}D_{i}\left(\boldsymbol{\Phi}\left(1,i,n\right)+\boldsymbol{\Phi}\left(2,i,n,k\right)\right)\varepsilon_{i}\right\Vert (iv) \\
 &+ \left\Vert \frac{1}{S_{n}}\sum_{i=1}^{n}D_{i}\mathcal{R}_{\mathcal{X},n,i}^k\varepsilon_{i}\right\Vert (v) .
\end{align*}
We have that 
(i) is obviously of order $O_{P}\left(\sqrt{p_{X}}q^{2}C(0,q)^{2}/n+q^2C(0,q)^2\mathscr{R}_X(q)/\sqrt{n}\right)$, and 
\begin{align*}
  \left \Vert(ii)\right\Vert & \leq C\left\Vert \frac{1}{S_{n}}\sum_{i=1}^{n}D_{j}\boldsymbol{X}_{j}\left(\boldsymbol{\Phi}\left(1,j,n\right)+\boldsymbol{\Phi}\left(2,j,n,k\right)\right)\right\Vert\left\Vert\frac{1}{S_{n}}\sum_{i=1}^{n}D_{i}\boldsymbol{\Phi}_{q}\left(Z_{0,i},X_{0,i}\right)\varepsilon_{i}\right\Vert\\
  & =O_P\left(\frac{p_Xq^2C(0,q)C(1,q)\Xi_{1,n}}{\sqrt{n}}\right), 
\end{align*}
\begin{align*}
    \left\Vert (iii) \right\Vert & \leq \left\Vert \frac{1}{S_{n}}\sum_{i=1}^{n}D_{j}\boldsymbol{X}_{j}\boldsymbol{\Phi}_{q}\left(Z_{0,j},X_{0,j}\right)^{\mathrm{T}}\right\Vert\left\Vert \boldsymbol{\Gamma}\left(1,n\right)+\boldsymbol{\Gamma}\left(2,n\right)+\boldsymbol{\Gamma}\left(3,n,k\right)\right\Vert\left\Vert\frac{1}{S_{n}}\sum_{i=1}^{n}D_{i}\boldsymbol{\Phi}_{q}\left(Z_{0,i},X_{0,i}\right)\varepsilon_{i}\right\Vert\\
    & =O_P\left(\frac{p_Xq^4C(0,q)^3C(1,q)\Xi_{1,n}}{\sqrt{n}}\right),
\end{align*}
\begin{align*}
    \left\Vert (iv) \right\Vert &\leq C\sqrt{p_X}qC(0,q)\left\Vert\left(\frac{1}{S_n}\sum_{i=1}^n \nabla_{u}\boldsymbol{\Phi}_{q}\left(Z_{0,i},X_{0,i}\right)\boldsymbol{Z}_{i}^{\mathrm{T}}\varepsilon_i\right)\Delta\widehat{\boldsymbol{\delta}} +\left(\frac{1}{S_n}\sum_{i=1}^n \nabla_{v}\boldsymbol{\Phi}_{q}\left(Z_{0,i},X_{0,i}\right)\boldsymbol{X}_{i}^{\mathrm{T}}\varepsilon_i\right)\Delta\widehat{\boldsymbol{\beta}}^k \right\Vert\\
&=O_P\left(\frac{p_Xq^2C(0,q)C(1,q)\Xi_{1,n}}{\sqrt{n}}\right), 
\end{align*}
and 
\[
\left\Vert(v)\right\Vert\leq \sup_{i,k\geq k_S(n,\gamma)} \left\Vert  \mathcal{R}_{\mathcal{X},n,i,k}  \right\Vert =O_{P}\left(p_X\sqrt{p_{X}}q^4C(0,q)^{3}\Xi_{1,n}^{2}\left(q^{2}C(0,q)C(1,q)^2+C(2,q)\right)\right).
\]
This finishes the proof. 
\end{proof}

\autoref{theorem3.2-2} can be proved if we combine the results in \autoref{lemma6}--\autoref{lemma8}.

\subsection{Proof of \autoref{corollary1}}
The  result is obvious so is left out.

\subsection{Proof of \autoref{theorem4}}

\begin{proof}
  Write $\boldsymbol{\varphi}_0 = (\boldsymbol{\varphi}_{c,0}^{\mathrm{T}},\boldsymbol{\varphi}_{d,0}^{\mathrm{T}})^{\mathrm{T}}$.   To show the point identification of $\boldsymbol{\varphi}_0$, we note that since \autoref{condition11} (iii) holds, $\nabla_{r_0}E(T|\boldsymbol{R}_{e})$ and $\nabla_{\boldsymbol{R}_{c}} E(T|\boldsymbol{R}_{e})$ can be consistently estimated uniformly for all interior point $\boldsymbol{R}_e\in\mathcal{R}_e$, so in the following analysis we can assume that   $E(T|\boldsymbol{R}_{e})$ is known for all all $\boldsymbol{R}_e\in\mathcal{R}_e$, and both $\nabla_{r_0} E(T|\boldsymbol{R}_{e})$,  and $\nabla_{\partial\boldsymbol{R}_{c}} E(T|\boldsymbol{R}_{e})$ are known for interior points  $\boldsymbol{R}_e\in\mathcal{R}_e$. According to \autoref{condition11} (iv) we can find interior point $\boldsymbol{R}_e^1$ such that $\nabla_{  r_0 } E(T|\boldsymbol{R}_e^1)\neq 0$. Note that $\nabla_{ r_{0}} E(T|\boldsymbol{R}_e^1) = \nabla_{w} F_W\left(r_{0}^1 + (\boldsymbol{R}^1)^{\mathrm{T}}\boldsymbol{\varphi}_0\right) $ and  
 $
\nabla_{r_{c, j} } E(T|\boldsymbol{R}_e^1)= \nabla_{w} F_W\left(r_{0}^1 + (\boldsymbol{R}^1)^{\mathrm{T}}\boldsymbol{\varphi}_0\right)  \varphi_{c,j,0} $ for $  1\leq j \leq p_R^c$,  
    where $p_R^c$ is the dimension of $\boldsymbol{R}_c$. This implies that we can identify $\varphi_{c,j,0}$ by 
    \[
    \varphi_{c,j,0} = \frac{\nabla_{ r_{c, j}} E(T|\boldsymbol{R}_e^1)}{\nabla_{ r_0} E(T|\boldsymbol{R}_e^1)  }, \ 1\leq j \leq p_R^c 
    \]
    To identify $\boldsymbol{\varphi}_{d}$, we first state a useful result. For any function $h(t)$ which is nondecreasing and has continuous derivative, if $\nabla_{} h(t)>0$ for $t = t_0$, then $h(t_0) = h(t_1)$ must imply $t_0 = t_1$. To prove such result, we note that 
    $h(t_1) - h(t_0) = \int_{0}^1 \nabla_{} h(t_0 + \tau (t_1 - t_0))d\tau (t_1 - t_0)$. Since $h$ has continuous  derivative,   $\nabla_{}h(t)$ is lower bounded by some positive constant for a neighborhood of $t_0$. Then since $\nabla  h(t)\geq 0$, we have that $\int_{0}^1 \nabla_{} h(t_0 + \tau (t_1 - t_0))d\tau>0$. So if $t_1>t_0$, we have that $h(t_1)>h(t_0)$, which leads to a contradiction. Given the above result, we note that when \autoref{condition11} (iv) and (v) hold, we have that $r_{0}^2 + (\boldsymbol{R}_{c}^2)^{\mathrm{T}}\boldsymbol{\varphi}_{c,0} - r_{0}^{j} - (\boldsymbol{R}_{c}^j)^{\mathrm{T}}\boldsymbol{\varphi}_{c,0} = (\boldsymbol{R}_d^j - \boldsymbol{R}_d^2)^{\mathrm{T}}\boldsymbol{\varphi}_{d,0}$ for $3\leq j \leq p_R^d+2$, so  
    \[
    \begin{pmatrix}
       r_{0}^2 + (\boldsymbol{R}_{c}^2)^{\mathrm{T}}\boldsymbol{\varphi}_{c,0} - r_{0}^{3} - (\boldsymbol{R}_{c}^3)^{\mathrm{T}}\boldsymbol{\varphi}_{c,0} \\
       \vdots\\
       r_{0}^2 + (\boldsymbol{R}_{c}^2)^{\mathrm{T}}\boldsymbol{\varphi}_{c,0} - r_{0}^{p_R^d+2} - (\boldsymbol{R}_{c}^{p_R^d+2})^{\mathrm{T}}\boldsymbol{\varphi}_{c,0}
       \end{pmatrix} 
       = \begin{pmatrix}
       (\boldsymbol{R}_d^3 - \boldsymbol{R}_d^2)^{\mathrm{T}} \\
       \vdots\\
       ( \boldsymbol{R}_d^{p_R^d+2} - \boldsymbol{R}_d^{2})^{\mathrm{T}}
    \end{pmatrix} \boldsymbol{\varphi}_{d,0}
    \]
    So nonsingularity of $(\boldsymbol{R}_d^3 -\boldsymbol{R}_d^2 , \cdots, \boldsymbol{R}_d^{p_R^{d}+2}-\boldsymbol{R}_d^2)$ guarantees identification of $\boldsymbol{\varphi}_{d,0}$. 

    When \autoref{condition12} further holds, using the previous method we  can identify $\boldsymbol{\delta}_0$ so we can deal with the case where $R_0$ and $Z_0$ are known.  For any $R_0$ and $Z_0$, $E(T|R_0) = F_W (R_0)$,  $E(D|T = 0, R_0, Z_0) = (F_U(Z_0) - F_{W,U}(R_0,Z_0))/(1 - F_W(R_0))$, where $F_U$ is the marginal CDF of $U$, and $E(D|T = 1, R_0, Z_0) =  F_{W,U}(R_0,Z_0+\tau_{1,0})/F_W(R_0)$. So for any interior point $(R_0, \boldsymbol{Z}^{\mathrm{T}})^{\mathrm{T}} \in \overline{\mathcal{Z}}_e$, $F_U(Z_0) - F_{W,U}(R_0,Z_0)$ and $F_{W,U}(R_0,Z_0+\tau_{1,0})$ can be identified by $E(D|T =0, {R}_0,{Z}_0)(1-E(T|R_0))$ and $E(D|T =1, {R}_0,{Z}_0) E(T|R_0)$. Also, note that 
\[
    -\nabla_{R_0} [E(D|T =0, {R}_0,{Z}_0)(1-E(T|R_0))]   = -\nabla_{R_0 } (F_U(Z_0)-F_{W,U}(R_0, Z_0)) = \nabla_{ w} F_{W,U}(R_0,Z_0), \]
    and 
     \[
    \nabla_{R_0 } [E(D|T =1, {R}_0,{Z}_0) E(T|R_0)] = \nabla_{R_0} F_{W,U}(R_0, Z_0+\tau_{1,0})  = \nabla_{w} F_{W,U}(R_0,Z_0+\tau_{1,0}), 
    \]
    Note that $\nabla_{w} F_{W,U}(w, u) $ is nondecreasing with respect to $u$ because $\nabla_{wu} F_{W,U}(w, u) $ is the joint density of $W$ and $U$ and is nonnegative. Then since \autoref{condition12}(vi) requires that $\nabla_{wu} F_{W,U}(R_0^{\prime},Z_0^{\prime}) >0$, and  $\nabla_{w} F_{W,U}(R_0^{\prime},Z_0^{\prime})  = \nabla_{w} F_{W,U}(R_0^{\prime},Z_0^{\prime\prime}+\tau_{1,0}) $, we have that $Z_0^{\prime} = Z_{0}^{\prime\prime} + \tau_{1,0}$ must hold because $\nabla_{wu} F_{W,U}(w,u) $ is continuous according to \autoref{condition12}(i). So $\tau_{1,0}$ is  identified by $\tau_{1,0} = Z_{0}^{\prime}  - Z_{0}^{\prime\prime}$. 

    Finally, similar to the identification of $\tau_{1,0}$, note that \[E(Y|T =0, D=1, {R}_0,{Z}_0, X_0)P(T=0, D=1|R_0, X_0) = F_{U,V}(Z_0, X_0) - F_{W,U,V}(R_0, Z_0,X_0)\]
    and
    \[E(Y|T =1, D=1, {R}_0,{Z}_0, X_0)P(T=1, D=1|R_0, X) = F_{W,U,V}(R_0, Z_0,X_0+\tau_{2,0}),\]
    where $F_{U,V}$ is the joint density of $U$ and $V$. So 
    \[
    -\nabla_{R_0,Z_0}\left( E(Y|T =0, D=1, {R}_0,{Z}_0, X_0)P(T=0, D=1|R_0, Z_0)\right) = \nabla_{wu} F_{W,U,V}(R_0, Z_0, X_0) ,
    \]
    and
    \[
    \nabla_{R_0,Z_0}\left( E(Y|T =1, D=1, {R}_0,{Z}_0, X_0)P(T=1, D=1|R_0, Z_0)\right) =  \nabla_{wu} F_{W,U,V}(R_0, Z_0, X_0 +\tau_{2,0}),
    \]
    So \autoref{condition13}(vi) implies that 
    $\nabla_{wu} F_{W,U,V}(R_0^{\prime}, Z_0^{\prime}, X_0^{\prime})  = \nabla_{wu} F_{W,U,V}(R_0^{\prime}, Z_0^{\prime}, X_0^{\prime\prime} +\tau_{2,0}) $. Since $\nabla_{wu} F_{W,U,V}(w, u, v) $ is nondecreasing with respect to $v$,  $\nabla_{wuv} F_{W,U,V}(R_0, Z_0,   X_0) $ is continuous, and $\nabla_{wuv} F_{W,U,V}(R_0^{\prime}, Z_0^{\prime},   X_0^{\prime}) $, we have  that $X_0^{\prime} = X_0^{\prime\prime} +\tau_{2,0}$.  So $\tau_{2,0}$ can be identified by $\tau_{2,0} =X_0^{\prime} -  X_0^{\prime\prime}$. 
\end{proof}

\end{spacing}
\end{document}